\documentstyle[11pt,epsfig]{article}
\newcommand{\beq}{\begin{equation}}
\newcommand{\eeq}{\end{equation}}
\newcommand{\beqa}{\begin{eqnarray}}
\newcommand{\eeqa}{\end{eqnarray}}
\newcommand{\beqan}{\begin{eqnarray*}}
\newcommand{\eeqan}{\end{eqnarray*}}
\newcommand{\ba}{\begin{array}}
\newcommand{\ea}{\end{array}}
\newcommand{\ben}{\begin{enumerate}}
\newcommand{\een}{\end{enumerate}}
\newcommand{\bfl}{\begin{flushleft}}
\newcommand{\efl}{\end{flushleft}}
\newcommand{\btab}{\begin{tabular}}
\newcommand{\etab}{\end{tabular}}
\newcommand{\bit}{\begin{itemize}}
\newcommand{\eit}{\end{itemize}}
\newcommand{\bdes}{\begin{description}}
\newcommand{\edes}{\end{description}}
\newcommand{\bdm}{\begin{displaymath}}
\newcommand{\edm}{\end{displaymath}}
\newcommand{\nl}{\nonumber \\}
\newcommand{\no}{\nonumber}
\newcommand{\ul}{\underline}
\newcommand{\ol}{\overline}
\newcommand{\ra}{\rightarrow}
\newcommand{\Ra}{\Rightarrow}
\newcommand{\ve}{\varepsilon}
\newcommand{\vp}{\varphi}

\newcommand{\dg}{\dagger}
\newcommand{\wt}{\widetilde}
\newcommand{\wh}{\widehat}

\newcommand{\Ha}{{\cal H}}

\newcommand{\cL}{{\cal L}}
\newcommand{\M}{{\cal M}}

\newcommand{\dfrac}{\displaystyle \frac}

\newcommand{\lets}{\stackrel{<}{_\sim}}
\newcommand{\del}{\partial}

\newcommand{\Fsl}{\not\!\!}

\newcommand{\toG}{\stackrel{G}{\to}}
\newcommand{\dis}{\displaystyle}
\newcommand{\mtiny}[1]{{\mbox{\tiny #1}}}

\begin{document}
\begin{titlepage}
\begin{flushright}
UWThPh-1994-49\\
Dec. 31, 1994
\end{flushright}

\vspace{2cm}

\begin{center}
{\LARGE \bf
CHIRAL PERTURBATION THEORY*}\\[50pt]
G. Ecker  \\
Institut f\"ur Theoretische Physik \\
Universit\"at Wien \\
Boltzmanngasse 5, A-1090 Wien, Austria
\vfill
{\bf Abstract} \\
\end{center}
\noindent
The main elements and methods of chiral perturbation theory,
the effective field theory of the Standard Model below the scale
of spontaneous chiral symmetry breaking, are summarized. Applications
to the interactions of mesons and baryons at low energies are reviewed,
with special emphasis on developments of the last three years. Among the
topics covered are the strong, electromagnetic and semileptonic weak
interactions of mesons at and beyond next--to--leading order in the
chiral expansion, nonleptonic weak interactions of mesons, virtual
photon corrections and the meson--baryon system. The discussion is limited to
processes at zero temperature, for infinite volume and with at most one baryon.

\vfill
\begin{center}
To appear in \\[5pt]
Progress in Particle and Nuclear Physics, Vol. 35 \\[5pt]
Pergamon Press, Oxford
\end{center}

\vfill
\noindent * Work supported in part by FWF (Austria), Project No. P09505--PHY
and by HCM, EEC--Contract No. CHRX--CT920026 (EURODA$\Phi$NE)
\end{titlepage}

\tableofcontents

\vspace*{1cm}

\renewcommand{\thesection}{\arabic{section}}
\renewcommand{\thesubsection}{\arabic{section}.\arabic{subsection}}
\renewcommand{\theequation}{\arabic{section}.\arabic{equation}}
\renewcommand{\thetable}{\arabic{section}.\arabic{table}}
\renewcommand{\thefigure}{\arabic{section}.\arabic{figure}}

\setcounter{equation}{0}
\setcounter{subsection}{0}
\setcounter{table}{0}
\setcounter{figure}{0}

\section{EFFECTIVE FIELD THEORIES}
\label{sec:EFT}
Effective field theories have become valuable tools in many branches
of physics, in particular in particle and nuclear physics. The notion
of effective field theories seems to suggest a basic difference
to fundamental field theories. In the same spirit, a distinction is often
made between (fundamental) renormalizable and (effective)
non--renormalizable quantum field theories. These traditional discriminations
tend to mask the common features. Firstly, all known quantum
field theories of particle and nuclear physics are probably
effective field theories in the sense that they are low--energy approximations
of some underlying, more  ``fundamental" theories. Secondly, all
quantum field theories in four space--time dimensions have to be
renormalized, whether they be renormalizable or
non--renormalizable in the standard terminology. The renormalization
procedure transfers the unknown structure of the theory at short distances
(high energies) to a few low--energy constants (LECs). As long
as we are only interested in phenomena at comparatively low
energies, all the short--distance structure is encoded in those LECs.

Although all quantum field theories have to be renormalized,
they differ in their sensitivity to the high--energy structure
of the underlying theory. It is convenient to distinguish between two
categories of quantum field theories.
\bdes
\item[i.] {Asymptotically free theories} \\
These ultraviolet stable theories are the only candidates for truly
fundamental quantum field theories. Nothing in their structure indicates a
limiting energy beyond which they can no more be applied. Most likely, they
are also the only ones that can be defined by constructive quantum
field theory in a mathematically complete way (Balaban, 1987, 1988, 1989).
\item[ii.]  {Ultraviolet unstable theories} \\
Such theories contain information about their limited validity. This
information may be of little practical use as for QED where the
limiting energy is much higher than the scale where QED is embedded in a more
fundamental theory, the Standard Model (Glashow, 1961;
Weinberg, 1967; Salam, 1968). The information may be of phenomenological
relevance as for the Higgs sector of the Standard Model: the mathematical
concept of ``triviality" becomes physical reality by yielding an
upper limit for the Higgs mass. The non--renormalizable theories
belong to this category. They differ from the renormalizable members
of this class only insofar as the loop expansion produces
new LECs at every order. The limitation of these theories to the
low--energy domain is more manifest than for the
renormalizable ones, but in principle similar. It probably does not even
make sense to construct ultraviolet unstable theories in axiomatic
field theory. They have to be modified at short distances both for
physical and for mathematical reasons. Nevertheless, it is both useful
and mathematically well defined to treat them in perturbation theory as
long as one is sufficiently below the problematic scale. The convergence
of the perturbation expansion is physically irrelevant. The real
issue is how well the asymptotic expansion works at low energies.
\edes

Effective field theories are the quantum field theoretical
implementation of the quantum ladder. As the energy increases and smaller
distances are probed, new degrees of freedom become relevant
that must be included in the theory. At the same time,
other fields may lose their status of fundamental fields as the
corresponding states are recognized as bound states of the new degrees
of freedom. Conversely, as the energy is lowered, some degrees of freedom are
frozen out and disappear from the accessible spectrum of states.
To model the effective field theory at low energies, we rely especially
on the symmetries of the ``fundamental" underlying theory, in addition
to the usual axioms of quantum field theory embodied in an
effective Lagrangian. This Lagrangian must contain {\it all} terms
allowed by the symmetries of the fundamental theory
for the given set of fields (Weinberg, 1979).
This completeness guarantees that the effective theory
is indeed the low--energy limit of the fundamental theory.

One can distinguish two kinds of effective field theories.
\subsection*{A. Decoupling effective field theories}
For energies below a certain scale $\Lambda$, all heavy (with respect
to $\Lambda$) degrees of freedom are integrated out leaving only
the light degrees of freedom in the effective theory. No light
particles are generated in the transition from the fundamental
to the effective level. The effective Lagrangian has the general
form
\beq
\cL_{\rm eff} = \cL_{d\leq 4} + \sum_{d>4} \frac{1}{\Lambda^{d-4}}
\sum_{i_d} g_{i_d} O_{i_d}  \label{eq:EFT}
\eeq
where $\cL_{d\leq 4}$ contains the potentially renormalizable terms
with operator dimension $d \leq 4$, the $g_{i_d}$ are dimensionless
coupling constants expected to be of $O(1)$,
and the $O_{i_d}$ are monomials
in the light fields with operator dimension $d$. At energies
much below $\Lambda$,
corrections due to the non--renormalizable parts ($d>4$) are suppressed
by powers of $E/\Lambda$. For $E/\Lambda$ small enough, $\cL_{d\leq 4}$
can be regarded as the ``fundamental'' Lagrangian at low energies.
There are many examples of this type:
QED for $E \ll m_e$ (Euler, 1936; Heisenberg and Euler, 1936), the Fermi
theory of weak interactions for $E \ll M_W$ and, although we do not
know the relevant scale in this case, the Standard Model itself.
There are many candidates for an underlying theory at smaller distances
(composite Higgs, SUSY, grand unification, superstrings,\dots).
With the exception of the Higgs sector, the Standard Model does
not provide any clues for the scale $\Lambda$. There is also no
experimental evidence for terms in the effective Lagrangian with
$d > 4$.
\subsection*{B. Non--decoupling effective field theories}
The transition from the fundamental to the effective level occurs through a
phase transition via the
spontaneous breakdown of a symmetry generating light
($M \ll \Lambda$) pseudo--Goldstone bosons. Since a spontaneously broken
symmetry relates processes with different numbers of Goldstone bosons,
the distinction between renormalizable ($d\leq 4$) and non--renormalizable
($d>4$) parts in the effective Lagrangian like in (\ref{eq:EFT})
becomes meaningless in general. With some exceptions (see Sect.~\ref
{subsec:nlr}), the effective Lagrangian of a non--decoupling theory is
generically non--renormalizable. Nevertheless, as already emphasized, such
Lagrangians define perfectly consistent quantum field theories
at sufficiently low energies. Instead of the operator dimension as
in (\ref{eq:EFT}), the number of derivatives of the fields distinguishes
successive terms in the Lagrangian.

The general structure of effective Lagrangians with spontaneously broken
symmetries is largely independent of the specific physical
realization. This is exemplified by two examples in particle physics,
but there are many more in condensed matter physics (Leutwyler, 1994b;
Fradkin, 1991).
\bdes
\item[a.] {The Standard Model without Higgs bosons} \\
Even if there is no fundamental Higgs boson, the gauge symmetry
$SU(2)\times U(1)$ can be spontaneously broken to $U(1)_{\rm em}$
(heavy--Higgs scenario). As a manifestation of the universality of Goldstone
boson interactions, the scattering of longitudinal gauge vector bosons
is in first approximation analogous to $\pi\pi$ scattering.
\item[b.] {The Standard Model for $E \ll \mbox{1 GeV}$} \\
At low energies, the relevant degrees of freedom of the Standard
Model are not quarks and gluons, but the pseudoscalar mesons
and other hadrons. The pseudoscalar mesons play a special role as
the pseudo--Goldstone bosons of spontaneously broken chiral
symmetry. The Standard Model in the hadronic sector
at low energies is described by a non--decoupling effective theory
called chiral perturbation theory (CHPT).
\edes
The similarities in these two examples are evident. Although the
renormalizable Higgs model is a part of the Standard Model,
nature has made a different choice
in the low--energy realization of the Standard Model. The analogue of the
Higgs model, the linear sigma model, is known to be only a toy model that
cannot be regarded as the effective theory of QCD at low energies. As will be
discussed in Sect.~\ref{sec:chi}, the constraints of
renormalizability are too restrictive for a realistic low--energy
approximation of QCD.

As the effective field theory of the Standard Model at low energies,
CHPT is becoming the common
language of nuclear and low--energy particle physics. Following
Weinberg (1990, 1991), CHPT is now also used in such genuine nuclear
domains as the nucleon--nucleon force where the traditional potential
approach is being merged with concepts of effective field theory.
Except for a brief excursion in Sect.~\ref{sec:baryons}, this review
covers applications of CHPT to low--energy reactions of mesons
and baryons with at most one incoming and one outgoing baryon
(at temperature $T=0$ and for infinite volume).
In Sect.~\ref{sec:chi}, the basic elements of chiral symmetry
and its spontaneous breaking are put together. The lowest--order Lagrangians
are listed for all those sectors that will be covered in this review.
Generalized CHPT is contrasted with the standard approach. The strong,
electromagnetic and semileptonic weak interactions to $O(p^4)$
in the chiral expansion are the subject of Sect.~\ref{sec:strong}.
Among the topics considered are the values of the light quark
masses, the values and the interpretation of the LECs of
$O(p^4)$ and the odd--intrinsic--parity sector. Recent phenomenological
applications to pion--pion scattering, semileptonic kaon decays,
meson decays to a lepton pair and $\eta$ decays are reviewed. The
nonleptonic weak interactions are treated in Sect.~\ref{sec:nonleptonic}.
To $O(p^4)$, both the dominant kaon decay modes to pions as well
as the large class of radiative nonleptonic decays are
discussed. The manifestation of the chiral anomaly in the nonleptonic
weak sector is examined. Sect.~\ref{sec:p6} summarizes recent
and ongoing work beyond next--to--leading order in the
meson sector. The calculation of the generating functional
of $O(p^6)$ is outlined, in
particular the two--loop diagrams appearing at that order. In addition
to the recent complete $O(p^6)$ calculation of $\gamma\gamma \to
\pi^0\pi^0$, attempts to determine the dominant higher--order
corrections for the decay $K_L \to \pi^0 \gamma\gamma$ are also discussed.
Recent work on virtual photons in the mesonic sector is described in Sect.~\ref
{sec:virtual}. Applications include electromagnetic corrections to
meson masses and isospin violating transitions like semileptonic
$\eta$ decays. Finally, Sect.~\ref{sec:baryons} tries to
summarize the large body of work that has been done in the meson--baryon
system during recent years. On the theoretical side, the
heavy baryon approach to CHPT and the renormalization of the generating
functional to $O(p^3)$ are discussed. Most of the detailed work
in this sector has been done in the framework of chiral $SU(2)$,
e.g., in pion--nucleon scattering and in the prospering field of photo--nucleon
reactions. The three--flavour meson--baryon sector [chiral $SU(3)$]
is also considered where most investigations have concentrated on the leading
non--analytic chiral corrections.

The purpose of this review is to give a self--contained introduction
to the methods and a summary of recent applications of CHPT. Concerning the
applications, the main emphasis will be on developments of the last three
years till the end of 1994. More material, both on the
foundations and methods as well as on applications of CHPT, can
be found in the following books, reviews, lecture notes and proceedings
of specialized workshops and conferences:
Bernstein and Holstein (1995), de Rafael (1995), Maiani et al. (1995, 1992),
Leutwyler (1994d, 1993, 1991),
Ecker (1994a, 1993a), Mei\ss ner (1994c, 1993, 1992), Bijnens (1993a),
Pich (1993), Donoghue et al. (1992a), Gasser (1990), Georgi (1984).

\setcounter{equation}{0}
\setcounter{subsection}{0}
\setcounter{table}{0}
\setcounter{figure}{0}

\section{CHIRAL SYMMETRY}
\label{sec:chi}
\subsection{Nonlinear realization}
\label{subsec:nlr}
The QCD Lagrangian with $N_f$ ($N_f=$ 2 or 3) massless quarks $q=(u,d,\dots)$
\beqa
{\cL}^0_{\rm QCD} &=& \ol{q} i \gamma^\mu\left(\partial_\mu + i g_s {\lambda_
\alpha\over 2} G^\alpha_\mu\right)q - {1\over 4}G^\alpha_{\mu\nu}
G^{\alpha\mu\nu} + \cL_{\mbox{\tiny heavy quarks}} \label{eq:QCD0}\\*
&=& \ol{q_L} i \Fsl{D} q_L + \ol{q_R} i \Fsl{D} q_R - {1\over 4}G^\alpha_
{\mu\nu}G^{\alpha\mu\nu} + \cL_{\mbox{\tiny heavy quarks}} \no \\*
q_{R,L} &=& {1\over 2}(1 \pm \gamma_5)q \no
\eeqa
has a global symmetry
$$
\underbrace{SU(N_f)_L \times SU(N_f)_R}_{\mbox{chiral group $G$}}
\times U(1)_V \times U(1)_A ~.
$$
At the effective hadronic level, the quark
number symmetry $U(1)_V$ is realized as baryon number. The axial
$U(1)_A$ is not a symmetry at the quantum level due to the  Abelian
anomaly ('t Hooft, 1976; Callan et al., 1976; Crewther, 1977).
The Noether currents of the chiral group G are
\beq
J_A^{a\mu} = \ol{q_A} \gamma^\mu {\lambda_a \over 2} q_A \qquad
(A=L,R ~; ~a=1,\dots, N_f^2-1) \label{eq:Ncurrents}
\eeq
with associated Noether charges
\beq
Q^a_A =  \int d^3x J^{a0}_A  ~.
\eeq

A classical symmetry can be realized in quantum field theory
in two different ways depending on how the vacuum responds
to a symmetry transformation.
All theoretical and phenomenological evidence suggests that the
chiral group $G$ is spontaneously broken to the vectorial subgroup
$SU(N_f)_V$. The axial generators of $G$ are non--linearly realized
and there are $N_f^2 - 1$ massless pseudoscalar Goldstone bosons
(Goldstone, 1961). There is a well--known procedure (Coleman et al.,
1969; Callan et al., 1969) how to realize a
spontaneously broken symmetry on quantum fields. In the special case
of chiral symmetry with its parity transformation, the Goldstone
fields can be collected in a unitary matrix field $U(\vp)$ transforming as
\beq
U(\vp) \toG g_R U(\vp) g_L^{-1} ~,\qquad
(g_L,g_R) \in G \label{Utr}
\eeq
under chiral rotations. There are different parametrizations of $U(\vp)$
corresponding to different choices of coordinates for the chiral
coset space $SU(N_f)_L\times SU(N_f)_R / SU(N_f)_V$. A
convenient choice is the exponential parametrization (for $N_f=3$)
\beq
U(\vp)=\exp{(i\lambda_a \vp^a/F)} ~,\qquad
{1 \over \sqrt{2}}\lambda_a \vp^a = \left( \ba{ccc}
\dfrac{\pi^0}{\sqrt{2}} + \dfrac{\eta_8}{\sqrt{6}} & \pi^+ &  K^+ \\*
\pi^- & -\dfrac{\pi^0}{\sqrt{2}} + \dfrac{\eta_8}{\sqrt{6}} &  K^0 \\*
K^- & \ol{K^0} & - \dfrac{2 \eta_8}{\sqrt{6}} \ea \right)~, \label{eq:Uphi}
\eeq
where $F$ will turn out to be the meson decay constant in the
chiral limit.

 A more basic quantity from a geometrical point of view
is another matrix field $u(\vp)$, the square root of $U(\vp)$
with the standard choice of coset coordinates. Its chiral transformation
\beq
u(\vp) \toG g_R u(\vp) h(g,\vp)^{-1}
= h(g,\vp) u(\vp) g_L^{-1} \label{eq:coset}
\eeq
introduces the so--called compensator field $h(g,\vp)$ representing an
element of the conserved subgroup $SU(N_f)_V$. For $g\in SU(N_f)_V$,
i.e. for $g_L = g_R$, the compensator $h(g)$ is a usual unitary representation
matrix, independent of the Goldstone fields $\vp$ (linear
representation \`a la Wigner--Weyl). For a proper chiral transformation
($g_L\not= g_R$), on the other hand, $h(g,\vp)$ does depend on $\vp$
(non--linear realization \`a la Nambu--Goldstone).

It is instructive to see the abstract quantities introduced
above emerge naturally in a specific example. The
simplest such example is the linear $\sigma$--model (Schwinger,
1957; Gell-Mann and L\'evy, 1960). It is
a seeming counterexample to the classification of Sect. \ref{sec:EFT}
because it is a renormalizable quantum field
theory describing the spontaneous breaking of chiral symmetry.
Rewriting it in the form of a non--decoupling
effective field theory will bring the ingredients of spontaneously broken
chiral symmetry
to the surface. The exercise will also demonstrate the price of
renormalizability: although it has the right symmetries by construction,
the linear $\sigma$--model is not general enough to describe the real world
(Gasser and Leutwyler, 1984). It serves the purpose of a toy model, but it
should not be mistaken for the effective field theory of QCD at low energies.

We rewrite the $\sigma$--model Lagrangian for the pion--nucleon system
\beq
\cL_\sigma = {1\over 2} \left(\del_\mu\sigma\del^\mu\sigma +
 \del_\mu\vec{\pi}\del^\mu\vec{\pi}\right) -{\lambda\over 4}
\left(\sigma^2 + \vec{\pi}^2 - v^2\right)^2 + \ol\psi
 ~i\not\!\partial\psi - g \ol\psi \left(\sigma + i\vec{\tau}\vec{\pi}
\gamma_5\right)\psi \label{eq:Lsig1}
\eeq
$$
\psi = { p \choose n}
$$
in the form
\beq
\cL_\sigma = {1\over 4}\langle \del_\mu\Sigma\del^\mu\Sigma \rangle
- {\lambda\over 16} \left( \langle \Sigma^\dg\Sigma \rangle - 2 v^2\right)
^2 + \ol{\psi_L} ~i\not\!\partial\psi_L + \ol{\psi_R} ~i\not\!\partial\psi_R
- g \ol{\psi_R} \Sigma\psi_L - g \ol{\psi_L} \Sigma^\dg\psi_R
\label{eq:Lsig2} \eeq
$$
\Sigma = \sigma{\bf 1} - i\vec{\tau}\vec{\pi} ~,\quad\qquad
\langle A \rangle = tr A
$$
to exhibit the chiral symmetry $G=SU(2)_L\times SU(2)_R$~:
$$
\psi_A \toG g_A \psi_A~,\qquad g_A \in SU(2)_A \quad(A=L,R)~,
\qquad \Sigma \toG g_R\Sigma g_L^{-1}~.
$$
For $v^2>0$, the chiral symmetry is spontaneously broken and the
``physical" fields are the massive field $\hat\sigma = \sigma - v$
and the Goldstone fields $\vec \pi$. The Lagrangian
with its non--derivative couplings for the fields $\vec\pi$ seems to be
at variance with the Goldstone theorem predicting a vanishing amplitude
whenever the momentum of a Goldstone boson goes to zero.

In order to make the Goldstone theorem manifest in the Lagrangian,
we perform a field transformation from the original fields
$\psi$, $\sigma$, $\vec\pi$ to a new set $\Psi$, $S$, $\vec\vp$
through a polar decomposition of the matrix field $\Sigma$:
\beq
\Sigma = (v+S)U(\vp) ~,\quad\qquad \Psi_L=u\psi_L ~,\qquad
\Psi_R=u^\dg \psi_R \label{eq:ftrans}
\eeq
$$
S^\dg = S ~,\qquad U^\dg=U^{-1} ~,\qquad \det U=1 ~,\qquad U=u^2~.
$$
{}From the chiral transformation properties of the coset elements $u(\vp)$ in
(\ref{eq:coset}) one finds
\beq
U\toG g_R U g_L^{-1}~, \qquad S\toG S~, \qquad \Psi_A\toG h(g,\vp)\Psi_A
\quad (A=L,R)~.
\eeq
In the new fields, the $\sigma$--model Lagrangian (\ref{eq:Lsig1})
takes the form
\beqa
\cL &=& {v^2\over 4}(1 + {S\over v})^2\langle u_\mu u^\mu \rangle\no\\
&+& \ol\Psi ~i \not\!\nabla \Psi + {1\over 2} \ol\Psi
\not\!u \gamma_5\Psi - g(v+S)\ol\Psi \Psi + \ldots \label{eq:Lsig3}
\eeqa
with a covariant derivative $\nabla = \del + \Gamma$ and
\beqa
u_\mu(\vp) &=& i (u^\dg \partial_\mu u - u \partial_\mu u^\dg)
= i u^\dg \partial_\mu U u^\dg \no\\
\Gamma_\mu(\vp) &=& \frac{1}{2} (u^\dg \partial_\mu u
+ u \partial_\mu u^\dg) ~.
\label{eq:Gu}
\eeqa
The Lagrangian (\ref{eq:Lsig3}) allows a clear separation
between the model independent Goldstone boson interactions
induced by spontaneous chiral symmetry breaking and the model dependent
part involving the scalar field $S$ [the kinetic term and the
self--couplings of the scalar field are omitted in (\ref{eq:Lsig3})].

We can draw the following conclusions:
\bdes
\item[i.]
The Goldstone theorem is now manifest at the Lagrangian level: the
Goldstone bosons $\vec\vp$ contained in the matrix fields $u_\mu(\vp)$,
$\Gamma_\mu(\vp)$ have derivative couplings only.
\item[ii.]
S--matrix elements are unchanged under the field
transformation (\ref{eq:ftrans}), but the Green functions are very
different. For instance, in the pseudoscalar meson sector the field $S$
does not contribute at all
at lowest order, $O(p^2)$, whereas $\hat\sigma$ exchange is essential
to repair the damage done by the non--derivative couplings of the
$\vec\pi$.
\item[iii.]
The manifest renormalizability of the Lagrangian (\ref{eq:Lsig1}) has
been traded for the manifest chiral structure of (\ref{eq:Lsig3}).
Of course, the Lagrangian (\ref{eq:Lsig3}) is still renormalizable,
but this renormali\-zability has its price. It requires specific
relations between various couplings that have nothing to do with
chiral symmetry and, which is worse, are not in agreement with
experiment. For instance, the model contains the Goldberger--Treiman
relation (Goldberger and Treiman, 1958) in the form ($m$ is the nucleon mass)
\beq
m = g v \equiv g_{\pi NN} F_\pi~.
\eeq
Thus, instead of the physical value $g_A=1.26$ for the axial--vector
coupling constant $g_A$ the model has $g_A=1$ [compare with the CHPT Lagrangian
(\ref{eq:piN1}) in Sect.~\ref{subsec:low}]. As already
emphasized, the problems with the linear $\sigma$--model are even more severe
in the meson sector (Gasser and Leutwyler, 1984).
\item[iv.]
The transformation of the linear into the nonlinear $\sigma$--model is
also a manifestation of the universality of the lowest--order effective
Lagrangian for Goldstone fields. Every spontaneously broken symmetry
of a Poincar\'e invariant quantum field theory will give rise to
a nonlinear $\sigma$--model for the interaction of Goldstone
bosons at lowest order. At this order, the only difference between
different physical realizations is the value of the low--energy constant
(LEC) $F$ ($F=v$ in the linear $\sigma$--model) defined by the Goldstone
matrix elements
\beq
\langle 0|J^a_\mu(0)|\vp^b(p)\rangle =
i \delta_{ab} F p_\mu
\label{Gme}
\eeq
for the spontaneously broken currents $J^a_\mu$ and corresponding
Goldstone bosons $\vp^a$. The situation is a little different in
non--relativistic quantum field theories (Leutwyler, 1994b;
Fradkin, 1991).
\edes

\subsection{Chiral symmetry breaking interactions}
\label{subsec:chisb}
The Lagrangian of the Standard Model is not chiral invariant. The chiral
symmetry of the strong interactions is broken by the electroweak interactions
generating in particular non--zero quark masses. The basic assumption
of CHPT is that the chiral limit constitutes a realistic starting point
for a systematic expansion in chiral symmetry breaking interactions.

It is necessary to distinguish between external and dynamical
fields. External fields do not propagate. They are introduced in
the CHPT framework to generate Green functions of quark currents.
Following Gasser and Leutwyler (1984, 1985a),
we extend the chiral invariant QCD Lagrangian (\ref{eq:QCD0}) by
coupling the quarks to external hermitian matrix fields
$v_\mu, a_\mu,s,p$:
\beq
\cL = \cL^0_{\rm QCD} + \bar q \gamma^\mu(v_\mu + a_\mu \gamma_5)q -
\bar q (s - ip \gamma_5)q ~.\label{eq:QCD}
\eeq
The external field method has two major advantages:
\bdes
\item[i.] External photons and $W$ boson fields are among the gauge fields
$v_\mu,a_\mu$ ($N_f = 3$):
\beqa
r_\mu  =  v_\mu + a_\mu & = & - eQ  A_\mu^{\rm ext} + \dots
\label{eq:gf} \\*
l_\mu  =  v_\mu - a_\mu & = &  - eQ  A_\mu^{\rm ext}
        - \dfrac{e}{\sqrt{2}\sin{\theta_W}} (W^{\rm ext,+}_\mu T_+ +
{\rm h.c.}) + \dots \no
\eeqa $$
Q = \dfrac{1}{3} \mbox{diag}(2,-1,-1), \qquad
T_+ = \left( \ba{ccc}
0 & V_{ud} & V_{us} \\
0 & 0 & 0 \\
0 & 0 & 0 \ea \right)  ~.$$
$Q$ is the quark charge matrix and the $V_{ij}$ are Kobayashi--Maskawa mixing
matrix elements. Green functions for electromagnetic and semileptonic weak
currents can be obtained as functional derivatives of a generating functional
$Z[v,a,s,p]$ (cf. Sect.~\ref{subsec:gf}) with respect to
external photon and $W$ boson fields.
\item[ii.] The scalar and pseudoscalar fields $s,p$ give rise to Green
functions of (pseudo)scalar quark currents, but they also provide a very
convenient way of incorporating explicit chiral symmetry breaking through the
quark masses. The physically interesting Green functions are
functional derivatives of the generating functional $Z[v,a,s,p]$ at
$$
v_\mu = a_\mu = p = 0
$$
and
\beq
s = \M_q = \mbox{diag }(m_u, m_d, \ldots)~.\label{eq:mq}
\eeq
The practical advantage is that $Z[v,a,s,p]$ can be calculated in a
manifestly chiral invariant way. The actual Green functions with
broken chiral symmetry are then obtained by taking appropriate functional
derivatives.
\edes

Inclusion of external fields promotes the global chiral symmetry $G$
to a local one:
\beqa
q & \toG & g_R \, \dfrac{1}{2} (1 + \gamma_5)q +
                  g_L \, \dfrac{1}{2} (1 - \gamma_5)q \no \\*
r_\mu  & \toG & g_R r_\mu g^{-1}_R +
                             i g_R \partial_\mu g^{-1}_R \no \\*
l_\mu  & \toG & g_L l_\mu g^{-1}_L +
                             i g_L \partial_\mu g^{-1}_L \no\\*
s + i p & \toG & g_R (s + i p) g^{-1}_L ~.\label{eq:local}
\eeqa
The local nature of $G$ requires the introduction of a covariant
derivative
\beq
D_\mu U = \partial_\mu U - i  r_\mu U + i U l_\mu ~, \qquad
D_\mu U \toG g_R D_\mu U g^{-1}_L~,
\eeq
and of associated non--Abelian field strength tensors
\beqa
  F_R^{\mu\nu} & = & \partial^\mu r^\nu -
                     \partial^\nu r^\mu -
                     i [r^\mu,r^\nu]   \\*
  F_L^{\mu\nu} & = & \partial^\mu l^\nu -
                     \partial^\nu l^\mu -
                     i [l^\mu,l^\nu] ~. \no
\eeqa

External fields do not have kinetic
parts. Consequently, the external gauge fields are not affected by the
spontaneous breakdown of $G$. On the fundamental level, the
electroweak gauge symmetry is broken in two steps, at the Fermi scale
and at the chiral symmetry breaking scale. Thus, there is in principle
a small mixing between $\bar q q$ and whichever fields are responsible for the
electroweak breaking at the Fermi scale (Higgs, technicolour, ...).
Via the Higgs--Kibble mechanism, three of those states become the
longitudinal components of $W$ and $Z$ bosons. Here, we are only
interested in the light orthogonal states, the pseudoscalar
pseudo--Goldstone bosons.

Green functions of electromagnetic and semileptonic weak currents appear
in physical amplitudes with either real photons (radiative transitions)
or virtual photons and $W$ bosons coupling to lepton pairs (electromagnetic
form factors, lepton pair production, semileptonic weak interactions).
Although introduced as non--propagating external fields, the
gauge fields $A_\mu^{\rm ext}$, $W_\mu^{\rm ext}$ may also be viewed
as dynamical fields in these cases turning into real photons or
lepton pairs.

\begin{figure}
\centerline{\epsfig{file=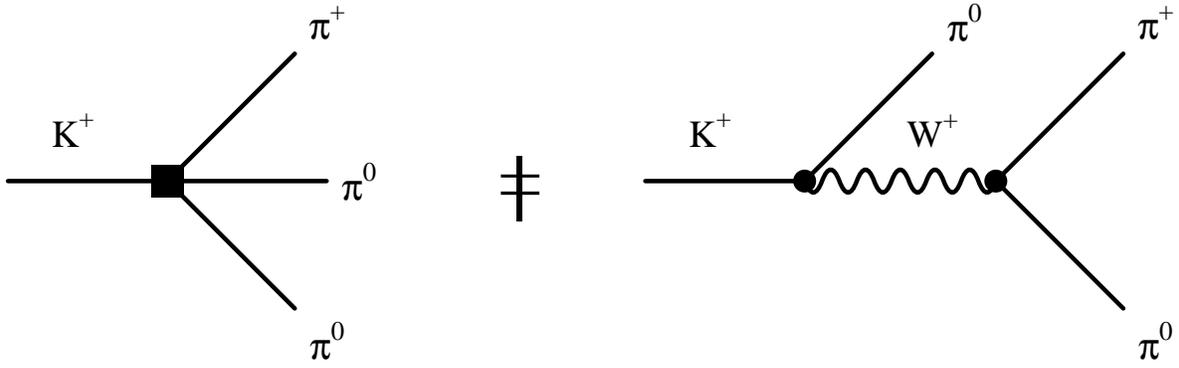,height=5cm}}
\caption{The nonleptonic $K\to 3 \pi$ amplitude cannot be obtained
from a contraction of two semileptonic Green functions.}\label{fig:K3pi}
\end{figure}

The situation is very much different if the gauge fields connect hadronic
degrees of freedom. In this case, the strong interactions
cannot be disentangled from the electroweak interactions any more.
In other words,
Green functions of quark currents are not sufficient to describe the
nonleptonic weak interactions or virtual photons. Consider as a
specific example the nonleptonic weak decay $K^+ \ra \pi^+ \pi^0 \pi^0$.
As indicated in Fig.~\ref{fig:K3pi}, the decay amplitude is not just
given by the $1W$--reducible contribution on the right--hand side.
The hadrons on both sides of the $W$ interact strongly with each
other. At the low energies relevant for CHPT, the correct procedure is
to first integrate out the $W$ together with the heavy quarks to arrive
at an effective Hamiltonian already at the quark level (Gilman and Wise, 1979):
\beq
\Ha_{\rm eff}^{\Delta S =1} = \frac{G_F}{\sqrt{2}} V_{ud} V_{us}^*
\sum_i C_i(\mu) Q_i + {\rm h.c.} \label{eq:Hnl}
\eeq
The $C_i(\mu)$ are Wilson coefficients depending on the QCD
renormalization scale $\mu$. The $Q_i$ are local four--quark operators
if we limit the operator product expansion (\ref{eq:Hnl}) to the
leading $d = 6$ operators. For the effective realization at the
hadronic level, the explicit form of the $Q_i$ (or of the Wilson
coefficients) is of no concern. All that is needed is the transformation
property of $\Ha_{\rm eff}^{\Delta S = 1}$ under chiral rotations:
\beq
\Ha_{\rm eff}^{\Delta S =1} \sim (8_L,1_R) + (27_L,1_R)~.
\label{eq:nldecomp}
\eeq

In order to calculate Green functions with a single insertion of
(\ref{eq:Hnl}), one introduces two spurion fields that transform
contragrediently to the two pieces in (\ref{eq:nldecomp}). In particular,
the octet spurion field $\lambda(x)$ transforms as
\beq
\lambda(x) \toG g_L \lambda(x) g^{-1}_L~.
\label{eq:spurion}
\eeq
Since we are only interested in weak amplitudes to first order in $G_F$,
the effective chiral Lagrangian for the nonleptonic weak interaction
will be constructed as the most general chiral invariant Lagrangian
linear in $\lambda(x)$ (and in the corresponding 27--plet spurion field).
Setting $\lambda(x)$ equal to the constant value
\beq
\lambda(x) = \frac{1}{2} (\lambda_6 - i \lambda_7)
\eeq
projects out the octet part of the $\Delta S = 1$ nonleptonic weak
interaction (and similarly for the 27--plet).

Unlike for the massive $W$ boson, integrating out the (massless)
photons cannot just be described
by a local operator at low energies. In a first step, the electromagnetic
field is made dynamical by including the appropriate kinetic term and by
enlarging the external vector field $v_\mu$ [cf. Eq.~(\ref{eq:gf})]:
\beq
v_\mu \ra v_\mu - e Q A_\mu~. \label{eq:dyn}
\eeq
CHPT then generates automatically all diagrams with virtual (and real)
photons. However, this cannot be the whole story. For instance, loop
diagrams with virtual photons will in general be divergent requiring
appropriate local counterterms. Restricting attention to single--photon
exchange, one must therefore add the most general chiral Lagrangian of
$O(e^2)$ that transforms as the product of two electromagnetic currents
under chiral rotations. For $N_f = 3$, the electromagnetic current is
pure octet. The relevant chiral Lagrangian for virtual photons is, in
addition to the replacement (\ref{eq:dyn}), given by the most general
chiral invariant Lagrangian that is bilinear in the spurion fields
$Q_L(x),Q_R(x)$ with transformation properties
\beq
Q_A(x) \toG g_A Q_A(x) g^{-1}_A, \qquad A = L,R~.
\eeq
Identifying the spurion fields with the quark charge matrix,
\beq
Q_A(x) = Q = \mbox{ diag}(2/3,-1/3,-1/3)~,
\eeq
gives rise to the effective chiral Lagrangian of $O(e^2)$ with the correct
transformation properties.

For the remainder of this review, the spurion fields will not appear
as explicit fields anymore, but will be set equal to their constant
values. The spurion notation can always be reintroduced to check the
proper chiral behaviour of the various effective Lagrangians.

\subsection{Effective chiral Lagrangians of lowest order}
\label{subsec:low}
CHPT is the low--energy effective field theory of the Standard Model. The
chiral Lagrangians are
organized in a derivative expansion based on the chiral counting rules
\beqa
U & \hspace{3cm} & O(p^0) \no \\*
D_\mu U, v_\mu, a_\mu & & O(p) \no \\*
F_{L,R}^{\mu\nu} & & O(p^2) ~.\label{eq:cc}
\eeqa
The baryon and the spurion fields have zero chiral dimension. CHPT is also
an expansion in quark masses around the chiral limit. In principle, one
can formulate CHPT as an independent expansion in both derivatives and
quark masses (cf. Sect.~\ref{subsec:GCHPT}). It is convenient,
however, to combine these two expansions in a single one by making use
of the relations between meson and quark masses. Standard CHPT is
defined by the simplest choice corresponding to the counting rule
[see Eq.~(\ref{eq:masses}) below]
\beqa
s,p & \hspace{3cm} & O(p^2) \label{eq:sp}
\eeqa
for the scalar and pseudoscalar external fields. The more general
framework of Generalized CHPT (Fuchs et al., 1991; Stern et al., 1993;
Knecht et al., 1993) will be discussed in Sect.~\ref{subsec:GCHPT}.

\subsubsection{Strong interactions of mesons}
The locally chiral invariant Lagrangian of lowest order describing the
strong, electromagnetic and semileptonic weak interactions of mesons
is given by (Gasser and Leutwyler, 1984, 1985a)
\beq
\cL_2 = \frac{F^2}{4} \langle D_\mu U D^\mu U^\dg + \chi U^\dg +
\chi^\dg U \rangle ~,\qquad
\chi = 2B(s + ip)~. \label{eq:L2}
\eeq
The two LECs of $O(p^2)$ are related to the pion decay constant and to
the quark condensate in the chiral limit:
\beq
F_\pi =  F[1 + O(m_q)] = 92.4 MeV
\eeq
$$\langle 0|\bar u u |0\rangle = - F^2 B[1 + O(m_q)] ~.
$$
Using for instance the parametrization (\ref{eq:Uphi}) for $U(\vp)$ and
setting the external scalar field equal to the quark mass matrix
[cf. Eq.~(\ref{eq:mq})], one can
immediately read off from (\ref{eq:L2}) the pseudoscalar meson masses
to leading order in $m_q$:
\beqa
M^2_{\pi^+} & = & 2 \hat{m} B \no \\
M^2_{\pi^0} & = & 2 \hat{m} B + O\left[{(m_u - m_d)^2\over m_s
- \hat{m}}\right] \no \\
M^2_{K^+} & = & (m_u + m_s)B \label{eq:masses} \\
M^2_{K^0} & = & (m_d + m_s)B \no \\
M^2_{\eta_8} & = & {2 \over 3}(\hat{m} + 2m_s)B +
O\left[{(m_u - m_d)^2\over m_s - \hat{m}}\right]  \no \\
\hat{m} & := & {1 \over 2}(m_u + m_d) ~.\no
\eeqa
With the quark condensate assumed to be non--vanishing in the
chiral limit ($B \neq 0$), these relations explain the chiral counting rule
(\ref{eq:sp}). Moreover, they give rise to several well--known relations:
\beq
F_\pi^2 M_\pi^2 = - 2 \hat{m} \langle 0|\bar u u|0\rangle  \qquad
\mbox{ (Gell-Mann et al., 1968)} \hspace*{2cm} \label{eq:GMOR}
\eeq\beq
B = \frac{M_\pi^2}{2 \hat{m}} = \frac{M_{K^+}^2}{m_s + m_u} =
\frac{M_{K^0}^2}{m_s + m_d} \qquad
\mbox{ (Gell-Mann et al., 1968; Weinberg, 1977)} \label{eq:ratios}
\eeq\beq
3M^2_{\eta_8} = 4 M_K^2 - M_\pi^2 \qquad
\mbox{ (Gell-Mann, 1957; Okubo, 1962)}. \label{eq:GMO}
\eeq

\subsubsection{Strong interactions of mesons and baryons}
The chiral Lagrangian starts at $O(p)$ (Gasser et al., 1988;
Krause, 1990):
\beq
\cL_{\pi N}^{(1)}  =  \bar \Psi(i \not\!\nabla - m + \frac{g_A}{2}
\not\!u \gamma_5) \Psi \qquad (N_f = 2) \label{eq:piN1}
\eeq\beq
\cL_{MB}^{(1)}  =  \langle \bar B(i \not\!\nabla - m)B + \frac{d}{2}
\bar B \gamma^\mu \{u_\mu,B\} + \frac{f}{2} \bar B \gamma^\mu \gamma_5
[u_\mu,B]\rangle \qquad (N_f = 3)~. \label{eq:MB1}
\eeq
The Lagrangian $\cL_{\pi N}^{(1)}$ is of the form expected from the
discussion of the linear $\sigma$--model in Sect.~\ref{subsec:nlr}.
The connection $\Gamma_\mu$ defining the covariant derivative and the
vielbein field $u_\mu$ now include the external gauge fields:
\beqa
\Gamma_\mu & = & \frac{1}{2} \{ u^\dg (\partial_\mu - i r_\mu)u +
u (\partial_\mu - i l_\mu)u^\dg\} \label{eq:conn}\\
u_\mu & = & i \{ u^\dg (\partial_\mu - i r_\mu)u -
u (\partial_\mu - i l_\mu) u^\dg\}\no~.
\eeqa
$B(x)$ is the octet of lowest--lying baryons in the usual matrix notation.

At $O(p)$, there are two (three) LECs for $N_f = 2~(3)$~: $m$ is the
nucleon (baryon) mass and $g_A$ is the nucleon axial--vector coupling
constant in the chiral limit. The axial $SU(3)$ coupling constants
$f,d$ are related to $g_A$ via
\beq
g_A = f + d~.
\eeq
With the usual definition of the $\pi N$ coupling constant $g_{\pi N N}$,
the Goldberger--Treiman relation (Goldberger and Treiman, 1958)
\beq
g_{\pi N N} = \frac{g_A m}{F}
\eeq
is an exact relation in the chiral limit.

\subsubsection{Nonleptonic weak interactions of mesons}
The effective chiral Lagrangian describing the $\Delta S =1$ nonleptonic
weak interaction of mesons starts at $O(G_F p^2)$ (Cronin, 1967):
\beq
\cL_2^{\Delta S=1} = G_8 \langle \lambda L_\mu L^\mu \rangle +
G_{27}\left(L_{\mu 23} L^\mu_{11} + \frac{2}{3} L_{\mu 21} L^\mu_{13}\right)
+ {\rm h.c.} \label{eq:G827}
\eeq
$$
\lambda = (\lambda_6 - i \lambda_7)/2 ~, \qquad
L_\mu = i F^2 U^\dg D_\mu U~.
$$
The two parts of this Lagrangian correspond to the two pieces of
$\Ha_{\rm eff}^{\Delta S=1}$ in (\ref{eq:nldecomp}). The octet and
27--plet coupling constants $G_8,G_{27}$ can be extracted from
$K \ra \pi \pi$ decay rates:
\beq
|G_8| \simeq 9 \cdot 10^{-6} \mbox{ GeV}^{-2}~, \qquad
G_{27}/G_8 \simeq 1/18~.
\eeq
The large disparity between these constants ($\Delta I = 1/2$ rule,
octet enhancement) is input for CHPT. In practice, this
implies that the 27--plet contribution can usually be neglected, unless
there is no octet amplitude for the transition in question.

\subsubsection{Mesons and virtual photons}
Since the spurion fields $Q_L,Q_R$ have vanishing chiral dimension, the
chiral Lagrangian of lowest order is of $O(e^2 p^0)$ and it consists
of a single term (Gasser and Leutwyler, unpublished; Ecker et al., 1989a):
\beq
\cL_{(e^2p^0)} = e^2 C \langle Q U Q U^\dg\rangle~.\label{eq:e2p0}
\eeq
This Lagrangian yields both Dashen's theorem \footnote{The
neutral mesons remain massless in the chiral limit.} (Dashen, 1969)
\beq
M^2_{\pi^+} = M^2_{K^+} = \frac{2 e^2 C}{F^2} + O(m_q)
\label{eq:Dashen}
\eeq
and Sutherland's theorem (Sutherland, 1966)
\beq
A(\eta \ra 3\pi) = O(m_u - m_d, e^2 m_q)~.
\eeq
The pion electromagnetic mass difference is dominated by (\ref{eq:e2p0})
because the corrections to (\ref{eq:Dashen}) are of $O[(m_u - m_d)^2]$
in this case. From the experimental mass difference (Review Part.
Prop., 1994) one extracts
\beq
C = 6 \cdot 10^{-5}\mbox{ GeV}^4~.
\eeq

\subsection{Generalized CHPT}
\label{subsec:GCHPT}
CHPT is based on a two--fold expansion. As a
low--energy effective field theory, it is an expansion in small
momenta. On the other hand, it is also an expansion in the quark masses
$m_q$ around the chiral limit. In full generality,
the effective chiral Lagrangian is of the form
\beq
\cL_{\rm eff} = \sum_{i,j} \cL_{ij}~, \qquad \qquad
\cL_{ij} = O(p^i m^j_q)~. \label{eq:gexp}
\eeq

The two expansions become related by expressing the pseudoscalar meson
masses in terms of the quark masses. If the quark condensate is
non--vanishing in the chiral limit, the squares of the meson masses
(\ref{eq:masses}) start out linear in $m_q$.
Assuming the linear terms to give the dominant contributions to the
meson masses, one arrives at the standard chiral counting with
$m_q = O(p^2)$ and
\beq
\cL_{\rm eff} = \sum_d \cL_d~, \qquad \qquad
\cL_d = \sum_{i + 2j = d} \cL_{ij}~.
\eeq

In addition to simplicity, there are at the moment two significant pieces
of evidence in favour of the standard counting. The first one is the
experimentally well satisfied Gell-Mann--Okubo relation (\ref{eq:GMO}),
which is modified in Generalized CHPT (see below).
There is also increasing evidence from lattice gauge theory that
the quark condensate is not only non--vanishing in the chiral limit but
also of the size expected in the standard picture. Some recent results
for both the quark condensate (extrapolated to the chiral limit)
and the pion decay constant obtained with
different methods are collected in Table \ref{tab:lattice}. Also shown
are the experimental value of $F_\pi$ and the standard value for
$\langle 0|\bar u u|0\rangle$ from the quark masses (Gasser
and Leutwyler, 1982).

\renewcommand{\arraystretch}{1.1}
\begin{table}
\caption{The pion decay constant $F_\pi$ and the
quark condensate $\langle 0| \bar u u|0\rangle$ from
various lattice calculations [see also Negele (1995)]. Also shown are the
experimental value of $F_\pi$ and the standard CHPT value for the quark
condensate (Gasser and Leutwyler, 1982). The scale dependent condensate
is not always given at the same scale; for details consult the original
literature.}
\label{tab:lattice}
$$
\begin{tabular}{|c|c|c|} \hline
 method & $F_\pi$ (MeV) & $[- \langle 0|\bar u u|0
\rangle]^{1/3}$ (MeV) \\ \hline
quenched Wilson fermions & $ 82 \pm 11$ & $\sim 260 $ \\
(Daniel et al., 1992; Weingarten, 1994) & & \\ \hline
dynamical staggered fermions & $94 - 105$ & $310 - 370$ \\
2 flavours (Fukugita et al., 1993) & & \\ \hline
dynamical staggered fermions & $75 \pm 10$ & $215 \pm 25$ \\
4 flavours (Altmeyer et al., 1993) & & \\ \hline
experiment (Review Part. Prop., 1994) & 92.4 &  \\
CHPT (Gasser and Leutwyler, 1982) & & $225 \pm 25$ \\ \hline
\end{tabular}
$$
\end{table}

Taken at face value, Table \ref{tab:lattice} suggests good agreement
between the standard picture and lattice results. With due caution in
view of the present limitations of lattice simulations, there is
strong evidence for the ratio
\beq
- \frac{\langle 0|\bar u u|0\rangle}{F^3} = \frac{B}{F}
\label{eq:B/F}
\eeq
to be substantially bigger than one.

Neither the Gell-Mann--Okubo relation nor
lattice gauge theory establish the standard picture once and for all.
If one is very cautious (and this caution will have its price), one may
want to include the case where $B$ is comparable to $F$, much smaller than
indicated in Table \ref{tab:lattice}.
Generalized CHPT (Fuchs et al., 1991; Stern et al., 1993; Knecht
et al., 1993) is a scheme that is adapted for small values of $B$.
It amounts to a different ordering of the  $\cL_{ij}$ in the
effective chiral Lagrangian (\ref{eq:gexp}). To keep track of the
different chiral counting, the parameter $B$ is formally declared to be
a quantity of $O(p)$ implying via (\ref{eq:masses}) that also $m_q$
is $O(p)$. The effective Lagrangian (\ref{eq:gexp}) is then written
as (Knecht and Stern, 1994b)
\beq
\cL_{\rm eff} = \sum_d \wt \cL_d~, \qquad \qquad
\wt \cL_d = \sum_{i+j+k = d} B^k \cL_{ij} ~.
\eeq
\paragraph{Remarks}
\bdes
\item[i.] The effective chiral Lagrangian is the same in Standard and in
Generalized CHPT. In the generalized picture, more terms appear at a
given order that are relegated to higher orders in the standard counting.
Obviously, this procedure increases the number of undetermined LECs
at any given order.
\item[ii.] In the standard framework, Lorentz invariance implies that
the chiral dimension $d$ increases in steps of two in the meson sector:
\beq
\cL_{\rm eff} = \cL_2 + \cL_4 + \cL_6 + \ldots
\eeq
In contrast, due to $m_q = O(p)$ the generalized expansion is of the form
\beq
\cL_{\rm eff} = \wt \cL_2 + \wt \cL_3 + \wt \cL_4 + \ldots
\eeq
\item[iii.] The generalized counting needs some accustoming. Both $m_q$
and $B$ are counted as $O(p)$, although
$B$ does not have to vanish in the chiral limit. Even though Generalized
CHPT is relevant for $B\simeq F$, $F$ must not be treated as a quantity
of $O(p)$.
\edes

A major difference between the two schemes already appears at lowest
order, $O(p^2)$. Instead of the standard Lagrangian $\cL_2$ in
(\ref{eq:L2}), the most general form of $\wt \cL_2$ is (Knecht and
Stern, 1994b)
\beqa
\label{eq:L2g}
\wt \cL_2 &=& \frac{F^2}{4} \{ \langle D_\mu U^\dg D^\mu U\rangle +
2B \langle \wt \chi U^\dg + \wt \chi^\dg U \rangle \no \\
&& \mbox{} + A_0 \langle (U^\dg \wt \chi)^2 +
(\wt \chi^\dg U)^2\rangle + Z^S_0 \langle \wt \chi U^\dg +
\wt \chi^\dg U\rangle^2 \no \\
&& \mbox{} + Z^P_0 \langle \wt \chi U^\dg - \wt \chi^\dg U\rangle^2 +
H_0 \langle \wt \chi^\dg \wt \chi\rangle \} \\
\wt \chi &=& s + ip = \frac{\chi}{2B} ~. \no
\eeqa
The difference between $\wt \cL_2$ and $\cL_2$ is part of $\cL_4$ [cf.
Eq.~(\ref{eq:L4})]: $A_0 \sim L_8$, $Z^S_0 \sim L_6$ (Zweig rule
suppressed), $Z^P_0 \sim L_7$, $H_0 \sim L_{12}$ (not directly
observable). The number of observable LECs increases from two in the
standard picture $(F,B)$ to five $(F,B,A_0,Z^S_0,Z^P_0)$ in the
generalized scheme. Consequently, there are fewer constraints among
physical quantities already at lowest order. Turning to the meson
masses, one finds that the quark mass ratio
$$
r = \dfrac{m_s}{\hat{m}}
$$
can no longer be expressed in terms of meson mass ratios only as
in (\ref{eq:ratios}). Instead,
Generalized CHPT suggests a range (Knecht and Stern, 1994b)
\beq
6 \simeq \frac{2M_K}{M_\pi} - 1 = r_1 \leq r \leq r_2 =
\frac{2M^2_K}{M^2_\pi} - 1 \simeq 26 \label{eq:r1r2}
\eeq
where $r = r_2$ corresponds to the standard picture $(A_0 = Z^S_0 =
Z^P_0 = 0)$, while $r = r_1$ is reached for $B = 0$ with
$M^2 \sim m^2_q$ at leading order.

As expected, another consequence of Generalized CHPT is a modification
of the Gell-Mann--Okubo relation (\ref{eq:GMO}) already at $O(p^2)$:
\beq
3 M^2_{\eta_8} + M^2_\pi - 4 M^2_K = 4 (m_s - \hat{m})^2
(A_0 + 2 Z^P_0) + O(p^3).
\eeq
There is no argument known (Knecht and Stern, 1994b) why
$A_0 + 2 Z^P_0 \simeq 0$. The Gell-Mann--Okubo relation is an accidental
relation in Generalized CHPT.

The proponents of Generalized CHPT face the task of providing
convincing evidence that the additional LECs introduced are indeed as
big as the generalized counting suggests. This is difficult for at
least two reasons:
\begin{itemize}
\item At every order in the derivative expansion, Standard and Generalized
CHPT yield the same results in the chiral limit. Explicit chiral symmetry
breaking effects are small in general and difficult to isolate.
\item It is tempting, but not necessarily significant, to find better
agreement with experiment with more free parameters. Only a comprehensive
analysis of many different processes can provide a consistent picture.
\end{itemize}

Such a comprehensive analysis is not yet available. The present state of
Generalized CHPT has been reviewed recently by Knecht and Stern (1994b)
where references to the original work can be found. There are mainly
two areas where more precise data are expected to shed
light on the size of the quark condensate and of
the quark mass ratios: low--energy $\pi\pi$ scattering (cf.
Sect.~\ref{subsec:phenp4}) and the coupling strength of the $\pi'$,
the first radial recurrence of the pion, to the axial--vector current.
The second question can in principle be answered by measuring angular
asymmetries at the percent level in the
decays $\tau\to 3 \pi \nu_\tau$ in the $\pi'$ region (Stern
et al., 1994).

\subsection{Generating functional of Green functions}
\label{subsec:gf}
To lowest order in the chiral expansion, all amplitudes are tree--level
amplitudes that can be read off directly from the lowest--order
effective chiral Lagrangians in Sect.~\ref{subsec:low}.
In order to have a systematic procedure for the calculation of higher orders,
it is useful to introduce the generating functional of connected
Green functions. Restricting the discussion for the moment to
the strong, electromagnetic and semileptonic weak interactions
of mesons, all amplitudes of interest can be obtained from the generating
functional of connected Green functions of quark currents $Z[v,a,s,p]$.
At the fundamental level it is defined in the usual way,
\beq
e^{\dis i Z[v,a,s,p]} = <0~{\rm out}|0~{\rm in}>_{v,a,s,p}~, \label{eq:Zdef}
\eeq
in terms of the vacuum transition amplitude in the presence of external fields
associated with the Lagrangian (\ref{eq:QCD}).

At the hadronic level, $Z[v,a,s,p]$ is calculated with an effective
chiral Lagrangian. Following Weinberg (1979), one writes down
the most general Lagrangian of pseudoscalar meson
fields (and possibly other degrees of freedom) that shares the symmetries
of the underlying theory. However, there is a loophole in this
line of reasoning. The chiral Lagrangian is obviously not uniquely
defined. For instance, adding a total derivative will not change the physical
content of the effective field theory. The question is then whether the
effective Lagrangian must really be locally chiral symmetric for the generating
functional $Z[v,a,s,p]$ to have this property. After all, the
local chiral symmetry has been the main ingredient for writing
down the most general effective chiral Lagrangians. The loophole was recently
closed by Leutwyler (1994a) who showed that the freedom of adding
total derivatives and performing meson field redefinitions can
always be used to bring the effective Lagrangian to a manifestly
locally chiral invariant form. The only exception to the theorem
is the chiral anomaly that requires a Wess--Zumino term (see
Sect.~\ref{sec:strong}), but the rest
is gauge invariant (see also D'Hoker and Weinberg, 1994). Therefore,
the generating functional $Z[v,a,s,p]$ may be calculated at the
hadronic level as
\beq
e^{\dis i Z[v,a,s,p]} =
\int [dU(\vp)] e^{\dis i \int d^4x {\cL}_{\rm eff}} ~,\label{eq:master}
\eeq
with the gauge invariant effective chiral Lagrangian written as an
expansion in derivatives and external fields according to the counting
rules (\ref{eq:cc}), (\ref{eq:sp}):
\beq
\cL_{\rm eff} = \cL_2 + \cL_4 + \dots \label{eq:Leff}
\eeq
The chiral expansion of $\cL_{\rm eff}$ induces a corresponding expansion
for the generating functional
\beq
Z = Z_2 + Z_4 + \dots
\eeq
At lowest order, the functional $Z_2$  is the classical action
\beq
Z_2[v,a,s,p] = \int d^4x \cL_2 (U,v,a,s,p) \label{eq:class}
\eeq
where
$$
U = U[v,a,s,p] $$
is to be understood as a functional of the external fields via the
equation of motion for $\cL_2$~:
\beq
\Box U U^\dg - U \Box U^\dg = \chi U^\dg - U \chi^\dg -
{1 \over 3}\langle\chi U^\dg - U \chi^\dg\rangle {\bf 1}~. \label{eq:EOM}
\eeq
As already noted, this amounts to reading off the relevant tree--level
amplitudes directly from the lowest--order Lagrangian (\ref{eq:L2}) by
using an explicit parametrization of $U(\vp)$ like (\ref{eq:Uphi}).

Why not stop here? In the sixties, many proponents of effective Lagrangians
argued that due to their non--renormalizability such Lagrangians only make
sense at tree level.
Today, we view effective field theories on almost the same footing as the
``fundamental" gauge theories. They admit a perfectly
consistent loop expansion that ensures all axiomatic properties of
a quantum field theory like unitarity and analyticity. Unitarity,
written schematically as
\beq
\Im m ~T \sim |T|^2~,
\eeq
requires that in general loop amplitudes start contributing at $O(p^4)$
in the meson sector, because the (real) amplitudes are at least $O(p^2)$.
The non--renormalizability manifests itself in the appearance of
additional terms that are not present in the lowest--order Lagrangian.
However, $\cL_{\rm eff}$ is already the most general chiral invariant
Lagrangian. Since the divergences can be absorbed by local counterterms
that exhibit the same symmetries as the initial Lagrangian
(Weinberg, 1960), $\cL_{\rm eff}$
automatically includes all terms needed for renormalization to
every order in the loop expansion. Although one needs a specific regularization
procedure for the loop amplitudes, the renormalized theory is
independent of the chosen regularization. In practice, a mass--independent
regularization scheme like dimensional regularization is best
suited for the purpose. It respects all the symmetries of the
chiral Lagrangians, both in the meson and in the
meson--baryon system (Sect.~\ref{sec:baryons}).

To keep track of the chiral counting, it is convenient to
define the chiral dimension of an amplitude. For the
case under consideration where only pseudoscalar mesons appear in
internal lines [cf. (\ref{eq:master})], the chiral dimension $D$
of a connected $L$--loop amplitude with
$N_d$ vertices of $O(p^d)$ ($d = 2,4,\ldots)$ is given by
(Weinberg, 1979)
\beq
D = 2L + 2 + \sum_d (d-2) N_d~, \qquad d = 2,4,\ldots \label{eq:DL}
\eeq
and therefore, up to $O(p^4)$ :
\beqa
D = 2 : & L = 0, \; d = 2 \qquad & Z_2 = \int d^4x \cL_2 \nl
D = 4 : & L = 0, \; d = 4 \qquad & Z_4^{\rm tree} = \int d^4x {\cal
L}_4+Z_{\rm WZW}\nl & L =1, \; d = 2 \qquad & Z_4^{(L=1)} \mbox{ for } \cL_2~.
\label{eq:D4}
\eeqa
$Z_{\rm WZW}$ is the Wess--Zumino--Witten functional (Wess and
Zumino, 1971; Witten, 1983) to be discussed in Sect.~\ref{subsec:anom}.

For a given amplitude, the chiral dimension $D$ increases with
$L$ according to Eq.~(\ref{eq:DL}). In order to reproduce the (fixed)
physical dimension of the amplitude, each loop produces a factor $1/F^2$.
Together with the geometric loop factor $(4\pi)^{-2}$, the loop expansion
suggests
\beq
4\pi F_\pi = 1.2 \mbox{ GeV} \label{eq:naive}
\eeq
as the natural scale of the chiral expansion (Manohar and
Georgi, 1984). A more refined analysis indicates
\beq
4\pi F_\pi/\sqrt{N_f}\label{eq:soph}
\eeq
as the relevant scale for $N_f$ light flavours (Soldate and
Sundrum, 1990; Chivukula et al., 1993). Restricting
the domain of applicability of CHPT to momenta $|p| \lets
O(M_K)$, the natural expansion parameter of chiral amplitudes
based on the naive estimate (\ref{eq:naive}) is
expected to be of the order
\beq
\frac{M_K^2}{16 \pi^2 F_\pi^2} = 0.18~. \label{eq:omag}
\eeq
The more sophisticated estimate (\ref{eq:soph}) would yield an additional
factor three for chiral $SU(3)$. In addition, these terms often
appear multiplied with chiral logarithms. It is
therefore no surprise that substantial higher--order
 corrections  in the chiral expansion are the rule rather than the exception
for chiral $SU(3)$. On the other hand, for
$SU(2)_L\times SU(2)_R$ and for momenta
$|p| \lets O(M_\pi)$ the chiral expansion is expected to converge
considerably faster.

The generalization to the meson--baryon system will be discussed
in Sect.~\ref{sec:baryons}. For the nonleptonic weak interactions
and for the treatment of virtual photons the generating functional
can easily be extended to depend also on the respective spurion fields.
Since one is in practice only interested in amplitudes to
first order in $G_F$ or $e^2$, one adds the corresponding effective
Lagrangians to $\cL_{\rm eff}$ in
(\ref{eq:Leff}) and calculates the amplitudes to $O(G_F)$ and $O(e^2)$,
respectively.

\setcounter{equation}{0}
\setcounter{subsection}{0}
\setcounter{table}{0}
\setcounter{figure}{0}

\section{STRONG INTERACTIONS OF MESONS}
\label{sec:strong}
\subsection{Generating functional of $O(p^4)$}
\label{subsec:p4}
The Green functions and amplitudes of lowest order (current
algebra level) can be obtained from the classical action
(\ref{eq:class}). At next--to--leading order, $O(p^4)$, the functional
$Z_4[v,a,s,p]$ consists of three parts (Gasser and Leutwyler, 1984, 1985a)
in accordance with (\ref{eq:D4}):
\bit
\item Tree diagrams with a single vertex from the effective chiral
Lagrangian $\cL_4$ of $O(p^4)$ and any number of vertices from $\cL_2$
via the equation of motion (\ref{eq:EOM}).
\item The one--loop functional for the lowest--order Lagrangian $\cL_2$.
\item The Wess--Zumino--Witten functional $Z_{\rm WZW}$ (Wess and Zumino,
1971; Witten, 1983) to account for the
chiral anomaly (Adler, 1969a; Bell and Jackiw, 1969; Bardeen, 1969).
The corresponding odd--intrinsic--parity sector will be discussed in
Sect.~\ref{subsec:anom}.
\eit

The effective chiral Lagrangian $\cL_4(U,v,a,s,p)$ for chiral
$SU(3)$ has the form (Gasser and Leutwyler, 1985a)
\beqa
{\cal L}_4 & = & L_1 \langle D_\mu U^\dg D^\mu U\rangle^2 +
                 L_2 \langle D_\mu U^\dg D_\nu U\rangle
                     \langle D^\mu U^\dg D^\nu U\rangle \no \\*
& & + L_3 \langle D_\mu U^\dg D^\mu U D_\nu U^\dg D^\nu U\rangle +
    L_4 \langle D_\mu U^\dg D^\mu U\rangle \langle \chi^\dg U +
    \chi U^\dg\rangle  \no \\*
& & +L_5 \langle D_\mu U^\dg D^\mu U(\chi^\dg U + U^\dg
\chi)\rangle
    +
    L_6 \langle \chi^\dg U + \chi U^\dg \rangle^2 +
    L_7 \langle \chi^\dg U - \chi U^\dg \rangle^2  \no \\*
& & + L_8 \langle \chi^\dg U \chi^\dg U +
 \chi U^\dg \chi U^\dg\rangle
    -i L_9 \langle F_R^{\mu\nu} D_\mu U D_\nu U^\dg +
      F_L^{\mu\nu} D_\mu U^\dg D_\nu U \rangle \no \\*
& & + L_{10} \langle U^\dg F_R^{\mu\nu} U F_{L\mu\nu}\rangle +
    L_{11} \langle F_{R\mu\nu} F_R^{\mu\nu} + F_{L\mu\nu} F_L^{\mu\nu}\rangle +
    L_{12} \langle \chi^\dg \chi \rangle~.
\label{eq:L4}
\eeqa
It is the most general Lagrangian of $O(p^4)$ that satisfies local chiral
invariance, Lorentz invariance, $P$ and $C$. Since the loop expansion
is a systematic expansion around the classical solution (see
also Sect.~\ref{sec:p6}), the equation of motion (\ref{eq:EOM}) has
been used to reduce the number of terms. A similar Lagrangian
can be written down for chiral $SU(2)$ (Gasser and Leutwyler, 1984).

Before turning to the physics behind the new LECs $L_1, \ldots, L_{12}$,
let us consider the one--loop functional $Z_4^{(L=1)}[v,a,s,p]$. Formally,
it can be given in closed form as
\beq
Z_4^{(L=1)} = \frac{i}{2} \ln \det D_2 = \frac{i}{2} \mbox{ tr }\ln D_2~,
\label{eq:Z4loop}
\eeq
where $D_2$ is a differential operator associated with the lowest--order
Lagrangian (\ref{eq:L2}). However, $Z_4^{(L=1)}$ is divergent and
must be properly defined by means of a regularization. In a mass--independent
regularization scheme like dimensional regularization, chiral power counting
(\ref{eq:DL}) shows that all one--loop divergences have chiral dimension
$D=4$. Since the divergent
part $Z^{(L=1)}_{4,{\rm div}}$ has the form of a local action with all the
symmetries of $\cL_2$ (Weinberg, 1960), the corresponding Lagrangian
$\cL^{(L=1)}_{4,{\rm div}}$ must be of the form (\ref{eq:L4}),
\beqa
\cL^{(L=1)}_{4,{\rm div}}& =& - \Lambda(\mu) \sum_{i=1}^{12} \Gamma_i P_i
\label{eq:L4div} \\
\Lambda(\mu) &=& \frac{\mu^{d-4}}{(4\pi)^2} \left\{\frac{1}{d-4} - \frac{1}{2}
[\ln 4\pi + 1 + \Gamma'(1)]\right\} \no
\eeqa
with coefficients  $\Gamma_i$ listed in Table \ref{tab:Li}.
The $P_i$ are the field monomials in the chiral Lagrangian
$\cL_4$ in (\ref{eq:L4}) and the explicit form of $\Lambda(\mu)$ holds
for dimensional regularization. The renormalization of the one--loop
functional (\ref{eq:Z4loop}) is then implemented
by defining renormalized coupling constants $L_i^r(\mu)$ as
\beq
L_i = L_i^r(\mu) + \Gamma_i \Lambda(\mu) \label{eq:Liren}
\eeq
in order to cancel the divergent piece (\ref{eq:L4div}) in the one--loop
functional. By construction, the sum
\beq
Z_4^{(L=1)} + \int d^4x \cL_4(L_i) = Z^{(L=1)}_{4,{\rm fin}}
(\mu) + \int d^4x \cL_4(L_i^r(\mu)) \label{eq:renorm}
\eeq
is both finite and independent of the arbitrary scale $\mu$. The separate
scale dependences of the loop and of the counterterm parts cancel
in all Green functions and therefore in all physical amplitudes.

\renewcommand{\arraystretch}{1.1}
\begin{table}[t]
\begin{center}
\caption{Phenomenological values and source for the renormalized coupling
constants $L^r_i(M_\rho)$, taken from Bijnens et al. (1994e).
The quantities $\Gamma_i$
in the fourth column determine the scale dependence of the $L^r_i(\mu)$
according to Eq.~(\protect\ref{eq:scale}). $L_{11}^r$ and $L_{12}^r$ are not
directly accessible to experiment.} \label{tab:Li}
\vspace{.5cm}
\begin{tabular}{|c||r|l|r|}  \hline
i & $L^r_i(M_\rho) \times 10^3$ & source & $\Gamma_i$ \\ \hline
  1  & 0.4 $\pm$ 0.3 & $K_{e4},\pi\pi\rightarrow\pi\pi$ & 3/32  \\
  2  & 1.35 $\pm$ 0.3 &  $K_{e4},\pi\pi\rightarrow\pi\pi$&  3/16  \\
  3  & $-$3.5 $\pm$ 1.1 &$K_{e4},\pi\pi\rightarrow\pi\pi$&  0     \\
  4  & $-$0.3 $\pm$ 0.5 & Zweig rule &  1/8  \\
  5  & 1.4 $\pm$ 0.5  & $F_K:F_\pi$ & 3/8  \\
  6  & $-$0.2 $\pm$ 0.3 & Zweig rule &  11/144  \\
  7  & $-$0.4 $\pm$ 0.2 &Gell-Mann--Okubo,$L_5,L_8$ & 0             \\
  8  & 0.9 $\pm$ 0.3 & \small{$M_{K^0}-M_{K^+},L_5,$}&
5/48 \\
     &               &   \small{ $(2m_s-m_u-m_d):(m_d-m_u)$}       & \\
 9  & 6.9 $\pm$ 0.7 & $\langle r^2\rangle^\pi_V$ & 1/4  \\
 10  & $-$5.5 $\pm$ 0.7& $\pi \rightarrow e \nu\gamma$  &  $-$ 1/4  \\
\hline
11   &               &                                & $-$1/8 \\
12   &               &                                & 5/24 \\
\hline
\end{tabular}
\end{center}
\end{table}

The renormalized coupling constants $L_i^r(\mu)$ are measurable LECs
that characterize QCD at $O(p^4)$. Although the total Green functions
of $O(p^4)$ are independent of the chosen regularization, the
split between loop and tree--level contributions does depend on the
regularization scheme via the definition of $\Lambda(\mu)$ in
(\ref{eq:L4div}). The scale dependence of the $L_i^r(\mu)$ follows from
Eq.~(\ref{eq:Liren}):
\beq
L_i^r(\mu_2) = L_i^r(\mu_1) + \frac{\Gamma_i}{(4\pi)^2} \ln
\frac{\mu_1}{\mu_2}~.\label{eq:scale}
\eeq
Eqs.~(\ref{eq:renorm}) and (\ref{eq:scale}) show that the $\Gamma_i$
are also the coefficients of the so--called chiral logs $\sim \ln{p^2/\mu^2}$
in the one--loop functional.

It is clear from the structure of the one--loop functional that the
chiral dimension $D=4$ does not imply that the relevant Green functions
are just fourth--order polynomials in external momenta and masses. Instead,
the chiral dimension has to do with the degree of homogeneity of
the amplitudes in momenta and masses. Consider the Feynman amplitude
$A$ for a general process with $D_F$ external photons and
$W$ bosons (semileptonic transitions). If we define in addition
to (\ref{eq:DL})
\beq
D = D_L + D_F  ~,
\label{eq:DF}
\eeq
then $D_L$ is the degree of homogeneity of the amplitude $A$ as
a function of external momenta ($p$) and meson masses ($M$):
\beq
A(p,M;C_i^r(\mu),\mu/M) = M^{D_L} \, A ( p/M , 1;C_i^r(\mu), \mu/M )  ~.
\label{eq:homog}
\eeq
The $C_i^r(\mu)$ denote renormalized LECs. In the meson sector at
$O(p^4)$, they are just the $L_i^r(\mu)$, but the structure (\ref{eq:homog})
is completely general.

\subsection{Low--energy constants of $O(p^4)$}
\label{subsec:Li}
The LECs parametrize the most general solutions of the chiral
Ward identities, but they are themselves not constrained by the symmetries.
They can be interpreted as describing the influence of all
degrees of freedom not explicitly contained in the chiral Lagrangians.
In principle, they are calculable quantities in the Standard Model. In
practice and for the time being, they are extracted from experimental
input or estimated with additional model dependent assumptions.

Comparing the lowest--order Lagrangian $\cL_2$ with $\cL_4$ in (\ref{eq:L4}),
chiral dimensional analysis suggests as an approximate upper bound
\beq
|L_i| \; \lets \; \frac{N_f}{4(4\pi)^2} \simeq 5 \cdot 10^{-3}
\label{eq:Lib}
\eeq
for $N_f = 3$ light flavours. The present values for the $L_i^r$ at
$\mu=M_\rho$ are displayed in Table \ref{tab:Li}.
Chiral dimensional analysis appears to be quite successful here:
only $L_3$, $L_9$ and $L_{10}$ are close to the upper bound (\ref{eq:Lib}).

The main new input in Table \ref{tab:Li} comes from a recent analysis
of $K_{e4}$ decays (Bijnens et al., 1994a). In particular,
incorporation of the $K_{e4}$ data (see also Sect.~\ref{subsub:semi})
allows for a test of the Zweig rule classification of $L_1, L_2,
L_3$ (Gasser and Leutwyler, 1985a):
\beqa
2L_1 - L_2 && O(1) \\
L_1, L_2, L_3 && O(N_c)~,
\eeqa
where $N_c$ is the number of colours.
The recent analysis of Bijnens et al. (1994a) finds
\beq
\frac{L_2 - 2L_1}{L_3} = - 0.17 \ba{c} + 0.12 \\ - 0. 22 \ea
\eeq
in support of the Zweig rule.

A promising first attempt to calculate the $L_i$ ``directly"
from QCD has been undertaken by Myint and Rebbi (1994). Their
work documents the feasibility of extracting LECs from the lattice.
The first analytic results obtained in the strong--coupling and
large--$N_c$ limits are not too realistic, but the announced numerical
analysis is expected to come closer to the real world.

The actual values of the $L_i$ can be understood in terms of meson
resonance exchange (Ecker et al., 1989a; Donoghue et al., 1989a).
The situation can be summarized as follows:
\begin{description}
\item[Chiral duality:] \mbox{}  \\
The $L_i^r(M_\rho)$ are practically saturated by resonance exchange.
\item[Chiral VMD:]  \mbox{} \\
Whenever spin--1 resonances can contribute at all ($i = 1,2,3,9,10$),
the $L_i^r(M_\rho)$ are almost completely dominated by $V$ and $A$ exchange.
\end{description}
With additional QCD--inspired assumptions, all $V,A$ couplings can be
expressed in terms of $F_\pi$ and $M_V \simeq M_\rho$ only
(Ecker et al., 1989b):
\beq
8 L^V_1 = 4 L^V_2 = - \frac{4}{3} L^V_3 = L^V_9 = - \frac{4}{3}
L_{10}^{V+A} = \frac{F^2_\pi}{2 M^2_V}~. \label{eq:mm}
\eeq
These relations are in good agreement
with the phenomenological values of the $L^r_i(M_\rho)$ and they
also explain why only $L_3$, $L_9$ and $L_{10}$ are close to the
naive upper limit (\ref{eq:Lib}).

Two objections are sometimes raised against the interpretation of
resonance dominated $L_i$.

\paragraph{Objection \#~1~:} Resonance exchange produces scale independent
LECs although they are in fact scale dependent.\\[5pt]
A more precise formulation of resonance dominance (Ecker et al., 1989a)
clarifies the situation:
decomposing the $L_i^r(\mu)$ into resonance contributions $L^R_i$ and
remainders $\wh L_i(\mu)$ carrying the scale dependence,
\beq L^r_i(\mu) = \sum_R L^R_i + \wh L_i(\mu)~,
\eeq
there is a range in $\mu$ (depending on the
renormalization scheme, as discussed in the previous subsection) with
\beq
|\wh L_i(\mu)| \ll |L^r_i(\mu)| \quad \forall \; i~.
\eeq

\paragraph{Objection \#~2~:} The resonance parameters are determined at
$p^2 = M^2_R$, but the $L_i$ describe physics for $p^2 \ll M^2_R$. \\[5pt]
Consider the $VV - AA$ two--point
functions as a specific example:
\beqa
\label{eq:2p}
\lefteqn{i \int d^4x \; e^{ip \cdot x} \langle 0|T\{ V^i_\mu(x) V^j_\nu
(0) - A_\mu^i(x) A_\nu^j(0)\} |0\rangle = } \no \\
&=& (p_\mu p_\nu - g_{\mu\nu} p^2) \Pi_{LR,ij}^{(1)}(p^2) +
p_\mu p_\nu \Pi_{LR,ij}^{(0)}(p^2) ~.
\eeqa
Any given model for the spin--1 resonances will produce a spectral function
$\Pi_{LR}^{(1)}$ of the following structure in the narrow--width
approximation (only $V$ exchange is considered here):
\beq
\Pi_{LR}^{(1)} (p^2) = \frac{P_V(p^2)^2}{M_V^2 - p^2} + P_c(p^2)~,
\eeq
where $P_V(p^2)$, $P_c(p^2)$ are model dependent polynomials
characterizing the off--shell behaviour of $V$ exchange. From resonance
decays (e.g., $\rho \ra e^+ e^-$ in this case) one can fix
$F_V = P_V(M_V^2)$, but unless $P_V(p^2)$ is a constant, the decay width
tells us nothing about $P_V(p^2)$ for $p^2 \neq M_V^2$ nor about $P_c(p^2)$.

The solution of the puzzle is provided by the connection between low and
high energies characteristic for a quantum field theory like QCD. In fact,
QCD requires a relation between the polynomials $P_V(p^2)$ and
$P_c(p^2)$ such that the off--shell behaviour of $P_V(p^2)$ is actually
irrelevant. The key to the solution is the unsubtracted dispersion
relation satisfied by $\Pi_{LR}^{(1)}$:
\beq
\Pi_{LR}^{(1)}(p^2) = \int_0^\infty \frac{ds}{s - p^2}
[\rho_V^{(1)}(s) - \rho_A^{(1)}(s)] ~.
\eeq
Again in the narrow--width approximation, $V$ exchange yields
\beq
\rho_V^{(1)}(s) = F_V^2 \; \delta(s - M_V^2)
\eeq
and therefore [recall $F_V = P_V(M_V^2)$]
\beq
\Pi_{LR}^{(1)}(p^2) = \frac{P_V(M_V^2)^2}{M_V^2 - p^2} =
\frac{P_V(p^2)^2}{M_V^2 - p^2} + P_c(p^2)~,
\eeq
fixing the counterterm polynomial $P_c(p^2)$ uniquely:
\beq
P_c(p^2) = (M_V^2 - p^2)^{-1} [P_V(M_V^2)^2 - P_V(p^2)^2]~.
\eeq

The general conclusion is that Green functions at small $p^2$ are
determined by the on--shell resonance parameters, {\em independently}
of higher--order couplings (Ecker et al., 1989b; Ecker, 1989d).
It is this universality that implies in particular the equivalence of
all realistic models for spin--1 meson resonances to $O(p^4)$
(Ecker et al., 1989b).

The symmetry breaking sector (involving the quark masses) is characterized
at $O(p^4)$ by the LECs $L_4$, $L_5$, $L_6$, $L_7$, $L_8$. These LECs
are only sensitive to
scalar (octet $S$ and singlet $S_1$) and pseudoscalar ($\eta'$) exchange.
In the case of $L_5$, only the scalar octet can contribute. Saturating
the unsubtracted dispersion relation for the scalar form factor
$\langle \pi | \bar u s| K\rangle$ with $S$ exchange, one finds
(Leutwyler, 1990)
\beq
L_5 \simeq \frac{F_\pi^2}{4 M_S^2} ~.
\eeq
The LEC $L_5$ governs $SU(3)$ breaking in the meson decay constants
(Gasser and Leutwyler, 1985a):
\beqa
\frac{F_K}{F_\pi} &=& 1 + \frac{4L_5^r}{F^2} (M_K^2 - M_\pi^2) +
\mbox{chiral logs} \no \\
&\simeq& 1 + \frac{M_K^2 - M_\pi^2}{M_S^2} + \mbox{chiral logs} \no \\
& \simeq & 1.22  \qquad {\rm (Review~Part.~Prop.,~1994)}~.
\eeqa
With $M_S \simeq M_{\eta'} \simeq 1~{\rm GeV}$, one understands why
$SU(3)$ breaking in the meson sector is generally of
$O(25\%)$, except for the pseudoscalar masses. It is
precisely in the symmetry breaking sector that Generalized CHPT follows
a different approach. For a discussion of the corresponding LECs
in that scheme, I refer to the review of Knecht and Stern (1994b).

There has been a lot of activity in recent years to obtain the LECs
from different models of hadronic interactions at low energies. Most
prominent and also most successful among them are different variants
of the Nambu--Jona-Lasinio model (Nambu and Jona-Lasinio, 1961). Those
developments are outside the scope of the present article. There
exist specialized reviews of the subject (de Rafael, 1995; Bijnens, 1994b;
Hatsuda and Kunihiro, 1994; Alkofer et al., 1994; Volkov, 1993) that can
also be consulted for further references.

\subsection{Light quark masses}
\label{subsec:mq}
The quark masses depend on the QCD
renormalization scale. Since the effective Lagrangians cannot depend on
this scale, the quark masses always appear multiplied by quantities
that transform contragrediently under changes of the renormalization
scale. The chiral Lagrangian (\ref{eq:Leff}) contains the quark masses
via the scalar field $\chi$ defined in (\ref{eq:L2}). As long as one
does not use direct or indirect information on $B$, one can
only extract ratios of quark masses.

\begin{figure}
\centerline{\epsfig{file=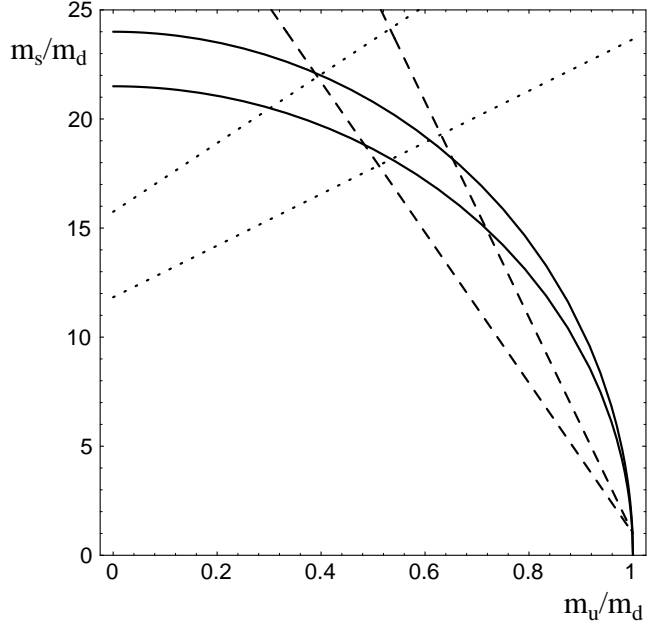,height=8cm}}
\caption{First quadrant of Leutwyler's ellipse (Leutwyler, 1990)
for $Q=24$ (upper curve) and $Q=21.5$ (lower curve). The dotted lines
correspond to $\Theta_{\eta \eta'}=-15^0$ (upper line) and $-25^0$
(lower line). The wedge between the two dashed lines corresponds
to $35 \leq R \leq 50$.}\label{fig:ellipse}
\end{figure}

The lowest--order mass formulas (\ref{eq:masses}) together with Dashen's
theorem (\ref{eq:Dashen}) lead to the Weinberg (1977) ratios
\beq
\frac{m_u}{m_d} = 0.55~, \qquad \qquad
\frac{m_s}{m_d} = 20.1 ~. \label{eq:mratio}
\eeq
These ratios are subject to higher--order corrections. The most important
ones are corrections of $O(p^4) = O(m_q^2)$ and $O(e^2 m_s)$. The
corrections of $O(p^4)$ were worked out by Gasser and Leutwyler (1985a)
who found that the ratios
\beqa
\frac{M_K^2}{M_\pi^2} &=& \frac{m_s + \hat{m}}{m_u + m_d} [1 + \Delta_M +
O(m_s^2)] \label{eq:MKpi} \\
\frac{(M_{K^0}^2 - M_{K^+}^2)_{\rm QCD}}{M_K^2 - M_\pi^2} &=&
\frac{m_d - m_u}{m_s - \hat{m}} [1 + \Delta_M + O(m_s^2)]
\eeqa
depend on the same correction $\Delta_M$ of $O(m_s)$:
\beq
\Delta_M = \frac{8(M_K^2 - M_\pi^2)}{F^2} (2 L_8^r - L_5^r) +
\mbox{ chiral logs~.} \label{eq:DelM}
\eeq
Using again (\ref{eq:Dashen}) to express $(M_{K^0}^2 -
M_{K^+}^2)_{\rm QCD}$ in terms of experimental mass differences
[$(M_{\pi^+}^2 - M_{\pi^0}^2)_{\rm QCD}$ is negligibly small], one obtains
a parameter--free relation for the ratio
\beq
Q^2 := \frac{m_s^2 - \hat{m}^2}{m_d^2 - m_u^2}: \label{eq:Qratio}
\eeq
\beq
Q^2 = \frac{M_K^2}{M_\pi^2} \cdot
\frac{M_K^2 - M_\pi^2}{M_{K^0}^2 - M_{K^+}^2 + M_{\pi^+}^2 - M_{\pi^0}^2}
\cdot \left[1 + O(m_s^2) + O\left( e^2 \frac{m_s}{m_d - m_u}
\right) \right].
\eeq
Without the corrections, this relation implies $Q \simeq 24$. As will be
discussed in Sect.~\ref{sec:virtual}, corrections of $O(e^2 m_s)$ to
Dashen's relation (\ref{eq:Dashen}) tend to decrease $Q$ by
approximately 10\%. Plotting $m_s/m_d$ versus $m_u/m_d$ leads to
Leutwyler's ellipse (Leutwyler, 1990) which is to a very good
approximation given by
\beq
\frac{1}{Q^2} \left( \frac{m_s}{m_d}\right)^2 + \left(
\frac{m_u}{m_d}\right)^2 = 1~.
\eeq
In Fig.~\ref{fig:ellipse}, the relevant quadrant of the ellipse is shown
for $Q = 24$ (upper curve) and $Q = 21.5$ (lower curve).

To the order considered, the quark mass ratios $m_s/m_d$ and $m_u/m_d$
cannot be determined separately from low--energy data alone due to an
accidental symmetry of $\cL_2 + \cL_4$ (Kaplan and Manohar, 1986). The
chiral Lagrangian is invariant to this order under the transformations
\\[5pt]
\centerline{
$m'_u = \alpha_1 m_u + \alpha_2 m_d m_s$ \qquad
 (and cyclic permutations)}
\beq
B' = B/\alpha_1, \qquad L'_6 = L_6 - \alpha, \qquad
L'_7 = L_7 - \alpha, \qquad L'_8 = L_8 + 2\alpha \label{eq:KM}
\eeq
$$
\alpha = \frac{\alpha_2 F^2}{32 \alpha_1 B}~.
$$
Consequently, all Green functions of vector and axial--vector currents
and thus all S--matrix elements are invariant under the
Kaplan--Manohar transformations (\ref{eq:KM}). Since $\Delta_M$ in
(\ref{eq:DelM}) is not invariant, it can only be fixed via Green
functions of (pseudo)scalar quark densities. Expressing $L_5$ and $L_8$
through the well--measured corrections to $F_K/F_\pi$ and to the
Gell-Mann--Okubo formula (\ref{eq:GMO}), $\Delta_M$ can also be written
in the form (Leutwyler, 1990)
\beq
\Delta_M = - \frac{32 (M_K^2 - M_\pi^2)}{F^2} \; L_7 - 0.33~,
\label{eq:DMnum}
\eeq
reducing the problem to a determination of the scale independent
LEC $L_7$.

Various arguments lead to a consistent picture for $L_7$ corresponding
to the value given in Table \ref{tab:Li}. Both at leading order in
$1/N_c$ (Gasser and Leutwyler, 1985a) and from saturation of spectral
function sum rules (Leutwyler, 1990), one finds
\beq
L_7 = - \frac{F^2}{48 M^2_{\eta'}}~.\label{eq:L7}
\eeq
$L_7$ also accounts for $\eta - \eta'$ mixing at $O(p^4)$ leading to
the relation (Gasser and Leutwyler, 1985a)
\beq
F^2(M^2_{\eta'} - M^2_\eta) \sin^2 \Theta_{\eta \eta'} =
- 24 L_7 (M^2_\eta - M^2_\pi)^2~.
\eeq
Experimentally, $\Theta_{\eta \eta'} \simeq - 20^0 $ (Review Part.
Prop., 1994). For a given value of $\Theta_{\eta \eta'}$,
Eq.~(\ref{eq:MKpi}) defines a linear relation between $m_s/m_d$ and
$m_u/m_d$. Allowing for higher--order chiral corrections, the dotted
lines in Fig.~\ref{fig:ellipse} correspond to $\Theta_{\eta \eta'} =
- 15^0$ (upper line) and $- 25^0$ (lower line), respectively. Using
$L_7$ from (\ref{eq:L7}), one obtains $\Delta_M \simeq -0.16$. Although
the destructive interference between the two terms in (\ref{eq:DMnum})
enhances the uncertainty, the conclusion is that $\Delta_M$ is indeed
small as expected for a higher--order correction.

There is completely independent information on the quark mass ratios
from mass splittings in the baryon octet. From an analysis of the
leading non--analytic corrections to the baryon masses of $O(m_q^{3/2})$
(Gasser, 1981; Gasser and Leutwyler, 1982), one can extract the ratio
\beq
R = \frac{m_s - \hat{m}}{m_d - m_u} = 43.5 \pm 3.2~.\label{eq:Rq}
\eeq
Information from $\rho - \omega$ mixing confirms this value
(Gasser and Leutwyler, 1982). Although electromagnetic corrections (Lebed and
Luty, 1994) are not included in these estimates, a rather conservative
range $35 \leq R \leq 50$ is used in Fig.~\ref{fig:ellipse} to define
the wedge between the two dashed lines. Independent information on $R$
from the branching ratio $\Gamma(\psi' \ra \psi \pi^0)/\Gamma(\psi'
\ra \psi \eta)$ is compatible with this range (Gasser and
Leutwyler, 1982; Donoghue and Wyler, 1992b; Luty and Sundrum, 1993a).

Altogether, a consistent picture emerges as shown in Fig.~\ref{fig:ellipse}.
The allowed values for the quark mass ratios are still
close to the Weinberg values (\ref{eq:mratio}).
Nevertheless, there are pertinacious claims in the literature
that $m_u = 0$ is a perfectly acceptable
value for the mass of the lightest quark. The issue at stake is the
problem of strong $CP$ violation [for a review, see Peccei (1989)] that would
find a natural solution for
$m_u = 0$. Let me therefore briefly review some arguments (Leutwyler,
1990, 1993, 1994c) that make this scenario very unrealistic.

For $m_u = 0$, one would be confronted with very large flavour
asymmetries in the matrix elements of (pseudo)scalar operators
jeopardizing the basic assumption of CHPT that the light quark masses can be
treated as perturbations. For instance, the lowest--order formula
for $M_{K^0} - M_{K^+}$ would be off by a factor four. Requiring the
higher--order terms in this mass difference not to exceed the
leading term, already leads to a lower bound $m_u/m_d > 1/3$. An often
quoted model with large flavour symmetry breaking and $m_u = 0$ is
due to Choi (1992, and references therein). However, Leutwyler
(1994c) has demonstrated that the model is in conflict with spontaneous chiral
symmetry breaking because it is incompatible with $B \neq 0$.
As the model is unable to account for a LEC of lowest order, it can hardly
be trusted for a calculation of higher--order corrections. Note
incidentally that for small $|B|$ Generalized CHPT (discussed in
Sect.~\ref{subsec:GCHPT}) requires a bigger value of $m_u$
than the standard scheme.

For the absolute magnitude of quark masses, the most reliable estimates
are based on QCD sum rules (e.g., Bijnens et al., 1994c; Jamin and M\"unz,
1994; Chetyrkin et al., 1994; Adami et al., 1993; Eletsky and Ioffe, 1993;
Dominguez et al., 1991; Narison, 1989; Dominguez and de Rafael, 1986).
Although the inherent uncertainties of the method are difficult to
quantify, the values from different sources are remarkably stable and
consistent with the standard values of Gasser and Leutwyler (1982).
The most recent determinations give for the running $\ol{\rm MS}$ masses
at a scale of 1 GeV
\beq
m_u + m_d = (12 \pm 2.5)~\mbox{MeV}\qquad (\mbox{Bijnens~et~al.,~1994c})
\eeq
and
\beq
m_s = \left\{\ba{ll}
(189 \pm 32)~\mbox{MeV}& \qquad (\mbox{Jamin~and~M\"unz,~1994}) \\
(171 \pm 15)~\mbox{MeV}& \qquad (\mbox{Chetyrkin~et~al.,~1994})
\ea \right.~.
\eeq
As for many other quantities of interest in the low--energy domain, the
final word is expected to come from lattice evaluations
(LATTICE 93, 1994).

\subsection{Odd--intrinsic--parity sector}
\label{subsec:anom}
The effective Lagrangian ${\cal L}_2 +{\cal L}_4$ is by construction
locally chiral invariant. Because
dimensional regularization preserves this symmetry, the one--loop
functional is invariant as well. On the other hand, the vacuum
transition amplitude (\ref{eq:Zdef}) is not invariant under the transformations
(\ref{eq:local}) due to the chiral anomaly.
The definition of the fermionic determinant responsible for the chiral
anomaly may be chosen such that the generating functional is invariant under
the transformations generated by the vector currents. For infinitesimal chiral
$SU(3)$ transformations
\beq
g_R(x)=1+i\alpha(x)+i\beta(x) ~, \quad g_L(x)=1+i\alpha(x)-i\beta(x) ~,
\eeq
the change in $Z$ then only involves the difference $\beta(x)$ between $g_R$
and $g_L$ (Bardeen, 1969; Wess and Zumino, 1971),
 \beqa\label{eq:anomaly}
\delta Z &=& -\int d^4x \langle \beta(x)\Omega(x)\rangle \nl
\Omega(x)&=&\frac{N_c}{16\pi^2}\epsilon^{\mu\nu\rho\sigma}\left
[v_{\mu\nu} v_{\rho\sigma} +\frac{4}{3}D_\mu a_\nu D_\rho a_\sigma
+\frac{2i}{3}\{v_{\mu\nu},a_\rho a_\sigma\} \right.\nl
&&\hspace{1.9cm}+\left.\frac{8i}{3}a_\rho
v_{\mu\nu}a_\sigma   +\frac{4}{3}a_\mu a_\nu a_\rho a_\sigma\right ]\nl
v_{\mu\nu}&=&\partial_\mu v_\nu -\partial_\nu v_\mu
-i[v_\mu,v_\nu]\nl
D_\mu a_\nu &=& \partial_\mu a_\nu -i[v_\mu,a_\nu]
\eeqa
$$ \varepsilon_{0123} = 1 ~.$$
Notice that $\Omega$ depends only on the external
fields $v_\mu$ and $a_\mu$, but not on the quark masses.
A functional $Z[U,l,r]$ that reproduces the chiral anomaly
was first constructed by Wess and Zumino (1971). For practical
purposes, it is useful to write it in the explicit form given by Witten
(1983):
\beqa
Z[U,l,r]_{\rm WZW} &=&-\dfrac{i N_c}{240 \pi^2}
\int_{M^5} d^5x \epsilon^{ijklm} \langle \Sigma^L_i
\Sigma^L_j \Sigma^L_k \Sigma^L_l \Sigma^L_m \rangle \label{eq:WZW} \\*
 & & - \dfrac{i N_c}{48 \pi^2} \int d^4 x
\varepsilon_{\mu \nu \rho \sigma}\left( W (U,l,r)^{\mu \nu
\rho \sigma} - W ({\bf 1},l,r)^{\mu \nu \rho \sigma} \right)
\nl
W (U,l,r)_{\mu \nu \rho \sigma} & = &
\langle U l_{\mu} l_{\nu} l_{\rho}U^{\dg} r_{\sigma}
+ \frac{1}{4} U l_{\mu} U^{\dg} r_{\nu} U l_\rho U^{\dg} r_{\sigma}
+ i U \partial_{\mu} l_{\nu} l_{\rho} U^{\dg} r_{\sigma}
\nl
& & + i \partial_{\mu} r_{\nu} U l_{\rho} U^{\dg} r_{\sigma}
- i \Sigma^L_{\mu} l_{\nu} U^{\dg} r_{\rho} U l_{\sigma}
+ \Sigma^L_{\mu} U^{\dg} \partial_{\nu} r_{\rho} U l_\sigma
\nl
& & -\Sigma^L_{\mu} \Sigma^L_{\nu} U^{\dg} r_{\rho} U l_{\sigma}
+ \Sigma^L_{\mu} l_{\nu} \partial_{\rho} l_{\sigma}
+ \Sigma^L_{\mu} \partial_{\nu} l_{\rho} l_{\sigma}  \\
& & - i \Sigma^L_{\mu} l_{\nu} l_{\rho} l_{\sigma}
+ \frac{1}{2} \Sigma^L_{\mu} l_{\nu} \Sigma^L_{\rho} l_{\sigma}
- i \Sigma^L_{\mu} \Sigma^L_{\nu} \Sigma^L_{\rho} l_{\sigma}\rangle
\nl
& & - \left( L \leftrightarrow R \right) ~,\no \eeqa
$$
\Sigma^L_\mu = U^{\dg} \partial_\mu U ~,\quad
\Sigma^R_\mu = U \partial_\mu U^{\dg} $$
where $\left( L \leftrightarrow R \right)$ stands for the interchange
$$
U \leftrightarrow U^\dg~, \qquad l_\mu \leftrightarrow r_\mu~~,
\qquad \Sigma^L_\mu \leftrightarrow \Sigma^R_\mu ~. $$
The first term in Eq.~(\ref{eq:WZW}) bears the mark of the
anomaly: this part of the action is local in {\it five} dimensions,
but it cannot be
written as a finite polynomial in $U$ and $\partial_\mu U$ in four dimensions.
This term involves at least five pseudoscalar fields. It
contributes to $K_{e5}$ decays, but its contribution
is proportional to the electron mass and therefore strongly suppressed
(Blaser, 1994). The convention used in Eq.~(\ref{eq:WZW}) ensures
that $Z[U,l,r]_{\rm WZW}$ conserves parity.

The Wess--Zumino--Witten functional contains all
anomalous contributions to electromagnetic and semileptonic weak meson
decays. However, this does not imply that there are no other contributions
to transitions of odd intrinsic parity, which are characterized by
an $\ve$ tensor in the amplitude. Clearly, up to and including
$O(p^4)$, only the chiral anomaly can contribute to such
amplitudes (at tree level). At next--to--leading order in the
odd--intrinsic--parity sector (one should avoid the misleading expression
``anomalous" sector), $O(p^6)$, there are two contributions analogous
to the even--parity sector at $O(p^4)$: the one--loop functional
with a single vertex from the Wess--Zumino--Witten Lagrangian
and tree--level diagrams from an effective chiral Lagrangian of $O(p^6)$
containing an $\ve$ tensor.

The renormalization of the one--loop functional was carried out
by several groups (Donoghue and Wyler, 1989b; Issler, 1989;
Bijnens et al., 1990b; Akhoury and Alfakih, 1991). Chiral power counting
shows that the coefficients of the Wess--Zumino--Witten functional
are not renormalized in higher orders. Since this functional satisfies
the anomalous Ward identity for the generating functional, all additional
higher--order terms in the odd--intrinsic--parity sector must be chiral
invariant. This can indeed be shown in an explicit regularization at
every order in the loop expansion (Issler, 1989). In the usual
way, the divergences of the one--loop functional are absorbed by
the coefficients of the chiral Lagrangian of $O(p^6)$ with odd intrinsic
parity. This Lagrangian has been written down by three groups with three
different results (Issler, 1989; Akhoury and Alfakih, 1991;
Fearing and Scherer, 1994). The most careful analysis is
due to Fearing and Scherer (1994) who find 32 independent terms.

It is rather clear that we will not be able to fix all 32 renormalized
LECs of $O(p^6)$ with odd intrinsic parity from experiment. Two approaches
have been pursued to arrive at (model dependent) predictions for
those coefficients: resonance saturation (Bijnens et al., 1990b; Pallante
and Petronzio, 1993; Moussallam, 1994) and the constituent quark
model based on the Nambu--Jona-Lasinio model (Bijnens, 1991).
Such predictions are essential to make contact with experiment
in the ``anomalous" sector. Among the observables investigated
are the slope parameters in the decays $P\to\gamma
\gamma^*$ ($P=\pi^0, \eta$). It turns out that the LECs of
$O(p^6)$ are absolutely necessary to understand the experimental
slopes because the loop amplitudes at reasonable values of the renormalization
scale are much too small. Both $V$ exchange (Bijnens et al.,
1990b) and the constituent quark model with $M_Q\simeq 250$ MeV (Bijnens,
1991) can account for the data. Another interesting area is the $P^3
\gamma$ complex, in particular $\gamma \pi^+ \to \pi^+ \pi^0$ and
$\eta \to \pi^+ \pi^- \gamma$. Again the finite parts are needed
although the loop contributions are not negligible in this case (Bijnens
et al., 1900b). For details I refer to the review by Bijnens
(1993a). There is also considerable interest now on the
experimental side to improve the existing measurements, especially
concerning the $\gamma\to 3 \pi$ transition (Miskimen, 1994; Moinester,
1994). Further applications of CHPT to $O(p^6)$ in the odd--intrinsic--parity
sector will be mentioned in the next subsection.

\subsection{Phenomenology at next--to--leading order}
\label{subsec:phenp4}

\subsubsection{$\pi\pi$ scattering}
\label{subsec:pipi}
$\pi\pi$ scattering is a traditional testing ground for spontaneously
broken chiral symmetry. It involves only the pseudo--Goldstone bosons
of chiral $SU(2)$. Near threshold, the chiral expansion for
$SU(2)$ is expected to converge rather rapidly because the
natural expansion parameter is of the order
\beq
\dfrac{4 M^2_\pi}{16 \pi^2 F^2_\pi} = 0.06~.
\eeq
Of course, chiral logarithms will often enhance this value.
Nevertheless, $\pi\pi$ scattering offers a crucial test of the
validity of the chiral expansion and of Standard CHPT in particular.
The elastic scattering of pions is also one of the more promising instances
to answer the questions raised by Generalized CHPT
(Sect.~\ref{subsec:GCHPT}).

The scattering amplitude for
\beq
\pi^a(p_a) + \pi^b(p_b) \to \pi^c(p_c) + \pi^d(p_d)
\eeq
is determined by a single scalar function $A(s,t,u)$ defined by the
isospin decomposition
\beqa
T_{ab,cd} &=& \delta_{ab}\delta_{cd} A(s,t,u) + \delta_{ac}\delta_{bd}
A(t,s,u) + \delta_{ad}\delta_{bc} A(u,t,s) \nl
A(s,t,u) &=& A(s,u,t) \qquad \mbox{[crossing]}
\eeqa
\beqa
s &=& (p_a + p_b)^2 = 4 (M^2_\pi + q^2) \nl
t &=& (p_a - p_c)^2 = -2 q^2 (1 - \cos \theta) \\
u &=& (p_a - p_d)^2 = -2 q^2 (1 + \cos \theta)~. \no
\eeqa
Here, $q$ and $\theta$ are the center--of--mass momentum and scattering angle.
The amplitudes $T^I(s,t)$ of definite isospin $(I = 0,1,2)$ in
the $s$--channel are decomposed into partial waves:
\beq
T^I(s,t)=32\pi\sum_{l=0}^{\infty}(2l+1)P_l(\cos{\theta})t_l^I (s)~.
\eeq
Unitarity implies that in the elastic region $4M_\pi^2 \leq s
\leq 16M_\pi^2$ the partial--wave amplitudes $t_l^I$ can be described
by real phase shifts $\delta_l^I$,
\beq
t_l^I(s)=\left(\dfrac{s}{s-4M_\pi^2}\right)^{1/2}\dfrac{1}{2i}
\{e^{2i\delta_l^I(s)}-1\}~.
\eeq
The behaviour of the partial waves  near threshold is of the form
\beq
\Re e\;t_l^I(s)=q^{2l}\{a_l^I +q^2 b_l^I +O(q^4)\}~.
\eeq
The quantities $a_l^I$ ($b_l^I$) are referred to as the $\pi\pi$
scattering lengths (slope parameters).

The amplitude $A(s,t,u)$ was evaluated to $O(p^4)$ in the framework of chiral
$SU(2)$ by Gasser and Leutwyler (1983, 1984). The threshold parameters
for $l\le 1$, $I\le 1$ are shown in Table \ref{tab:pipi} taken from a recent
compilation of Gasser (1994a). The results are compared with the lowest--order
predictions (Weinberg, 1966) and with experiment (Nagels et al., 1979).
As emphasized by Gasser (1994a), the $S$--wave scattering lengths
vanish in the chiral limit and are therefore especially
sensitive to electromagnetic corrections. Although these corrections have
not been calculated yet, one can get a feeling for their relevance
by using $M_{\pi^0}$ instead of the conventionally used $M_{\pi^+}$.
This lowers the value for $a_0^0$ by 0.016, comparable to the expected
size of higher--order chiral corrections.

\begin{table}
\begin{center}
\caption{ Threshold parameters in units of $M_{\pi^+}$ taken
from Gasser (1994a). The errors of the $O(p^4)$ results do not account
for higher--order corrections.}
\label{tab:pipi}
\vspace{1em}
\begin{tabular}{|llll|} \hline
 &$O(p^2)$&$O(p^4)$&Experiment\\ \hline
$a_0^0\qquad$ & $0.16\qquad$ & $0.20\pm0.005\qquad$ &$0.26\pm0.05
\qquad$\\
$b_0^0$&$0.18$&$0.25\pm0.02$&$0.25\pm0.03$\\
$a_1^1$&$0.030$&$0.038\pm0.003$&$0.038\pm0.002$\\
$b_1^1$&   &$(5\pm3)\times10^{-3}$&    \\ \hline
\end{tabular}
\end{center}
\end{table}

\begin{figure}
 \centerline{\epsfig{file=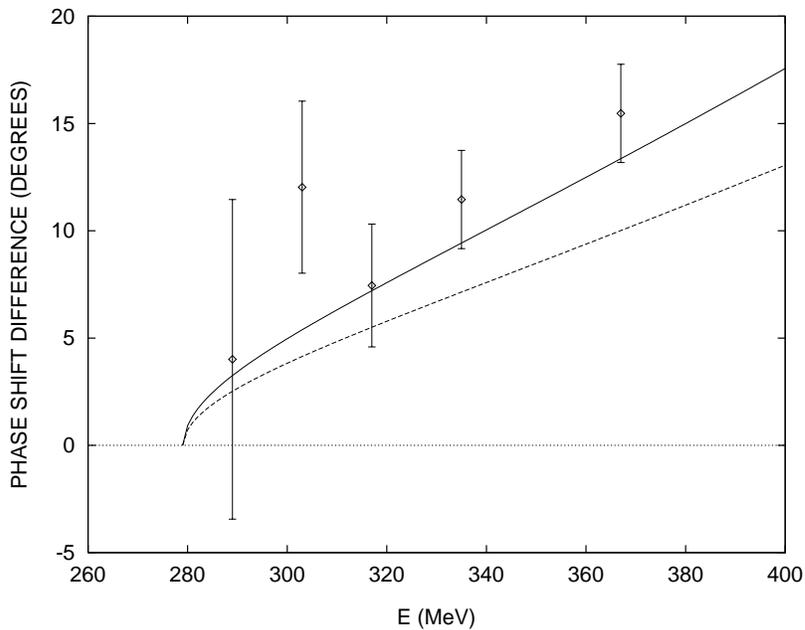,height=8cm}}
\caption{The phase shift difference $\delta_0^0-\delta_1^1$ as a
function of $E = \protect\sqrt s$ taken from Gasser (1994a). The data are from
Rosselet et al. (1977). The solid line stands for the CHPT result at
one--loop accuracy, whereas the dashed line displays the leading--order
term.} \label{fig:pipi}
\end{figure}

The phase shifts $\delta_l^I$ may be calculated from the partial--wave
amplitudes as (Gasser and Mei\ss ner, 1991a)
\beq
\delta_l^I(s)=(1-4M_\pi^2/s)^{1/2}\Re e\;t_l^I(s) + O(p^6)~.
\eeq
In Fig.~\ref{fig:pipi}, the phase shift difference $\delta_0^0 -
\delta_1^1$ is compared with $K_{e4}$ data from Rosselet et al. (1977).
Both in Fig.~\ref{fig:pipi} and in Table \ref{tab:pipi}, one
sees a clear improvement of the $O(p^4)$ predictions compared to
the lowest--order results.

At the present level of accuracy, CHPT to $O(p^4)$ is completely
compatible with the experimental information. However,
neglecting errors, one may be tempted to perceive a tendency for both
the scattering length $a_0^0$ and the phase shift difference $\delta_0^0
-\delta_1^1$ to be a little on the low side in comparison with experiment.
Clearly, one has to wait for more precise data, e.g., from $K_{e4}$
experiments at the $\Phi$ factory DA$\Phi$NE under construction
in Frascati (Maiani et al., 1992, 1995). A proposed experiment
at CERN (Czapek et al., 1992) measuring the lifetime of $\pi^+\pi^-$
atoms would be a completely independent source of information
for the difference $|a_0^0 - a_0^2|$.

On the theoretical side, attempts to calculate $A(s,t,u)$ to
$O(p^6)$ are under way (Bijnens et al., in preparation;
Colangelo, 1994; see also Sect.~\ref{sec:p6}). In addition, elastic
pion--pion scattering has been studied within the framework
of Generalized CHPT (Stern et al., 1993; Knecht et al., 1994c). Although
the amplitude $A(s,t,u)$ has been calculated to $O(p^4)$
in the generalized counting, the main message can
already be extracted from the lowest--order amplitude of
$O(p^2)$. In the notation of Sect.~\ref{subsec:GCHPT}, the amplitude
of $O(p^2)$ is given by
\beqa
A(s,t,u) &=& {1\over F^2_\pi} \left(s - {4\over 3}M^2_\pi\right) +
\alpha \dfrac{M^2_\pi}{3 F^2_\pi}\\*
\alpha &=& 1 + 6 \dfrac{r_2 -r}{r^2 -1} \left(1 + \dfrac{2 Z_0^S}{A_0}\right)
{}~.
\eeqa
Ignoring the Zweig suppressed parameter $Z_0^S$, it is easy to
check that both the scattering length $a_0^0$ and the phase shift
difference $\delta_0^0 - \delta_1^1$ increase as $r$ decreases from
its standard value $r_2$ (at tree level) towards its lower limit
$r_1$ given in (\ref{eq:r1r2}). For $r\simeq 10$, the
predictions agree nearly perfectly with experiment even if the errors
were considerably smaller. This conclusion is not much changed
by the $O(p^4)$ calculation especially since quite a number of new
unknown LECs enter for which only rough order--of--magnitude
estimates can be given (Knecht et al., 1994c). In any case,
Generalized CHPT can only make its point
if there are noticeable differences to the standard scenario already at
lowest order. It is interesting, although not completely
unrelated to $\pi\pi$ scattering, that a similar analysis of $\gamma
\gamma \to \pi^0\pi^0$ also favours $r\simeq 10$ (Knecht et al.,
1994a). We shall come back to this process in Sect.~\ref{sec:p6}
within the standard approach where the experimental data can be fully
understood on the basis of a calculation to $O(p^6)$ (Bellucci
et al., 1994).

\subsubsection{Semileptonic $K$ decays}
\label{subsub:semi}
All semileptonic $K$ decays, including radiative transitions,
that can be investigated experimentally in the near future
have now been calculated to $O(p^4)$ [for an up--to--date review,
see Bijnens et al. (1994d)]. Here, I want to review briefly the
recent work of Bijnens et al. (1994a) on $K_{l4}$ decays.

The decays
\beq
K \to \pi\pi \ell \nu \qquad (\ell=e, \mu)
\eeq
are described by four form factors $F$, $G$, $H$, $R$.
In CHPT, one sets up an energy expansion of the form
\beq
A= \dfrac{M_K}{F_\pi}\left\{ A_0 + A_2 + A_4 + \dots\right\}
\qquad (A=F,G,H,R)
\eeq
where $A_n$ is a quantity of $O(p^n)$ due to the generating functional
of $O(p^{n+2})$. The difference in the chiral dimension is of course
due to the external $W$ field and to the momenta in the definition
of the transition amplitude in terms of form factors.

The form factor
$H$ involves an $\ve$ tensor and therefore three momenta in the
amplitude. Consequently, $H_0=0$ and the leading contribution is
due to the Wess--Zumino--Witten functional.  For the decay $K^+\to
\pi^+ \pi^- \ell^+ \nu_\ell$, it is given by (Bijnens, 1990a;
Riggenbach et al., 1991)
\beq
      H = -\dfrac{\sqrt{2} M^3_K}{8 \pi^2 F^3_\pi} = -2.7 ~,
      \label{eq:H2}
\eeq
in excellent agreement with the experimental value $H=-2.68 \pm 0.68$
(Rosselet et al., 1977). The agreement persists if corrections of
$O(p^6)$, i.e. $H_4$, are included. The loop and the local
contributions (estimated with vector meson exchange, cf.
Sect.~\ref{subsec:anom}) almost compensate the renormalization
of the kaon decay constant, leaving the prediction (\ref{eq:H2}) practically
unchanged (Ametller et al., 1993).

The form factor $R$ was calculated to $O(p^4)$ by Bijnens et al.
(1994a). Effectively, it only contributes to $K_{\mu 3}$ decays because
its contribution to the rate is suppressed by a factor $m^2_\ell$.
The corrections of $O(p^4)$ for the form factors $F$, $G$ had
been calculated before (Bijnens, 1990a; Riggenbach et al., 1991) and
were found to be substantial. In fact, the experimental results from
$K_{e3}$ decays are 30 $\div$ 50 $\%$ above the leading contributions
$F_0$, $G_0$. In order to reduce the uncertainties of the chiral
prediction, Bijnens et al. (1994a) have estimated the
size of higher--order contributions to the form factor $F$ that
is dominantly $S$--wave. Following Donoghue et al. (1990), they write
down a dispersion relation for the $S$--wave projection of the relevant
amplitude and fix the subtraction constants via CHPT. Then they perform
a partial unitarization by employing an Omn\'es--type representation
for the partial--wave amplitude.
It turns out that the effects of final state interactions are sizable
in this case because they are related to the $I=0$ $S$--wave $\pi\pi$
phase shift.
This procedure leads both to a reduction of uncertainties
in the LECs $L_1$, $L_2$, $L_3$ and to improved predictions for
the form factor $G$ and for the total decay rates. In Table \ref{tab:Kl4},
I reproduce the results of fits (Bijnens et al., 1994a) to
$K_{e4}$ form factors and $\pi\pi$ threshold parameters to determine
the $L_i$ ($i=$1,2,3). One observes that inclusion of the unitarity
corrections produces a noticeable shift especially for $L_1$.
Moreover, the overall description of the $\pi\pi$ scattering data is
improved in this way, in particular for the $D$--wave scattering lengths.
The results in column 5 give rise to the new values of $L_1$, $L_2$,
$L_3$ in Table \ref{tab:Li}, taking into account theoretical error
estimates in addition to the purely statistical errors in Table \ref
{tab:Kl4}.

\begin{table}\centering
\caption{\label{tab:Kl4}
Results of  fits with  one--loop and unitarized  form factors, respectively
(Bijnens et al., 1994a). The
errors quoted for the $L^r_i$ are statistical only. The $L^r_i$ are given in
units of $10^{-3}$ at the scale $\mu=M_\rho$, the scattering lengths $a^I_l$
and the slopes $b_l^I$ in appropriate powers of $M_{\pi^+}$.}
\begin{center}
\vspace{.5cm}
\begin{tabular}{|c|cc|cc|c|}
\hline
         & \multicolumn{2}{c|}{ $K_{e4}$ data
alone}&\multicolumn{2}{c|}{$K_{e4}$  and $ \pi \pi$ data} & experiment\\
         & one--loop        & unitarized & one--loop & unitarized
&Nagels et al. (1979) \\ \hline
$L^r_1$& $0.65 \pm 0.27$&$0.36\pm0.26$  & $0.60\pm0.24$  &$0.37\pm0.23$ & \\
$L^r_2$& $1.63 \pm 0.28$ &$1.35\pm0.27$ &$1.50\pm0.23$    &$1.35\pm0.23$
&\\ $L_3$&$-3.4\pm1.0$   &$-3.4\pm1.0$   &$-3.3\pm0.86$ &$-3.5\pm0.85$  & \\
\hline

$ a_0^0$& $0.20$& $0.20$ &$ 0.20$ & $0.20$       &$0.26\pm0.05$\\

 $b_0^0$ & $0.26$ & $0.25$& $0.26$ & $0.25$&$0.25\pm0.03$\\

$-$10 $a_0^2$ &$0.40$ &$0.41$ &$0.40$ &$0.41$ &$0.28\pm0.12$\\

$-$10 $b_0^2$     &$0.67$ &$0.72$ &$0.68$ &$0.72$ &$0.82\pm0.08$\\

10 $a_1^1$     &$0.36$  &$0.37$  &$0.36$  &$0.37$  &$0.38\pm0.02$\\

$10^2 b_1^1$     &$0.44 $ &$0.47$  &$0.43$  &$0.48$  &\\

$10^2 a_2^0$     &$0.22 $  &$0.18$   &$0.21$   &$0.18$   &$0.17\pm0.03$ \\

$10^3 a_2^2$     &$0.39 $   &$0.21 $   &$0.37 $   &$0.20 $   &$0.13\pm0.3$\\
\hline
$\chi^2/N_\mtiny{DOF}$& $0/0$&$0/0$&$8.8/7$&$4.9/7$&  \\
\hline
\end{tabular}
\end{center}
\end{table}

Once the leading partial waves are known from, e.g.,
$K^+\rightarrow \pi^+\pi^-e^+\nu_e$ decays, the chiral representation allows
one to predict the remaining rates with rather small uncertainties
(Bijnens et al., 1994a). In the near future, substantial higher
precision can be expected in $K_{l4}$ experiments, e.g., at the $\Phi$
factory DA$\Phi$NE (Maiani et al., 1992, 1995). This precision will be
especially welcome for extracting the $\pi\pi$ phase shift difference
$\delta^0_0 - \delta^1_1$ and for determining the corresponding threshold
parameters with small uncertainties (see also Sects.~\ref
{subsec:pipi} and \ref{subsec:GCHPT}).

\subsubsection{The decays {\bf $P \to \ell^+ \ell^-$} ~({\bf $P=\pi^0, \eta,
K^0$;~ $\ell=e, \mu$})}
\label{subsec:ll}
The decays of pseudoscalar mesons into a lepton pair proceed through
two--photon intermediate states and via local $P \ell^+ \ell^-$ operators.
The latter have not been included in the effective Lagrangians discussed
so far, but it is straightforward to do so.

The amplitude for the decay of a neutral spin--0 meson $P$ of
mass $M$ into a lepton pair has the general form
\beq
T(P \ra \ell^+ \ell^-) = \bar u(iB + A\gamma_5) v
\eeq
with a decay rate
\beq
\Gamma(P \ra \ell^+ \ell^-) = \frac{M\beta_\ell}{8\pi}
(|A|^2 + \beta^2_\ell |B|^2), \qquad
\beta_\ell = \left( 1 - \frac{4 m_\ell^2}{M^2} \right)^{1/2}.
\eeq
In the limit of $CP$ conservation, the
transitions for $P=\pi^0, \eta, K^0_2$ proceed via the $S$--wave
amplitude $A$, while the decay with $P=K^0_1$ is determined by the
$P$--wave amplitude $B$.

For $P=\pi^0$ or $\eta$, the strong and electromagnetic
interactions are relevant. The effective Lagrangian for
$P \to \ell^+ \ell^-$ of lowest chiral dimension has the form (Savage et
al., 1992)
\beq
\cL (P \to \ell^+ \ell^-) = \dfrac{3 i \alpha^2}{32 \pi^2} \bar \ell \gamma^\mu
\gamma_5 \ell\left\{2 \chi_1 \langle Q^2 U \del_\mu U^\dg \rangle
+ \chi_2 \langle Q(UQ\del_\mu U^\dg - \del_\mu UQU^\dg)\rangle\right\}
\label{eq:SLW}
\eeq
where $\ell=e$ or $\mu$ and $Q$ is the quark charge matrix defined in
Eq.~(\ref{eq:gf}). The decay amplitudes depend on the sum
$\chi_1 + \chi_2$. In fact, this coefficient is needed to renormalize
the divergent one--loop contribution with a two--photon intermediate
state and an anomalous $P \gamma\gamma$ vertex (Savage et al., 1992).
Thus, one can determine the renormalized LEC $\chi_1^r + \chi_2^r$
from one decay rate to get absolute predictions for the
remaining ones. Taking the recently measured branching ratio (Kessler
et al., 1993; Review Part. Prop., 1994) $BR(\eta \to \mu^+ \mu^-)=
(5.7 \pm 0.8)\cdot 10^{-6}$ as input, Savage et al. (1992) predicted
\beqa
 BR(\pi^0 \to e^+ e^-) &=& (7 \pm 1)\cdot 10^{-8}\\
 BR(\eta \to e^+ e^-) &=& (5 \pm 1)\cdot 10^{-9}\no
\eeqa
compared with the present experimental values (Review Part. Prop.,
1994)
\beqa
 BR(\pi^0 \to e^+ e^-)_{\rm exp} &=& (7.5 \pm 2.0)\cdot 10^{-8}\\
 BR(\eta \to e^+ e^-)_{\rm exp} &<& 3\cdot 10^{-4} \qquad (90\% {\rm~c.l.})
{}~.\no
\eeqa

The nonleptonic weak interactions (cf. Sect.~\ref{sec:nonleptonic}) are
responsible for the decays $K^0 \to \ell^+ \ell^-$. $CPS$ invariance
(Bernard et al., 1985; Crewther, 1986; Leurer, 1988)
and chiral symmetry dictate the lowest--order coupling
(Ecker and Pich, 1991)
\beq
\del^\mu K_2^0 \bar \ell \gamma_\mu \gamma_5 \ell \label{eq:Kll}
\eeq
contributing only to the $S$--wave amplitude $A$. There is in fact
a well--known short--distance contribution of this type for
$K_L \to \ell^+ \ell^-$ (Gaillard and Lee, 1974) with an important top--quark
contribution. However, this decay is dominated by the absorptive part of
the transition $K_L \to \gamma^* \gamma^* \ra \ell^+ \ell^-$~:
\beq
|\mbox{$\Im m$}~A_2^{\gamma\gamma}| = \dfrac{\alpha m_\ell}{4\beta_\ell M_K}
\ln \dfrac{1 + \beta_\ell}{1 - \beta_\ell}
\left[ \dfrac{64\pi \Gamma(K_L \ra 2\gamma)}{M_K}\right]^{1/2}
\eeq
or
\beq
BR(K_L \to \ell^+ \ell^-)_{\rm abs}
\simeq \dfrac{\alpha^2 m_\ell^2}{2\beta_\ell M_K^2}
\left(\ln \dfrac{1 + \beta_\ell}{1 - \beta_\ell} \right)^2 B(K_L \ra 2\gamma)~.
\eeq
The experimental branching ratio (Review Part. Prop., 1994)
$BR(K_L \ra 2\gamma) = (5.73 \pm 0.27) \cdot 10^{-4}$ yields for
$\ell = \mu$
\beq
|\mbox{$\Im m$}~A_2^{\gamma\gamma}| = (2.21 \pm 0.05) \cdot 10^{-12}
\eeq
and
\beq
BR(K_L \ra \mu^+ \mu^-)_{\rm abs} = (6.85 \pm 0.32) \cdot 10^{-9}~.
\eeq
Comparing with the latest measurements
\beq
BR(K_L \ra \mu^+ \mu^-) = \left\{ \ba{lr}
(7.9\pm 0.6 \pm 0.3) \cdot 10^{-9} & (\mbox{Akagi~et~al.,~1991})\\
(7.0\pm 0.5)\cdot 10^{-9} & (\mbox{Heinson~et~al.,~1991})\\
(7.4\pm 0.4)\cdot 10^{-9} & (\mbox{Review~Part.~Prop.,~1994})
\ea \right.~,
\eeq
one finds that the two--photon absorptive part (unitarity bound) nearly
saturates the total rate.

The dispersive part for the two--photon intermediate state is
model dependent. There are various models in the literature
(Bergstr\"om et al., 1983, 1990; B\'elanger and Geng, 1991; Ko, 1992)
that make different predictions for the dispersive part. As
long as not even the sign of the interference between the dispersive
and the short--distance part is definitely established, the decay
$K_L\to \mu^+\mu^-$
is of limited use for the determination of the CKM mixing matrix
element $V_{td}$. For $K_L\to e^+ e^-$, the unitarity bound
due to the two--photon absorptive part is $BR(K_L\to e^+ e^-)=
3\cdot 10^{-12}$.

The decays $K_S\to \ell^+ \ell^-$ are theoretically interesting because the
lowest--order amplitude in CHPT is unambiguously calculable
(Ecker and Pich, 1991). It is given by a two--loop diagram describing the
transition $K_1^0 \ra \gamma^* \gamma^* \ra \ell^+ \ell^-$. The loop amplitude
must be finite because, unlike (\ref{eq:Kll}), there is no corresponding
counterterm for $K^0_1$.
Normalizing to the rate for $K_S\to \gamma\gamma$, one obtains
the relative branching ratios (Ecker and Pich, 1991)
\beq
\dfrac{\Gamma(K_S \ra \mu^+ \mu^-)}{\Gamma(K_S \ra \gamma \gamma)} =
2 \cdot 10^{-6}~, \qquad \qquad
\dfrac{\Gamma(K_S \ra e^+ e^-)}{\Gamma(K_S \ra \gamma \gamma)} =
8 \cdot 10^{-9} ~,
\eeq
well below the present experimental upper limits (Review Part.
Prop., 1994).

\subsubsection{$\eta$ decays}
\label{subsec:eta}
Although there are still several open questions in $\eta$ decays,
progress has recently been made concerning the rates of the
dominant hadronic decay modes $\pi^+ \pi^- \pi^0$ and $3 \pi^0$.
To $O(p^4)$, the rate for the charged channel can be written (Gasser
and Leutwyler, 1985c) as
\beq
\Gamma (\eta \to \pi^+ \pi^- \pi^0) = \dfrac{\Gamma_0}{Q^4}~,
\qquad \Gamma_0 \simeq 51 ~\mbox{MeV}~,\label{eq:eta3pi}
\eeq
where $\Gamma_0$ depends only on pseudoscalar masses and decay constants
and on $L_3$ (cf. Table \ref{tab:Li}). Using Dashen's theorem to
calculate $Q \simeq 24$ (see Sect.~\ref{subsec:mq}) and accounting
for possible higher--order chiral corrections, Gasser and Leutwyler
(1985c) found $\Gamma (\eta \to \pi^+ \pi^- \pi^0) = (160 \pm
50)$ eV. In comparison, the present experimental value for this rate
is  (283 $\pm$ 26) eV (Review Part. Prop., 1994).

This long--standing problem has come closer to a solution through
the recent calculations of corrections to Dashen's theorem discussed
in Sect.~\ref{subsec:appvirt}. For instance, taking $Q=21$ in
(\ref{eq:eta3pi}) leads to $\Gamma (\eta \to \pi^+ \pi^- \pi^0) = 262$ eV.
Conversely, using the experimental width in (\ref{eq:eta3pi}) gives
$Q=20.6 \pm 0.5$, compatible with the estimates in
Sect.~\ref{subsec:appvirt}. Before declaring the $\eta \to 3 \pi$ problem
to be solved, some more work is certainly needed. The effect of
final state interactions beyond $O(p^4)$ should be estimated
(Anisovich, 1993; Kambor et al., 1994b). In addition,
including the relatively large $\eta - \eta^\prime$ mixing beyond $O(p^4)$
(where it only enters via the LEC $L_7$) could modify the picture (Pich, 1990).

A step in this direction has recently been taken by Moussallam (1994) who
investigated $O(p^6)$ corrections to the decays $P \to \gamma\gamma$
($P=\pi^0, \eta, \eta^\prime$). He argues that in order to perform the
large $N_c$ expansion consistently with the chiral expansion one
should treat terms of $O(1/N_c)$ effectively as $O(p^2)$.
Working to $O(p^4)$ according to this general counting with the
$\eta^\prime$ as an explicit field in the chiral Lagrangian, he finds
$\eta_8 - \eta_0$ mixing both in the kinetic and in the mass terms.
Consequently, the diagonalization requires two different mixing angles
instead of the familiar single one of previous treatments. In order to
obtain agreement with the experimental $P \to \gamma\gamma$ widths, a
local term of $O(p^6)$ involving the external (pseudo)scalar field and
therefore the quark mass matrix is needed. A small increase of $\Gamma (\pi^0
\to 2 \gamma)$ is also predicted (Moussallam, 1994), but much higher
experimental precision would be required to detect such a small correction.

The decay $\eta \ra \pi^0 \gamma \gamma$ is an interesting
example of a transition where the naive chiral dimensional analysis
fails dramatically. The decay
has many things in common with $\gamma \gamma \ra \pi^0 \pi^0$
(cf. Sect.~\ref{subsec:ggpipi}):
there are no contributions from either $\cL_2$ or $\cL_4$ and
the amplitude is given exclusively by the finite one--loop contribution.
However, the corresponding rate (Ecker, 1989c)
\beq
\Gamma (\eta \ra \pi^0\gamma\gamma)_{\rm loop} = 0.35 \cdot 10^{-2}\mbox{ eV}
\eeq
is about a factor 230 lower than the (single) experimental measurement
(Binon et al., 1982)
\beq
\Gamma (\eta \ra \pi^0 \gamma\gamma)_{\rm exp} = (0.82 \pm 0.17)
\mbox{ eV}~. \label{eq:etaexp}
\eeq
The reason for this discrepancy is well understood: the
usually dominating pion--loop amplitude is suppressed
by  $m_u - m_d$, while the kaon--loop contribution is small anyway. Not even
the  two--loop amplitude is of much help (Ecker, 1989c).
The dominant contribution arises at $O(p^6)$ due to vector meson
exchange and gives at least the right order of magnitude (Cheng,
1967). Ametller et al. (1992) have put the
relevant contributions together including effects of $O(p^8)$
and higher. Assuming positive interference between the significant
contributions, they conclude that the theoretical answer for the rate is
probably still lower than the experimental result (\ref{eq:etaexp}), but
the discrepancy is no more than two standard deviations. Additional
resonance contributions have been investigated by Ko (1993a).

There are many other $\eta$ decays worth investigating. The decays
$\eta \to \gamma \gamma^*$ and $\eta \to \pi^+ \pi^- \gamma$ are
considered in Sect.~\ref{subsec:anom}.
The transitions $\eta \to \ell^+ \ell^-$ are discussed in the previous
subsection and the semileptonic $\eta_{l3}$ decays are analysed in
Sect.~\ref{subsec:appvirt}. Further calculations of
semileptonic $\eta$ decays, all below $10^{-13}$ in branching ratio,
can be found in a recent paper of Bramon and Shabalin (1994). Ko (1993b)
has given the following estimates for channels that could soon be
(or have already been) detected:
\beqa
BR(\eta \to \pi^+ \pi^- e^+ e^-) & \simeq & 7 \cdot 10^{-4} \nl
BR(\eta \to \pi^+ \pi^- \mu^+ \mu^-) & \simeq & 6 \cdot 10^{-8} \\
BR(\eta \to \pi^0 \pi^0 e^+ e^-) & \simeq & 7 \cdot 10^{-8} \no~.
\eeqa
With a special model incorporating vector mesons, Picciotto and
Richardson (1993) obtained $BR(\eta \to \pi^+ \pi^- e^+ e^-)$ =
$(3.2 \pm 0.3)\cdot 10^{-4}$.

\setcounter{equation}{0}
\setcounter{subsection}{0}
\setcounter{table}{0}
\setcounter{figure}{0}

\section{NONLEPTONIC WEAK INTERACTIONS OF MESONS}
\label{sec:nonleptonic}
\subsection{Generating functional of $O(G_F p^4)$}
\label{subsec:GFp4}
At tree level, the effective Lagrangian $\cL_2^{\Delta S =1}$ in
(\ref{eq:G827}) gives rise to the current algebra relations between
$K\to 2 \pi$ and $K\to 3 \pi$ amplitudes. Even more so than in the strong
sector (see especially Sect.~\ref{subsec:Krad}), it is necessary to go
beyond the leading order in the nonleptonic sector for a meaningful comparison
with experiment.

As in the strong sector, the nonleptonic weak amplitudes of $O(G_F p^4)$
consist in general of several parts, which may contain arbitrary
tree structures associated with the lowest--order strong
Lagrangian $\cL_2$ :
\begin{enumerate}
\item[i.] Tree--level amplitudes from the effective chiral
Lagrangian $\cL_4^{\Delta S=1}$ of $O(G_F p^4)$ with the
transformation properties of a nonleptonic weak Lagrangian.
\item[ii.] One--loop amplitudes from diagrams with a single vertex
of $\cL_2^{\Delta S=1}$ in the loop.
\item[iii.] Reducible tree--level amplitudes with a single vertex from
$\cL_2^{\Delta S=1}$ and with a single vertex from $\cL_4$ in
(\ref{eq:L4}) or from the anomalous action (\ref{eq:WZW}).
\item[iv.] One--loop amplitudes of the reducible type, consisting
of a strong loop diagram connected to a vertex of $\cL_2^{\Delta S=1}$
by a single meson line. A frequently occurring diagram of this type
contains an external $K-\pi$ or $K-\eta$ transition as weak vertex,
possibly with one or two photons (generalized ``pole diagrams''). The
calculation of such diagrams can be simplified by
rediagonalizing the kinetic and mass terms of $\cL_2 + \cL_2^{\Delta
S=1}$ [``weak rotation'' (Ecker et al., 1987b, 1988)].
\end{enumerate}

The most general weak chiral Lagrangian of $O(G_F p^4)$ with the
appropriate $(8_L,1_R)$ and $(27_L,1_R)$ transformation properties
is quite involved. Using $CPS$ invariance (Bernard et al., 1985; Crewther,
1986; Leurer, 1988), the lowest--order equation of motion and the
Cayley--Hamilton theorem for 3--dimensional matrices, 35 independent
structures (plus 2 contact terms involving external fields only)
remain in the octet sector alone (Kambor et al., 1990; Ecker, 1990b;
Esposito-Far\`ese, 1991). However, many of those couplings contribute
only to transitions that involve an external $W$ boson in addition
to the nonleptonic weak transition.
Restricting attention to those terms where the only external gauge fields
are photons (or $Z$ bosons), 22 relevant
octet terms remain (Ecker et al., 1993b). A similar reduction takes
place in the 27--plet sector. The octet Lagrangian can be written as
\beqa
\cL_4^{\Delta S=1} &=& G_8 F^2 \sum_{i} N_i W_i \nl
&=& G_8 F^2 \{
N_1\; \langle \lambda D_\mu U^\dg D^\mu U D_\nu U^\dg D^\nu U \rangle
+ N_2\; \langle \lambda D_\mu U^\dg D^\nu U D_\nu U^\dg D^\mu U
\rangle \nl
&& + N_3\; \langle \lambda D_\mu U^\dg D_\nu U\rangle \langle D^\mu U^\dg D^\nu
U \rangle +
N_4\; \langle \lambda D_\mu U^\dg U\rangle \langle U^\dg D^\mu U D_\nu
U^\dg D^\nu U \rangle \nl
&& + N_5\; \langle \lambda \{ U^\dg \chi + \chi^\dg U, D_\mu U^\dg D^\mu U\}
\rangle +
N_6\; \langle \lambda D_\mu U^\dg U\rangle \langle U^\dg D^\mu U
(U^\dg \chi + \chi^\dg U)\rangle \nl
&& + N_7\; \langle \lambda (U^\dg \chi + \chi^\dg U)\rangle \langle D_\mu U^\dg
D^\mu U \rangle +
N_8\; \langle \lambda D_\mu U^\dg D^\mu U\rangle \langle  U^\dg \chi +
\chi^\dg U \rangle \nl
&& + N_9\; \langle \lambda [U^\dg \chi - \chi^\dg U, D_\mu U^\dg D^\mu U]
\rangle +
N_{10}\; \langle \lambda (U^\dg \chi + \chi^\dg U)^2\rangle \nl
&& + N_{11}\; \langle \lambda (U^\dg \chi + \chi^\dg U)\rangle \langle U^\dg
\chi + \chi^\dg U\rangle +
N_{12}\; \langle \lambda (U^\dg \chi - \chi^\dg U)^2\rangle \label{eq:L4w}\\
&& + N_{13}\; \langle \lambda (U^\dg \chi - \chi^\dg U)\rangle \langle U^\dg
\chi
- \chi^\dg U\rangle +
N_{14}\; i \langle \lambda \{ F_L^{\mu\nu} + U^\dg F_R^{\mu\nu} U,
D_\mu U^\dg D_\nu U\} \rangle \nl
&& + N_{15}\; i \langle \lambda D_\mu U^\dg (U F_L^{\mu\nu} U^\dg +
F_R^{\mu\nu}) D_\nu U \rangle +
N_{16}\; i \langle \lambda \{ F_L^{\mu\nu} - U^\dg F_R^{\mu\nu} U,
D_\mu U^\dg D_\nu U\} \rangle \nl
&& + N_{17}\; i \langle \lambda D_\mu U^\dg (U F_L^{\mu\nu} U^\dg -
F_R^{\mu\nu}) D_\nu U \rangle +
2 N_{18}\; \langle \lambda (F_L^{\mu\nu} U^\dg F_{R\mu\nu} U +
U^\dg F_{R\mu\nu} U F_L^{\mu\nu}) \rangle \nl
&& + N_{28}\; i \ve_{\mu\nu\rho\sigma} \langle \lambda D^\mu U^\dg U \rangle
\langle U^\dg D^\nu U D^\rho U^\dg D^\sigma U \rangle +
2 N_{29}\; \langle \lambda [U^\dg \wt F_R^{\mu\nu} U, D_\mu U^\dg D_\nu U]
\rangle \nl
&& + N_{30}\; \langle \lambda U^\dg D_\mu U \rangle \langle (\wt F_L^{\mu\nu} +
U^\dg \wt F_R^{\mu\nu} U) D_\nu U^\dg U \rangle \nl
&& + N_{31}\; \langle \lambda U^\dg D_\mu U \rangle \langle (\wt F_L^{\mu\nu} -
U^\dg \wt F_R^{\mu\nu} U) D_\nu U^\dg U \rangle \} \nl
&& + {\rm h.c.} +\dots \no
\eeqa
$$
\lambda = \frac{1}{2} (\lambda_6 - i \lambda_7)~, \qquad
\wt F_{A\mu\nu} = \ve_{\mu\nu\rho\sigma} F_A^{\rho\sigma}
\quad (A=L,R)~,
$$
with dimensionless coupling constants $N_i$. Only the 22 relevant octet
operators $W_i$ are listed in (\ref{eq:L4w}).

At first sight, it would seem that the predictive power of a
completely general chiral analysis using only symmetry constraints
is very limited. However, especially for the radiative $K$ decays
to be discussed in Sect.~\ref{subsec:Krad}, only a small subset of
the terms in the Lagrangian (\ref{eq:L4w}) contribute to the amplitudes.
Also for the dominant decay modes $K\to 2\pi, 3 \pi$, only a few
combinations of the $N_i$ contribute as will be discussed in
Sect.~\ref{subsec:Kpipi}.

As in the strong sector, the weak loop amplitudes
are in general divergent. Since the strong one--loop functional
of $O(p^4)$ has already been renormalized, the sum of the
reducible amplitudes (items iii and iv of the previous classification)
is finite and scale independent. The remaining renormalization
of the irreducible weak amplitudes (i, ii) proceeds in exactly
the same way as in Sect.~\ref{subsec:p4}.
Corresponding to Eq.~(\ref{eq:Liren}), the weak
couplings $N_i$ are decomposed as
\beq
N_i = N_i^r(\mu) + Z_i \Lambda(\mu)~.  \label{eq:Zi}
\eeq
The constants $Z_i$ are chosen to absorb
the one--loop divergences in the amplitudes (Kambor et al., 1990;
Ecker, 1990; Esposito-Far\`ese, 1991; Ecker et al., 1993).
The renormalized coupling constants
$N_i^r(\mu)$ are measurable quantities. As in the strong sector, the
scale dependences of the coupling constant and of the loop amplitude
cancel in the total amplitudes. The final amplitudes of $O(G_F p^4)$ are
finite and scale independent.

\subsection{Low--energy constants of $O(G_F p^4)$}
\label{subsec:Ni}
Both from a phenomenological and from a theoretical point of view,
the situation of the weak LECs of $O(G_F p^4)$ is much less advanced
than in the strong sector. The chiral duality observed for the $L_i$
is at least motivation enough to investigate the contributions
of meson resonances also to the weak nonleptonic Lagrangian.

However, it is clear from the outset that resonance exchange
cannot be the whole story in this case. For instance, there are genuine weak
short--distance contributions which have no equivalent in the strong
sector, such as the electromagnetic penguin
 contribution to the weak coupling constants $N_{14}$, $N_{16}$ and
$N_{18}$ (Ecker et al., 1988; Bruno and Prades, 1993). Another important
contribution to the weak constants is due to the chiral anomaly.
As will be discussed in Sect.~\ref{subsec:anomweak}, the chiral anomaly
contributes to and probably dominates the ``magnetic"
coupling constants $N_{28},\dots, N_{31}$.
Nevertheless, it is useful to compile the possible contributions
of meson resonances to the $N_i$. Such a general investigation of resonance
effects in the weak chiral Lagrangian was performed by Ecker et al. (1993b).
The result is as expected: one can understand which type of
resonances can in principle contribute to the different $N_i$,
but more quantitative predictions are impossible as long as
the weak resonance couplings are unknown.

A more predictive framework is based on the idea of
factorization [see Pich and de Rafael (1991) and references therein].
To leading order in $1/N_c$ and in the QCD coupling constant
$\alpha_s$, a generic four--quark operator with $V,A$ structure
is realized as a product of chiral currents, e.g.,
\beq
\ol{q}_{lL}\gamma^{\mu} q_{kL} \ol{q}_{jL}\gamma_{\mu} q_{iL}
 \Ra J_{lk}^{\mu} J_{\mu,ji}
\label{eq:fac}
\eeq
$$
J_\mu = \frac{\delta S_{\rm strong}}{\delta l^\mu} =
J_\mu^{(1)} + J_\mu^{(3)} + \ldots ~,
$$
where $S_{\rm strong}$ is the chiral action for the strong interactions
and $l^\mu$ is the octet of left--chiral external gauge fields.
The left--chiral current $J_\mu$ is decomposed into pieces of chiral
dimension $D = 1,3,\ldots$ with
\beq
J_\mu^{(1)} = - \frac{i}{2} \; F^2 U^\dg D_\mu U ~.
\eeq
The current of $O(p^3)$ has both a normal part depending on the $L_i$
and a parameter--free anomalous part
due to the Wess--Zumino--Witten functional (\ref{eq:WZW}). Here,
I only keep the part with even intrinsic parity and refer to
Sect.~\ref{subsec:anomweak} for a treatment of the chiral anomaly in
nonleptonic weak interactions.

The factorization model (Cheng, 1990; Bijnens et al., 1992;
Ecker et al., 1993b) is defined by the following
Lagrangian of $O(G_F p^4)$,
\beq
\cL^{\Delta S=1}_{4,{\rm FM}} = 4 k_f G_8 \langle \lambda_6 \{ J_\mu^{(1)},
J^{(3)\mu}\}\rangle ~, \label{eq:FM}
\eeq
with a fudge factor $k_f = O(1)$ to allow for a different overall scale
compared to $\cL_2^{\Delta S =1}$.
If factorization is to make sense, the scale factor $k_f$ is expected
to  be $O(1)$. The basic factorization relation (\ref{eq:fac})
contains no  reference to the QCD scale $\mu$. Consequently, the
non--leading corrections must provide the necessary $\mu$ dependence
(Pich and de Rafael, 1991) to compensate the scale
dependence of the Wilson coefficients $C_i(\mu)$ in
the effective Hamiltonian (\ref{eq:Hnl}). A minimal
improvement  of naive factorization ($k_f = 1$) is to put those
corrections into the fudge factor $k_f$ in Eq.~(\ref{eq:FM}). There is
of course no guarantee that all terms in (\ref{eq:FM}) have the same
scale factor.  The FM as defined here assumes this to be a reasonable
approximation with
$k_f$ left as a free parameter. A systematic improvement of this
assumption has been performed by Bruno and Prades (1993) using the method
of the effective action (Espriu et al., 1990). Another model
with the same structure of the weak Lagrangian of $O(G_F p^4)$
as in (\ref{eq:FM}) is the weak deformation model (Ecker et al., 1990a)
that predicts in addition $k_f = 1/2$. Although factorization
is in principle independent of resonance exchange, the factorization
model (\ref{eq:FM}) can also be viewed as a special model for the contributions
of resonances to the weak LECs $N_i$, at least to the extent that the
$L_i$ are saturated by resonance exchange (chiral duality).
We come back to the factorization model in the
following subsections. At present, the model can
accommodate the available experimental information, but it has
not really been put to a decisive test yet.

\subsection{$K\to 2 \pi,~3 \pi$}
\label{subsec:Kpipi}
Kambor et al. (1991) have calculated the dominant decay modes
$K \to 2\pi,3\pi$ to next--to--leading order in CHPT. In addition to the
two isospin amplitudes $A_0(\Delta I = 1/2)$ and $A_2(\Delta I = 3/2)$
for $K \ra 2\pi$, there are 10 measurable quantities in the standard
expansion of the $K \to 3\pi$ amplitudes to fourth order in the momenta
(Devlin and Dickey, 1979):
\beqa
\Delta I = \frac{1}{2} &:& \qquad \alpha_1,\beta_1,\zeta_1,\xi_1 \nl
\Delta I = \frac{3}{2} &:& \qquad \alpha_3,\beta_3,\gamma_3,\zeta_3,
\xi_3,\xi'_3~.
\eeqa

The current algebra analysis of $O(G_F p^2)$ predicts 7 observable quantities
in terms of the two parameters $G_8$, $G_{27}$ [the quadratic slope
parameters $\zeta_i$, $\xi_i$, $\xi'_3$ vanish to $O(G_F p^2)$]. It has been
known for a long time that those predictions show the right qualitative
trend only. There are sizable discrepancies between theory and experiment
at the current algebra level.

The agreement between CHPT and experiment is improved
substantially to $O(G_F p^4)$ (Kambor et al., 1991). One may worry that this
improvement is mainly due to the unknown LECs $N_i$ in the amplitudes.
A priori, there are 13 weak LECs ($N_1$,\dots, $N_{13}$) that could contribute
in the octet amplitudes. Taking a closer look at
the corresponding operators $W_i$ in Eq.~(\ref{eq:L4w}), one finds that
$W_4$ and $W_6$ can only contribute to transitions involving at least
five pseudoscalar fields. Likewise, $W_1$ and $W_2$ are
identical up to and including $O(\vp^4)$ so that only the sum $N_1+N_2$
appears in observable transition amplitudes. A further reduction
of counterterms occurs if one neglects terms of $O(M^2_\pi)$.
In addition to the tree--level (only for $A_0$, $\alpha_1$ and $\beta_1$)
and loop contributions, the octet amplitudes have the following local
terms of $O(G_F p^4)$ (Kambor et al., 1991):
\beqa
A_0 &=& -\dfrac{2 \sqrt{6}G_8 M^2_K(M^2_K - M^2_\pi)}{F_K} k_1\no \\*
\alpha_1 &=& - \dfrac{2 G_8 M^4_K}{9 F_K F_\pi} \left\{3 k_1 - k_2
+8(2 L_1 + 2 L_2 + L_3)\right\}\nl
\beta_1 &=& - \dfrac{G_8 M^2_K M^2_\pi}{3 F_K F_\pi} \left\{k_3 - 6 k_1
- 8(2 L_1 -  L_2 + L_3 - 12 L_4)\right\}\label{eq:Kpipi}\\
\zeta_1 &=& - \dfrac{G_8 M^4_\pi}{2 F_K F_\pi} \left\{k_2
- 8(2 L_1 + 2 L_2 + L_3)\right\}\nl
\xi_1 &=& - \dfrac{G_8 M^4_\pi}{2 F_K F_\pi} \left\{k_3
- 8(2 L_1 - L_2 + L_3)\right\}~.\no
\eeqa
The $k_i$ are combinations of the $N_i$ in (\ref{eq:L4w}):
\beqa
k_1 &=& -N_5 + 2 N_7 - 2 N_8 - N_9 \nl
k_2 &=& N_1 + N_2 + 2 N_3 \nl
k_3 &=& N_1 + N_2 - N_3~. \label{eq:ki}
\eeqa
All LECs in Eqs.~(\ref{eq:Kpipi}) and (\ref{eq:ki}) are the renormalized
ones, taken at a scale $\mu$.
Treating the strong couplings $L_i$ as known input [no errors
were included for the $L_i$ in the analysis of Kambor et al. (1991)],
one can exhibit the results in a concise way by eliminating the unknown
weak parameters $k_i$ (and the corresponding 27--plet couplings).
The resulting five relations can be formulated as predictions for the
slope parameters (Kambor et al., 1992). The predictions are compared with the
experimental values in Table~\ref{tab:slope}. Except for $\zeta_3$, which
is expected to be rather sensitive to isospin violating radiative
corrections, the agreement is quite impressive.

\begin{table}
\caption{Predicted and measured values of the quadratic slope parameters
in the $K \ra 3\pi$ amplitudes, all given in units of $10^{-8}$. The
table is taken from Kambor et al. (1992) and is based on the
$O(G_F p^4)$  calculation of Kambor et al. (1991).}
$$
\begin{tabular}{|c|c|c|} \hline
parameter & prediction & exp. value \\ \hline
$\zeta_1$ & $-0.47 \pm 0.18$ & $-0.47 \pm 0.15$ \\
$\xi_1$ & $-1.58 \pm 0.19$ & $-1.51 \pm 0.30$ \\
$\zeta_3$ & $-0.011 \pm 0.006$ & $-0.21 \pm 0.08$ \\
$\xi_3$ & $0.092 \pm 0.030$ & $-0.12 \pm 0.17$ \\
$\xi'_3$ & $-0.033 \pm 0.077$ & $-0.21 \pm 0.51$ \\ \hline
\end{tabular}
$$ \label{tab:slope}
\end{table}

One can try to understand the renormalized
values of the $k_i$ in (\ref{eq:ki}) from the fit of Kambor et
al. (1991) in terms of a model for the weak LECs. In the factorization
model discussed in the last subsection, only scalar exchange contributes
to the $k_i$ (Ecker et al., 1993b). The usually dominant $V$ exchange
drops out in the specific combinations (\ref{eq:ki}) (Isidori and
Pugliese, 1992). The factorization model predicts in particular
a negative $k_1$ and $k_2=k_3 > 0$. Taking the fitted values of the $k_i$ at
the scale $\mu=M_\rho$, these predictions are not satisfied
(Ecker et al., 1993b). However, it is probably premature to discard the
factorization model on the basis of this evidence. As one sees from
the amplitudes in (\ref{eq:Kpipi}), the strong LECs $L_i$ enter
with rather big coefficients. As already remarked, the errors for the
$k_i$ in the fit of Kambor et al. (1991) do not include errors
of the $L_i$ (see Table \ref{tab:Li}).  Moreover, the coefficient
$k_1$ is strongly scale dependent so that it is rather ambiguous
at which scale the factorization model should be compared with
the $k_i^r(\mu)$. In conclusion, the decays $K\to 2 \pi,~3 \pi$
are probably not the best place to test the factorization model,
at least not at the present level of accuracy. The radiative decays of
Sect.~\ref{subsec:Krad} are in principle (there is not enough experimental
information at the moment) better suited for this purpose,
because in that case $V$ and $A$ exchange contribute to the relevant
weak LECs (Ecker et al., 1993b).

\subsection{Chiral anomaly in nonleptonic kaon decays}
\label{subsec:anomweak}
The contributions of the chiral anomaly to strong, electromagnetic and
semileptonic weak amplitudes can be expressed in terms of the
Wess--Zumino--Witten functional (\ref{eq:WZW}). However,
the chiral anomaly also contributes to nonleptonic weak amplitudes
starting at $O(G_F p^4)$. Two different manifestations of the anomaly
can be distinguished.

The reducible anomalous amplitudes (type iii
in the classification of Sect.~\ref{subsec:GFp4}) arise from the
contraction of meson lines between a weak $\Delta S = 1$ Green function and
the WZW functional. At $O(G_F p^4)$, there can only be one such contraction
and the weak vertex must be due to the lowest--order nonleptonic Lagrangian
$\cL_2^{\Delta S=1}$ in Eq.~(\ref{eq:G827}).
Since $\cL_2^{\Delta S=1}$ contains bilinear terms in the meson fields,
the so--called pole contributions to anomalous nonleptonic amplitudes
can be given in closed form by a simultaneous diagonalization
(Ecker et al., 1987b, 1988) of the kinetic parts of the Lagrangians $\cL_2$ and
$\cL_2^{\Delta S=1}$. The corresponding local Lagrangian
(octet part only) is (Ecker et al., 1992):
\beq
\cL_{\rm an}^{\Delta S=1} =  \dfrac{ieG_8}{8\pi^2F} \wt F^{\mu\nu}
\partial_\mu \pi^0 K^+ \stackrel{\leftrightarrow}{D_\nu} \pi^- +
\dfrac{\alpha G_8}{6\pi F} \wt F^{\mu\nu} F_{\mu\nu}
\left(K^+ \pi^- \pi^0 - \dfrac{1}{\sqrt{2}} K^0 \pi^+ \pi^-\right) +
{\rm h.c.}
\label{eq:Law}
\eeq
Here
$F_{\mu\nu} = \partial_\mu A_\nu - \partial_\nu A_\mu$ is the
electromagnetic field strength tensor, $\wt F_{\mu\nu} = \ve_{\mu\nu
\rho\sigma} F^{\rho\sigma}$ its dual and
$D_\mu \vp^\pm = (\partial_\mu \pm ieA_\mu)\vp^\pm$ denotes the
covariant derivative with respect to electromagnetism.

There are also other reducible anomalous amplitudes. A generic example
is provided by a nonleptonic Green function where an external
$\pi^0$ or $\eta$ makes an anomalous transition to two photons.
Such transitions are the dominant $O(G_F p^4)$ contributions to the decays
$K_S \ra \pi^0\gamma\gamma$ (Ecker et al., 1987b)  and $K_L \ra \pi^0\pi^0
\gamma\gamma$ (Dykstra et al., 1991; Funck and Kambor, 1993). All
reducible anomalous amplitudes of $O(G_F p^4)$ are
proportional to $G_8$ in the octet limit. No other unknown parameters
are involved.

The second manifestation of the anomaly in nonleptonic weak amplitudes
arises diagrammatically from the contraction of the $W$ boson field
between a strong Green function on one side and the WZW functional on
the other side (Cheng, 1990). However, such diagrams cannot be taken
literally at a typical hadronic scale because of the presence of strongly
interacting fields on both sides of the $W$ (see the corresponding
discussion in Sect.~\ref{subsec:chisb} and Fig.~\ref{fig:K3pi}).
Instead, one must first
integrate out the $W$ together with the heavy quark fields. The
operators appearing in the operator product expansion must then be
realized at the bosonic level in the presence of the anomaly.

\begin{table}
\caption{A complete list of local anomalous nonleptonic weak $K$ decay
amplitudes of $O(G_F p^4)$ in the limit of $CP$ conservation.} \label{tab:anom}
$$
\begin{tabular}{|l|cccccc|} \hline
Transition & $\cL_{\rm an}^{\Delta S=1}$ & $W_{28}$ & $W_{29}$ & $W_{30}$
& $W_{31}$ & exp. \\ \hline
$K^+ \ra \pi^+ \pi^0 \gamma$       & x &   & x & x &   & x \\
$K^+ \ra \pi^+ \pi^0 \gamma\gamma$ & x &   & x & x &   &    \\
$K_L \ra \pi^+ \pi^- \gamma$       &   &   & x &   & x & x  \\
$K_L \ra \pi^+ \pi^- \gamma\gamma$ & x &   & x &   & x &    \\
$K^+ \ra \pi^+ \pi^0 \pi^0\gamma$  &   &   & x & x &   & x  \\
$K^+ \ra \pi^+ \pi^0 \pi^0 \gamma\gamma$ & & & x & x & & \\
$K^+ \ra \pi^+ \pi^+ \pi^- \gamma$ &   &   & x &   & x & x \\
$K^+ \ra \pi^+ \pi^+ \pi^- \gamma\gamma$ & & & x & & x & \\
$K_L \ra \pi^+ \pi^- \pi^0\gamma$ & & x & x & x && \\
$K_S \ra \pi^+ \pi^- \pi^0\gamma(\gamma)$ & & & x & x & x &  \\ \hline
\end{tabular}
$$
\end{table}

The bosonization of four--quark
operators in the odd--intrinsic--parity sector was investigated by
Bijnens et al. (1992). As in the even--intrinsic--parity sector, the bosonized
four--quark operators contain factorizable (leading in $1/N_c$)
and non--factorizable parts (non--leading in $1/N_c$).
Due to the non--renormalization theorem (Adler and Bardeen, 1969) of the
chiral anomaly, the factorizable contributions of $O(G_F p^4)$ can be
calculated exactly (Bijnens et al., 1992) in terms of the
anomalous part of the left--chiral
current of $O(p^3)$. It turns out that the factorizable
contributions produce already all the possible relevant octet operators
proportional to the $\ve$ tensor [$W_{28}$, $W_{29}$, $W_{30}$ and
$W_{31}$ in Eq.~(\ref{eq:L4w})].
The non--factorizable parts automatically have the right octet
transformation property (they do not get any contribution from the anomaly)
and are therefore also of the form $W_{28}, \dots$~, $W_{31}$.
Altogether, the $\Delta S=1$ effective Lagrangian in
the odd--intrinsic--parity sector of $O(G_F p^4)$ can be characterized by the
coefficients (Bijnens et al., 1992)
\beq
\ba{ll}
N_{28} = \dfrac{a_1}{8\pi^2} \qquad \qquad &
N_{29} = \dfrac{a_2}{32\pi^2} \\[10pt]
N_{30} = \dfrac{3a_3}{16\pi^2} \qquad \qquad &
N_{31} = \dfrac{a_4}{16\pi^2} ~,
\ea  \label{eq:Nan}
\eeq
where the dimensionless coefficients $a_i$ are expected to be positive
and probably (Bijnens et al., 1992) smaller than one [$a_i=1$
corresponds to $k_f= 1$ in (\ref{eq:FM})].

In Table \ref{tab:anom} all kinematically allowed nonleptonic
$K$ decays are listed that are sensitive either to the
anomalous Lagrangian $\cL_{\rm an}^{\Delta S=1}$ in
(\ref{eq:Law}) or to the direct terms of $O(G_F p^4)$ via (\ref{eq:Nan}).
The transitions with either three pions and/or two
photons in the final state are in general also subject to non--local
reducible anomalous contributions. In the nonleptonic weak sector,
the chiral anomaly contributes only to {\em radiative} $K$ decays.

\subsection{Radiative kaon decays}
\label{subsec:Krad}
Among rare $K$ decays [for general reviews, see, e.g., Ecker et
al. (1995), Ritchie and Wojcicki (1993), Littenberg and Valencia (1993),
Winstein and Wolfenstein (1993), Battiston et al. (1992)], the radiative
transitions are especially interesting for CHPT. Almost all radiative
decays have important long--distance contributions. Moreover,
they have the special property that they are ``trivial" to lowest order,
$O(G_F p^2)$, in the following sense:
\ben
\item Nonleptonic $K$ decay amplitudes with any number of real
or virtual photons and with at most one pion in the final state
vanish at $O(G_F p^2)$ (Ecker et al., 1987a, 1988):
\beq
A(K\ra [\pi] \gamma^* \dots \gamma^*) = 0 \qquad \mbox{ at  }O(G_F p^2)~.
\label{eq:pig*}
\eeq
\item The amplitudes for two pions and any number of real or virtual photons
in the final state factorize at $O(G_F p^2)$ into the on--shell amplitude
for the corresponding $K\ra \pi\pi$ decay and a generalized
bremsstrahlung amplitude independent of the specific decay
(de Rafael, 1989; Ecker et al., 1992, 1994b):
\beq
A(K\ra \pi\pi \gamma^* \dots \gamma^*)= A(K\ra \pi\pi) A_{\rm brems}~.
\label{eq:brems}
\eeq
\item A similar statement holds for the decays $K\ra 3 \pi
\gamma$: the amplitude of $O(G_F p^2)$ is completely determined by the
corresponding non--radiative decay $K\ra 3 \pi$ (D'Ambrosio et al.,
in preparation).
\een

Therefore, the leading non--trivial aspects of radiative nonleptonic kaon
decays are described by the generating functional of $O(G_F p^4)$
discussed in Sect.~\ref{subsec:GFp4}. An important question is whether
all the coupling constants $N_i$ can be measured in radiative
$K$ decay experiments. We list in Table~\ref{tab:Ni} all the nonleptonic
radiative transitions to which the $N_i$ contribute. There are other
decays not sensitive to the $N_i$ that are either given by finite one--loop
amplitudes and/or anomalous contributions
at $O(G_F p^4)$ ($K_S\ra \gamma^* \gamma^*$, $K^0\ra \pi^0 \gamma \gamma$,
$K^0\ra \pi^0\pi^0 \gamma\gamma$, $K_L\ra \pi^+\pi^-\gamma [\gamma]$)
or which vanish even at $O(G_F p^4)$ ($K_L\ra \gamma^*\gamma^*$,
$K^0\ra \pi^0\pi^0\gamma$).

\begin{table}
\caption{Decay modes to which the coupling constants $N_i$ contribute.
For the $3 \pi$ final states, only the single photon channels are
listed. For the neutral modes, the letters $L$ or $S$ in brackets
distinguish between $K_L$ and $K_S$ initial states in the limit of
$CP$ conservation.}
\label{tab:Ni}
$$
\begin{tabular}{|c|c|c|c|} \hline
$\pi$ & $2 \pi$ & $3 \pi$ & $N_i$ \\ \hline
$\pi^+ \gamma^*$ & & & $N_{14}^r - N_{15}^r$\\
$\pi^0 \gamma^*~(S)$ & $\pi^0\pi^0\gamma^*~(L)$ & & $2 N_{14}^r + N_{15}^r$\\
$\pi^+ \gamma\gamma$ & $\pi^+\pi^0\gamma\gamma$ & & $N_{14} - N_{15}
 -2 N_{18}$ \\
 & $\pi^+\pi^-\gamma\gamma~(S)$ & & $N_{14} - N_{15} -2 N_{18}$ \\
 & $\pi^+\pi^0\gamma$ & $\pi^+\pi^+\pi^-\gamma$ & $N_{14}-N_{15}-N_{16}
 -N_{17}$ \\
 & $\pi^+ \pi^- \gamma~(S)$ & $\pi^+\pi^0\pi^0\gamma$ & $N_{14}-N_{15}-N_{16}
 -N_{17}$ \\
 & & $\pi^+\pi^-\pi^0\gamma~(L)$ & $N_{14}-N_{15}-N_{16} -N_{17}$ \\
 & & $\pi^+\pi^-\pi^0\gamma~(S)$ & $7(N_{14}^r-N_{16}^r)+ 5(N_{15}^r
 + N_{17}^r)$ \\
 & $\pi^+\pi^-\gamma~(L)$ & $\pi^+\pi^-\pi^0\gamma~(S)$ & $N_{29} + N_{31}$ \\
 & & $\pi^+\pi^+\pi^-\gamma$ & $N_{29} + N_{31}$ \\
 & $\pi^+\pi^0\gamma$ & $\pi^+\pi^0\pi^0\gamma$ & $3 N_{29} - N_{30}$ \\
 & & $\pi^+\pi^-\pi^0\gamma~(S)$ & $2 (N_{29} + N_{31}) + 3 N_{29} -
N_{30}$ \\
 & & $\pi^+\pi^-\pi^0\gamma~(L)$ & $6 N_{28} - 4 N_{30}+ 3 N_{29} - N_{30}$
\\ \hline
\end{tabular}
$$
\end{table}

The information contained in Table~\ref{tab:Ni} leads to the following
conclusions:
\bit
\item Read horizontally, one finds all parameter--free relations
between radiative amplitudes of $O(G_F p^4)$. If in the last column the
renormalized constants $N_i^r$ are displayed, the corresponding
decays have divergent one--loop amplitudes. The other modes have
finite loop amplitudes.
\item Read vertically, we infer that from decays with at most two pions
in the final state only the following combinations of counterterm
coupling constants can in principle be extracted:
\beq
N_{14},\, N_{15},\, N_{16}+N_{17},\, N_{18},\, N_{29}+N_{31},\,
3 N_{29}-N_{30}~.
\eeq
\item Decays with three pions in the final state are needed to determine
$N_{16}$ and $N_{17}$ separately and the combination $3 N_{28} -
2 N_{30}$.
\item Whereas all \lq\lq electric" constants $N_{14},\dots,N_{18}$
can in principle be determined phenomenologically, this is not
the case for the \lq\lq magnetic" constants (the corresponding
operators $W_i$ contain an $\ve$ tensor): only three combinations of
the four constants $N_{28},\dots,N_{31}$ appear in measurable
amplitudes.
\eit

Except for the radiative decays with three pions in the final state
(D'Ambrosio et al., in preparation), all amplitudes appearing in
Table~\ref{tab:Ni} have been completely calculated to $O(G_F p^4)$.
In the near future, all decays with at most two pions in the
final state will be measured or remeasured
at dedicated facilities like the $\Phi$ factory DA$\Phi$NE in Frascati
(Maiani et al., 1992, 1995).
As long as experiments are only sensitive to the lowest--order
bremsstrahlung contributions for decays of the type $K\ra 3 \pi \gamma$,
but not to the interesting $O(G_F p^4)$ parts, the phenomenological
determination of the $N_i$ will remain incomplete. Nevertheless,
even without the full information available, one will be able both
to check some of the parameter--free low--energy theorems of $O(G_Fp^4)$
implicitly given in Table~\ref{tab:Ni} and to
test the predictions of various models for the coupling constants
such as the factorization model (\ref{eq:FM}).

The phenomenology of radiative $K$ decays to $O(G_F p^4)$ has been
reviewed recently by D'Ambro\-sio et al. (1994) for the physics program
foreseen at DA$\Phi$NE. A general review of rare $K$ decays
in the Standard Model is in preparation (Ecker et al., 1995). Instead of
repeating the contents of those articles here, I will describe in
Sect.~\ref{subsec:phenp6} attempts to go beyond $O(G_F p^4)$
for two specific decay channels.

\setcounter{equation}{0}
\setcounter{subsection}{0}
\setcounter{table}{0}
\setcounter{figure}{0}

\section{CHPT BEYOND NEXT--TO--LEADING ORDER}
\label{sec:p6}
\subsection{Generating functional of $O(p^6)$}
\label{subsec:2loop}
The generating functional $Z_6[v,a,s,p]$ consists of two parts. It contains
the next--to--leading order contributions of odd intrinsic parity discussed in
Sect.~\ref{subsec:anom}. Here, I want to briefly describe ongoing work
(Bijnens et al., in preparation) in the even--intrinsic--parity sector
beyond next--to--leading order involving in particular two--loop diagrams.

To simplify the notation, I denote in this subsection the external fields
$v_\mu, a_\mu,s,p$ collectively as $j_i$. The Goldstone fields will be
denoted $\vp_i$ as before, but I disregard that they enter the chiral
Lagrangian through the matrix field $U(\vp)$. Thus, the effective chiral
action is written
\beq
S[\vp,j] = S_2[\vp,j] + S_4[\vp,j] + S_6[\vp,j] + \ldots
\eeq
and the generating functional takes the form
\beq
Z[j] = Z_2[j] + Z_4[j] + Z_6[j] + \ldots~.
\eeq

The loop expansion can be viewed as a systematic expansion in $\hbar$
around the classical solution $\vp_{\rm cl}[j]$. For the calculation
of $Z_6$, the question arises whether the classical solution corresponds
to the extremum of $S_2$ as in the calculation of $Z_4$, or whether one
has to evaluate the ``classical'' solution from the effective
Lagrangian $\cL_2 + \cL_4$, i.e. including $O(p^4)$. There are more
than one good reason to use the same classical solution for the
calculation of $Z$ to any order in the chiral expansion. In other words,
we always expand around $\vp_{\rm cl}[j]$ that is defined by the
{\em lowest--order\/} Lagrangian $\cL_2$,
\beq
\left. \frac{\delta S_2[\vp,j]}{\delta \vp_i} \right|_{\vp=\vp_{\rm cl}}
= 0 \qquad \Ra \qquad \vp_{\rm cl}[j]~, \label{eq:EOM2}
\eeq
which is nothing but the equation of motion (\ref{eq:EOM}) in the present
notation.

\begin{figure}
\centerline{\epsfig{file=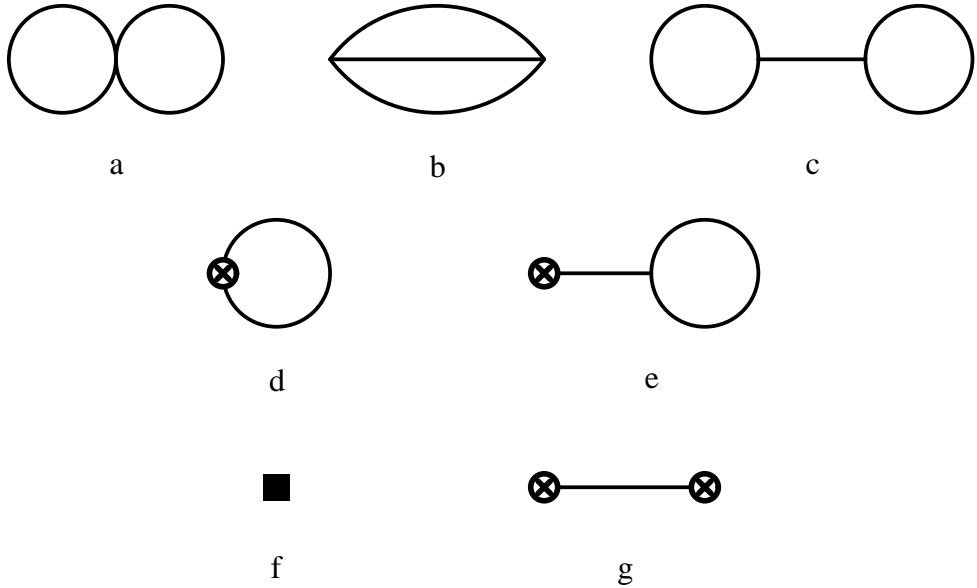,height=8cm}}
\caption{Diagrams contributing to the generating functional of $O(p^6)$.
The propagators and vertices carry the full tree structure associated
with the lowest--order Lagrangian $\cL_2$. Normal vertices are from
$\cL_2$, crossed circles denote vertices from $\cL_4$ and the square in diagram
f stands for a vertex from $\cL_6$.}\label{fig:p6}
\end{figure}

One major practical advantage of this procedure is its compatibility with
chiral counting. The diagrams in Fig.~\ref{fig:p6}
give rise to functionals of $\vp_{\rm cl}[j]$. Since $\vp_{\rm cl}[j]$
carries the full tree structure associated with $\cL_2$ (see the discussion
in Sect.~\ref{subsec:gf}), all vertices and propagators in
Fig.~\ref{fig:p6} have the same full tree structure attached to them.
As long as $\vp_{\rm cl}[j]$ is defined as in Eq.~(\ref{eq:EOM2}),
attaching the trees to those diagrams does not modify the chiral dimension
of the diagrams. This would no longer be true if $\vp_{\rm cl}[j]$
would be derived from the effective action $S_2 + S_4$ because
vertices and propagators in the trees would contain both $O(p^2)$ and
$O(p^4)$ pieces. The propagators appearing in Fig.~\ref{fig:p6}
are the full propagators (as functionals of $\vp_{\rm cl}[j]$) for the
lowest--order Lagrangian $\cL_2$. Thus, the (matrix) propagator is the
inverse of
\beq
\left. \frac{\delta^2 S_2 [\vp,j]}{\delta \vp_i \delta \vp_j}
\right|_{\vp = \vp_{\rm cl}[j]}~. \label{eq:prop}
\eeq

\begin{figure}
\centerline{\epsfig{file=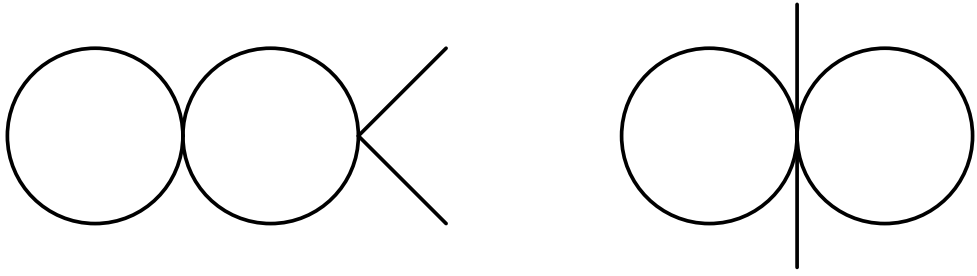,height=3.5cm}}
\caption{Butterfly diagrams contributing to $\gamma\gamma \to \pi\pi$.
The photons must be appended on every charged meson line and
on every vertex containing charged meson lines.}\label{fig:ggp6}
\end{figure}

Fig.~\ref{fig:p6} shows all the diagrams occurring in the calculation of
$Z_6[j]$. They contribute with different weights
to $Z_6$ (Bijnens et al., in preparation).
The normal vertices in Fig.~\ref{fig:p6}
are from $\cL_2$. For instance, the vertex in the butterfly
diagram a is given by
\beq
\left.\frac{\delta^4 S_2}{\delta \vp_i \delta \vp_j \delta \vp_k
\delta \vp_\ell} \right|_{\vp = \vp_{\rm cl}[j]}~.
\eeq
Expanding both this vertex and the two propagators in powers of $j$
leads to the relevant Green functions. To give an example, the
butterfly graph contains the diagrams in Fig.~\ref{fig:ggp6}
contributing to $\gamma \gamma \ra \pi \pi$ to be considered in the
next subsection.

The vertices denoted by crossed circles are from $\cL_4$. They are given by
\beq
\left. \frac{\delta S_4}{\delta \vp_i}\right|_{\vp = \vp_{\rm cl}}
\qquad \mbox{and} \qquad
\left. \frac{\delta^2 S_4}{\delta \vp_i \delta \vp_j}
\right|_{\vp = \vp_{\rm cl}}
\eeq
in diagrams d, e and g. Finally, the square vertex (diagram f) stands for
tree diagrams associated with the chiral Lagrangian $\cL_6$ of
$O(p^6)$. This Lagrangian has recently been constructed by Fearing
and Scherer (1994). After a careful consideration of different relations
between all possible monomials of $O(p^6)$, they arrive at 111 terms
for $\cL_6$. Unless further relations will be found, $\cL_6$ would
therefore contain 111 independent LECs. It is rather obvious that it
will not be possible to determine all of them from experiment.
Nevertheless, there is no reason for putting one's head in the sand as the
next subsection will demonstrate.

To carry out the renormalization program at $O(p^6)$, one has to calculate
the divergent part of $Z_6$ associated with the loop diagrams of
Fig.~\ref{fig:p6}. Just like $\cL^{(L=1)}_{\rm 4,div}$ in
(\ref{eq:L4div}), the divergence functional $Z_{\rm 6,div}$
carries direct physical information. At $O(p^4)$, the coefficients
$\Gamma_i$ of the simple poles in $d-4$ in (\ref{eq:L4div})
determine the chiral logs. Similarly, the residues of the double poles
in $d-4$ in $Z_{\rm 6,div}$ yield directly the squares of
chiral logs (Bijnens et al., in preparation; Colangelo, 1994).
Especially for observables where a full calculation to $O(p^6)$ is not
available, the leading (squares of) chiral logs could give an
indication of the possible size of $O(p^6)$ corrections.

Renormalization at $O(p^6)$ is in principle straightforward. The fact that
[recall Eq.~(\ref{eq:renorm})]
\beq
Z_4^{(L=1)} [j] + S_4 [\vp_{\rm cl}]
\eeq
is already finite and independent of the renormalization scale has
 the following consequences:
\begin{enumerate}
\item[i.] The sum of the one--particle--reducible diagrams c, e, g in
Fig.~\ref{fig:p6} is finite and scale independent.
\item[ii.] The sum of the irreducible loop diagrams a, b, d is free of
subdivergences.
\een

Therefore, the divergent part $Z_{\rm 6,div}$ takes the
form of a local action (that has again all the symmetries of $\cL_2$) with
divergent coefficients. The corresponding divergence Lagrangian
$\cL_{\rm 6,div}$ must then be a special case of the Lagrangian $\cL_6$
of Fearing and Scherer (1994), the most general effective chiral
Lagrangian of $O(p^6)$. Splitting the corresponding LECs in the same way
as in Eq.~(\ref{eq:Liren}) into finite renormalized and divergent parts,
the total functional $Z_6[j]$ is rendered finite and scale independent.

Setting up the loop expansion in the way described brings up
an interesting question. It turns out that using the
equation of motion (\ref{eq:EOM2}) or (\ref{eq:EOM}) in the construction
of $\cL_4$ (Gasser and Leutwyler, 1985a)
does not commute with calculating the vertices from $\cL_4$
appearing in diagrams d, e, g in Fig.~\ref{fig:p6}. Recalling once more
Eq.~(\ref{eq:EOM2}), the most general local action
$S_{\rm 4,gen} [\vp,j]$ is of the form (dropping the external fields
$j$ for ease of notation)
\beqa
S_{\rm 4,gen}[\vp] &=& S_4[\vp] + \wh S_4 [\vp] \\
\wh S_4[\vp] &=& c_{2,i}[\vp] S_{2,i}[\vp] + \frac{1}{2} c_{0,ij}[\vp]
S_{2,i}[\vp] S_{2,j}[\vp] \no
\eeqa
$$
S_{2,i}[\vp] := \frac{\delta S_2[\vp,j]}{\delta \vp_i}~,
$$
where $c_n[\vp]$ $(n = 0,2$) are local functionals of $O(p^n)$.
Because of (\ref{eq:EOM2}) we have
\beq
\wh S_4 [\vp_{\rm cl}] = 0
\eeq
and thus the functional $Z_4$ of $O(p^4)$ is insensitive to
$\wh S_4[\vp]$. However, $\wh S_4$ does contribute to the diagrams d, e, g
because
\beq
\left. \frac{\delta \wh S_4}{\delta \vp_i}\right|_{\vp = \vp_{\rm cl}}
\neq 0~, \qquad
\left. \frac{\delta^2 \wh S_4}{\delta \vp_i \delta \vp_j}
\right|_{\vp = \vp_{\rm cl}} \neq 0~.
\eeq
On the other hand, there must not be any measurable influence of
$\wh S_4[\vp]$ because the ``couplings'' $c_{0,ij}$ and $c_{2,i}$
are arbitrary and cannot be fixed at $O(p^4)$. A careful investigation
of the effect of $\wh S_4$ in diagrams d, e, g solves this dilemma:
the difference to setting $\wh S_4 \equiv 0$ from the beginning is a
{\em local\/} functional of $O(p^6)$ that can always be absorbed by
redefining the coefficients in $\cL_6$ [LECs of $O(p^6)$] because
$\cL_6$ is by construction the most general effective chiral
Lagrangian of $O(p^6)$.

For more details and concrete results I refer to forthcoming publications
(Bijnens et al., in preparation). As a final remark of this subsection,
let me address a question that has stirred up a considerable amount of
dust in the literature. Can one use the equation of motion and which
equation of motion for the calculation of Green functions and
amplitudes in theories governed by effective Lagrangians? I have gone
to some length in this subsection to make the answer self--evident:
setting up the loop expansion as an expansion around the classical
solution, one {\em must\/} in fact use the lowest--order equation of
motion (\ref{eq:EOM2}) to derive the relevant Green functions from the
generating functional. I have no illusions that the dust has
settled now.

\subsection{$\gamma\gamma\to \pi^0\pi^0$}
\label{subsec:ggpipi}
The production of pion pairs in $\gamma\gamma$ collisions was studied
to $O(p^4)$ by Bijnens and Cornet (1988a) and by Donoghue et al.
(1988). In the case of charged pion pair production,
inclusion of the amplitude of $O(p^4)$ (Bijnens and Cornet, 1988a) brings the
total amplitude into good agreement with experiment in the low--energy
region. Moreover, the correction of $O(p^4)$ to the Born term is
of the expected order of magnitude.

For the reaction $\gamma\gamma\to \pi^0\pi^0$, there is no contribution
of $O(p^2)$ and the amplitude of $O(p^4)$ is a pure one--loop amplitude
(Bijnens and Cornet, 1988a; Donoghue et al., 1988). The chiral prediction
disagrees with the Crystal Ball data (Marsiske et al., 1990) even near
threshold. On the other hand, dispersion theoretic calculations can
reproduce the behaviour of the cross section near threshold (e.g.,
Morgan and Pennington, 1991; Donoghue and Holstein, 1993a;
Dobado and Pel\'aez, 1993; Truong, 1993).

Bellucci et al. (1994) have taken up the challenge to
demonstrate that CHPT can explain the observed cross section by
including the amplitude of $O(p^6)$.
The matrix element for pion pair production
\beq
\gamma(q_1) \gamma(q_2) \to \pi^0(p_1) \pi^0(p_2)
\eeq
is given by
\beqa
T & = & e^2 \ve^\mu_1(q_1) \ve^\nu_2(q_2) V_{\mu \nu} \label{eq:ggkin}\\
V_{\mu \nu} &=& A(s,t,u) T_{1\mu \nu} + B(s,t,u) T_{2\mu \nu} \nl
 T_{1\mu \nu}& =& \frac{s}{2} g_{\mu \nu} - q_{1\nu} q_{2\mu} \nl
 T_{2\mu \nu} &=&  2 s \Delta_\mu \Delta_\nu -\nu^2 g_{\mu \nu}
  - 2 \nu (q_{1\nu} \Delta_\mu - q_{2\mu} \Delta_\nu) \no
\eeqa
 where
\beqa
 s&=& (q_1 + q_2)^2~,\; t = (p_1 -q_1)^2~, \; u = (p_2 - q_1)^2 \nl
 \nu &=& t-u ~,\; \Delta = p_1 - p_2 ~.\no
\eeqa
In terms of helicity amplitudes
\beqa
 H_{++} &=& A + 2 (4M^2_\pi - s) B \nl
 H_{+-} &=&  \frac{8(M_\pi^4- tu)}{s} B ~,
\eeqa
the differential cross section for unpolarized photons in the
center--of--mass system is
\beq
 \dfrac{d \sigma}{d \Omega}(\gamma \gamma \to \pi^0 \pi^0) =
\frac{e^4s}{1024\pi^2} (1-4 M^2_\pi/s)^{1/2}
(\mid H_{++} \mid^2 + \mid H_{+-} \mid^2)~.
\eeq
The helicity components $H_{++}$ and $H_{+-}$ correspond to
photon helicity differences $\lambda = 0,2$, respectively.

The leading term in the chiral expansion of $V^{\mu\nu}$ is
generated  by one--loop graphs (Bijnens and Cornet,
1988a; Donoghue et al., 1988) contributing only to the amplitude $A$:
\beqa
\label{eq:le1}
H_{++}^{(L=1)}&=&\frac{(M_\pi^2-s)F(s/M^2_\pi)}{4\pi^2 sF_\pi^2}
{}~,\;\; H_{+-}^{(L=1)}=0\\
 F(z)& =& \Biggl\{
   \ba{ll}
      1 - \dfrac{4}{z} \arcsin^{2}{(\sqrt{z}/2)}
           \qquad  & z\le 4 \\
      1 + \dfrac{1}{z}\left( \ln\dfrac{1-\sqrt{1-4/z}}{1+ \sqrt{1-4/z}} +
       i\pi \right)^2  & z\ge 4
   \Biggr. ~.\ea\no
\eeqa

The amplitude of $O(p^6)$ is generated by the diagrams of Fig.~\ref
{fig:p6} with two photons and two neutral pions appended to
vertices and propagators in all possible ways (see
Fig.~\ref{fig:ggp6} for some examples). The explicit expressions
for the helicity amplitudes $H_{+\pm}$ can be found in Bellucci
et al. (1994). The amplitudes depend on the phenomenologically known LECs
of $O(p^4)$ and on three additional LECs $h_\pm^r,h_s^r$
from ${\cal{L}}_6$. The latter have not yet been
determined in a systematic way, but they were estimated by Bellucci
et al. (1994) in the standard manner (Gasser and Leutwyler,
1984; Ecker et al., 1989a) using resonance exchange with $J^{PC}=1^{--},
1^{+-},0^{++},2^{++}$.

The cross section for $\gamma \gamma \to \pi^0 \pi^0$
is displayed in Fig.~\ref{fig:ggpp} as a function
of the center--of--mass energy  $E = \sqrt{s}$
for $|\cos \theta| \leq 0.8$ as determined in the
 Crystal Ball experiment (Marsiske et al., 1990).
The complete result of $O(p^6)$ (full curve) substantially improves
the cross section of $O(p^4)$ and is in perfect agreement with
experiment even at relatively high energies.
It also agrees with the dispersive calculation of Pennington (1992).

\begin{figure}
\centerline{\epsfig{file=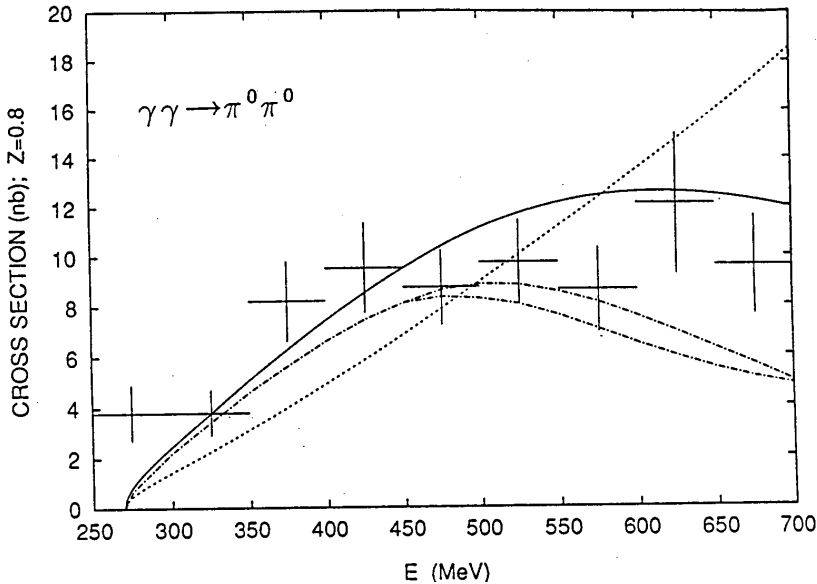,height=8cm}}
\caption{The cross section $\sigma(|\cos\theta|\leq Z)$ as a function
of the center--of--mass energy $E$ at $Z=0.8$, together with the data
 from the Crystal Ball experiment (Marsiske et al., 1990). The solid curve is
the full two--loop result (Bellucci et al., 1994) and the dashed curve
corresponds to the one--loop calculation (Bijnens and Cornet, 1988a; Donoghue
et al., 1988). The band denoted by the dash--dotted curves is the result
of the dispersive calculation by Pennington (1992). The figure
is taken from Bellucci et al. (1994).}
\label{fig:ggpp}
\end{figure}

It is important to emphasize that the LECs  $h_{\pm}^r$
and $h_s^r$ contribute very little to the cross section below
$E=450$MeV. The enhancement in the cross section is
mainly	due to $\pi \pi$ rescattering and to the renormalization of the
pion decay constant. Bellucci et al. (1994) also note that
the two--loop corrections are not unduly large; their size is similar to
the corresponding next--to--leading order correction in the isospin zero
$\pi \pi$ scattering amplitude (Gasser and Leutwyler, 1984)
and in the scalar form factor of the pion (Gasser and Mei\ss ner, 1991b).

The calculation of Bellucci et al. (1994) leads also to a much
more reliable determination than previously available of the neutral pion
polarizabilities that govern the behaviour of the amplitude for Compton
scattering $\gamma\pi^0 \to \gamma\pi^0$ near threshold.
Especially the sum of the electric and magnetic polarizabilities
is very sensitive to the amplitude of $O(p^6)$ because this combination
vanishes at $O(p^4)$. In contrast to the cross section for pair production
at low energies, the polarizabilities depend however in an essential way
on the LECs $h_\pm^r$. For a discussion of the
experimental situation I refer to the specialized literature
(Gasser, 1994b; Baldini and Bellucci, 1994; Portol\'{e}s and
Pennington, 1994).

\subsection{Phenomenology beyond next--to--leading order}
\label{subsec:phenp6}
In addition to $\gamma\gamma \to \pi^0 \pi^0$,
scalar and vector form factors of the pion (Gasser and Mei\ss ner, 1991b)
and vector current propagators (Golowich and Kambor, 1995)
have been completely calculated to $O(p^6)$
(even intrinsic parity). Calculations of $\gamma\gamma \to \pi^+
\pi^-$ (B\"urgi, work in progress) and elastic $\pi\pi$ scattering
(Bijnens et al., work in progress) are under way.

Moreover, many attempts have been undertaken to
control at least the dominant effects beyond next--to--leading order.
To appreciate the general strategy, one should recall that CHPT
as a quantum field theory accounts
both for the requirements of unitarity and analyticity and for
the most general local contributions to the amplitudes. In many cases,
one can set up dispersion relations for the amplitudes in
question to account for the proper analytic structure and use
CHPT to yield the most general set of subtraction constants (polynomials).
The advantage of this approach is that one can often improve
(Truong, 1988, 1991a, 1991b; Donoghue et al., 1990) upon the perturbative
implementation of unitarity of CHPT, even if one does not include
all terms of higher orders in the chiral expansion. At the same time, one
tries to locate the dominant local contributions, usually via resonance
exchange. The resulting amplitudes improve the known next--to--leading
order results and have a chance to capture the main gist of higher--order
effects. However, it must be emphasized that their
theoretical status is not the same as that of a complete CHPT calculation
to a given order in the chiral expansion.

If such a procedure may be useful for the strong interactions of
mesons, it becomes almost unavoidable for the nonleptonic weak interactions
of mesons. It is not clear whether it will ever be possible to
determine all LECs of $O(G_F p^4)$ (cf. Sect.~\ref
{sec:nonleptonic}), let alone higher--order couplings. On the
other hand, there are clear experimental indications for some nonleptonic
weak transitions that the amplitudes of $O(G_F p^4)$ are not
sufficient to describe the data. This is not too surprising in
view of the natural magnitude of higher--order corrections (\ref{eq:omag}) in
chiral $SU(3)$.

I shall concentrate here on the decays $K_L\to \pi^0\gamma\gamma$
and $K_S\to \gamma\gamma$
and refer to specialized reviews (D'Ambrosio et al., 1994; de Rafael, 1995;
Ecker et al., 1995) for more complete coverage. The transition $K_L\to
\pi^0\gamma\gamma$ shares, not surprisingly, many features with the
scattering process $\gamma\gamma \to \pi^0\pi^0$ of the last subsection.
There is no amplitude of $O(G_F p^2)$ and there are no local contributions
of $O(G_F p^4)$ making the amplitude a pure one--loop effect at this
order. More specifically, the transition
$$
K_L(p)\to \pi^0(p')\gamma (q_1) \gamma (q_2)
$$
is again characterized by two invariant amplitudes $A(y,z)$, $B(y,z)$
[cf. Eq.~(\ref{eq:ggkin})] with
\beq
y=p\cdot (q_1-q_2)/M_K^2~,\qquad z=(q_1+q_2)^2/M_K^2~.
\label{eq:yz}
\eeq
The physical region in the variables y and z
is given by the inequalities
\beq
|y|\le {1\over2} \lambda^{1/2}(1,z,r_\pi^2)~,\qquad\qquad
0\le z \le (1-r_\pi)^2 ~,
\label{eq:phsp}
\eeq
$$
r_\pi = M_\pi / M_K~,\qquad\qquad
\lambda(x,y,z) = x^2 + y^2 + z^2 - 2 (xy + yz + zx)~.
$$
The double differential decay rate for unpolarized photons is
\beq
 {d^2 \Gamma \over dy \ dz} = {M_K \over 2^9\pi^3} \left
\{z^2 \vert A+B \vert ^2 +
\left[y^2-{1\over 4}\lambda(1,z,r_\pi^2) \right]^2
\vert B \vert ^2 \right\}~.
\label{eq:klpggc}
\eeq

At $O(G_F p^4)$, the amplitude $B$ vanishes as in the case of
$\gamma\gamma \to \pi^0\pi^0$. The one--loop diagrams give rise
(Ecker et al., 1987b; Cappiello and
D'Ambrosio, 1988) to an amplitude $A$ of the form (octet part only)
\beq
A(z) = {{G_8\alpha M_K^2} \over \pi }
\left[(1-{r_\pi^2\over z}) F\left({z \over r_\pi^2}\right)-
(1-{{r_\pi^2}\over z}-{1 \over z}) F(z) \right]
\label{eq:klpggd}
\eeq
with the function $F(z)$ defined in Eq.~(\ref{eq:le1}).
The contribution proportional to $F(z)$ is due to the kaon loop,  while
$F({z / r_\pi^2})$ is generated by
the pion loop and has an absorptive part because the pions can be on--shell.
Except for the fact that the amplitudes in (\ref{eq:le1}) were calculated
in the framework of chiral $SU(2)$, the analytic structure of the
amplitude $A$ is of course the same in both transitions.
The kaon--loop contribution is actually much smaller than the
pion one as is the contribution of the 27--plet (Cappiello et al., 1993).

\begin{figure}
    \begin{center}
       \setlength{\unitlength}{1truecm}
       \begin{picture}(7.0,8.0)
       \put(-2.4,-4.8){\includegraphics{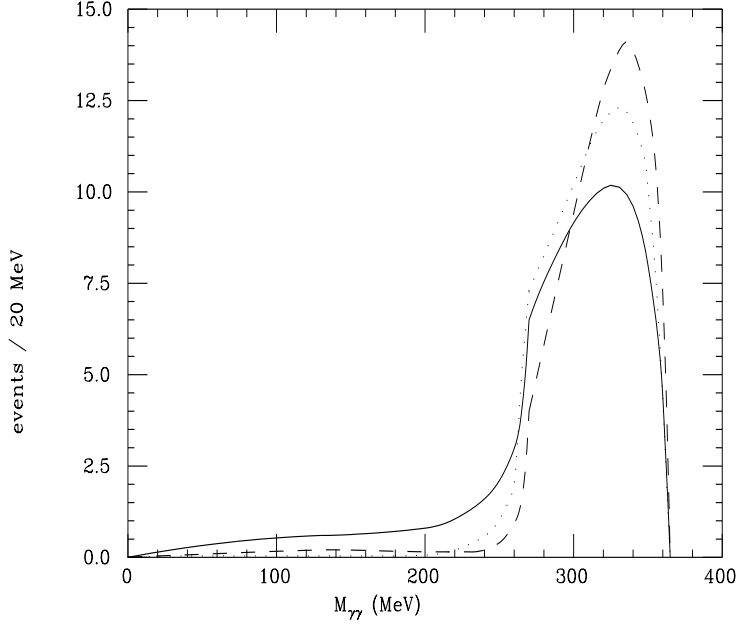}}
       \end{picture}
    \end{center}
    \caption{Theoretical predictions for the $2\gamma$--invariant--mass
distribution in $K_L \rightarrow \pi^0 \gamma\gamma$.
The dotted curve is the $O(G_F p^4)$ contribution,
the dashed and full curves correspond
to the $O(G_F p^6)$ calculation of Cohen et al. (1993)
with $a_V=0$ and $a_V=-0.9$, respectively. The spectra are normalized to
the $50$ unambiguous events of NA31 (cf. Fig.
\protect{\ref{fig:klspexp}}).}
    \protect\label{fig:klspth}
\end{figure}

The z--spectrum for a $y$--independent amplitude $A$ is given by
\beq
{ d\Gamma \over dz} = {M_K \over 2^{10} \pi^3 } z^2
\lambda^{1/2}(1,z,r_\pi^2)\vert A(z)\vert ^2 ~.
\eeq
The predicted branching ratio is
(Ecker et al., 1987b; Cappiello and D'Ambrosio, 1988)
\beq
BR(K_L \to \pi ^0 \gamma \gamma)=6.8\cdot{10}^{-7}~.
\label{eq:klpggp4}
\eeq
The $O(G_F p^4)$ prediction for $\displaystyle{
{d\Gamma\over dz}(K_L \rightarrow \pi^0  \gamma\gamma)}$ is shown as the
dotted curve in Fig.~\ref{fig:klspth}. The spectrum is very characteristic:
a  peak in the absorptive region ($K_L\to \pi \pi\pi\to \pi \gamma \gamma$)
 and a negligible contribution at low $z$ ($M_{\gamma\gamma}=M_K
\sqrt{z}$). Moreover, the prediction
agrees well with the experimental spectrum (Barr et al., 1992) shown
in Fig.~\ref{fig:klspexp}. On the other hand, the experimental value
for the rate is at least a factor two bigger than (\ref{eq:klpggp4}):
\beq
BR(K_L \to \pi^0 \gamma \gamma)=\left\{\ba{lr}
 (1.7 \pm 0.2 \pm 0.2 )\cdot{10}^{-6}&\qquad [{\rm Barr~et~al.~(NA31)},
1990, 1992] \\
 (1.86\pm 0.60\pm 0.60)\cdot{10}^{-6}&\qquad [{\rm Papadimitriou~et~al.
{}~(E731)}, 1991] \ea \right.~.\label{eq:brkl}
\eeq
Here, there is a difference to $\gamma\gamma \to \pi^0\pi^0$.
For $K_L \to \pi^0 \gamma\gamma$,
the shape of the spectrum is in good agreement with experiment,
but the total number of events predicted is definitely too low.
Is there a way to understand this discrepancy without ruining the
agreement in the spectrum? Short of a complete calculation to $O(G_F p^6)$,
many efforts have been made to grasp at least the main features of
the underlying physics. For rather obvious reasons, I describe here
the approach of Cohen et al. (1993) and refer to the already mentioned
specialized reviews (D'Ambrosio et al., 1994; de Rafael, 1995;
Ecker et al., 1995) for a more complete list of relevant papers.
\begin{figure}
    \begin{center}
       \setlength{\unitlength}{1truecm}
       \begin{picture}(7.0,8.0)
       \put(-2.4,-4.8){\includegraphics{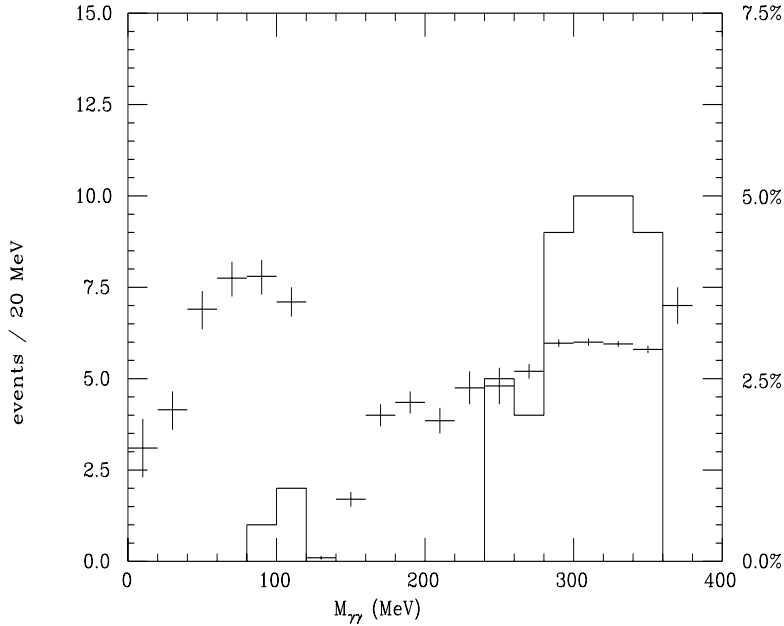}}
       \end{picture}
    \end{center}
    \caption{$2\gamma$--invariant--mass distribution (Barr et al., 1992) for
unambiguous $K_L \rightarrow \pi^0 \gamma\gamma$ candidates
(solid histogram). The crosses indicate the acceptance. }
    \protect\label{fig:klspexp}
\end{figure}

Since we expect the major contribution to the amplitude to come from
the two--pion intermediate states (as clearly indicated by the
experimental spectrum), we should try to get a more realistic answer
for this contribution. One indication in the right direction  is the
observation (Donoghue, 1991) that the octet coupling $G_8$ appearing in
(\ref{eq:klpggd}) has been  extracted from $K_S\to 2\pi$  underestimating
the $K_L\to 3\pi$ amplitudes  by 20$\% \div 30\%$. Therefore,
one expects an increase of the rate for $K_L\to \pi^0\gamma\gamma$
if one includes the $O(G_F p^4)$ corrections to $K_L\to 3\pi$ for
the weak vertex in the loop diagrams. The explicit calculation
(Cappiello et al., 1993; Cohen et al., 1993) shows indeed
an increase of the decay rate
by some $25\%$, but also a modification of the spectrum. In fact, the resulting
spectrum (dashed curve in Fig.~\ref{fig:klspth}) is even more strongly
peaked at large $z$ than found experimentally (see Fig.~\ref{fig:klspexp}).
This statement must be made more precise because the loop
amplitude is actually divergent now. Thus, there are certainly local
contributions of $O(G_F p^6)$ (and higher) that must be considered.
Cohen et al. (1993) adopted the successful recipe from the strongly
interacting sector: take the scale dependent loop amplitude at a
scale of the order of $M_\rho$ and saturate the LECs of $O(G_F p^6)$
with resonance exchange. As discussed in Sect.~\ref{subsec:Ni}, there
is however a problem with resonance exchange in the weak sector that
does not occur for purely strong interactions: the weak couplings
of resonances are essentially unknown. Nevertheless, one can parametrize the
$O(G_F p^6)$ contributions to $K_L\rightarrow \pi^0 \gamma \gamma$
by an effective vector coupling $a_V$ (Ecker et al., 1990a):
\beq
A={G_8 M_K^2\alpha\over \pi } a_V(3-z+r_{\pi}^2) ~,\qquad\qquad
B=-{2G_8 M_K^2\alpha\over \pi } a_V \label{klppge}~.
\eeq
Thus, $V$ exchange generates a $B$ amplitude changing the $O(G_F p^4)$
spectrum, particularly in the region of  small $z$. Even if one accepts
our ignorance about the size of $a_V$ related to vector meson exchange,
one can confront the corrected amplitudes $A$ and $B$ with the experimental
data. Since there is now a single free parameter $a_V$, one can
at least check if both rate and spectrum can be described by this
one--parameter amplitude. Choosing for instance $a_V \simeq
-0.9$ to reproduce the experimental spectrum, one obtains (Cohen et al.,
1993) indeed good agreement with the experimental branching ratios
(\ref{eq:brkl}).
A more complete unitarization of the $\pi \pi$ intermediate states
and inclusion of the experimental
$\gamma\gamma\to \pi\pi$ amplitudes increases the
$K_L\to \pi^0\gamma\gamma$ width by another 10\% (Kambor and
Holstein, 1994a).

There are additional checks whether the above approach really captures
the main physics. One argument in favour is provided by the factorization
model discussed in Sect.~\ref{subsec:Ni} that actually predicts
a negative value for the parameter $a_V$ of the right magnitude.
However, there is a more crucial test. The same mechanism applied to the
decay $K_S \to \gamma\gamma$ predicts that the branching ratio of
$O(G_F p^4)$ (D'Ambrosio and Espriu, 1986; Goity, 1987) should be essentially
unchanged:
\beq
 BR(K_S \rightarrow \gamma \gamma)=2.1\cdot{10}^{-6}~.
\label{eq:ksggbr}
\eeq
The ``unitarity corrections" are in this case already absorbed in
the coupling constant $G_8$ determined
from $K \to 2 \pi$ decays and there are no $V$ or $A$ exchange contributions.
The present experimental value
\beq
BR({K}_{S}\rightarrow 2 \gamma)=(2.4 \pm 1.2)\cdot{10}^{-6}
\label{eq:ksggb}
\eeq
is in agreement with the prediction (\ref{eq:ksggbr}), but
better precision is clearly needed to corroborate this
agreement. The $\Phi$ factory DA$\Phi$NE at Frascati is expected to
detect about 4000 events per year (D'Ambrosio et al., 1994) and
will improve the precision of the experimental rate considerably.

\setcounter{equation}{0}
\setcounter{subsection}{0}
\setcounter{table}{0}
\setcounter{figure}{0}

\section{MESONS AND VIRTUAL PHOTONS}
\label{sec:virtual}
\subsection{Generating functional of $O(e^2 p^2)$}
\label{subsec:e2p2}
As outlined in Sect.~\ref{subsec:chisb}, the photon can be
introduced as a dynamical degree of freedom into the CHPT framework.
The relevant lowest--order Lagrangian has the form (Ecker et al., 1989a)
\beq
\cL_2^Q = \frac{F^2}{4} \langle D_\mu U D^\mu U^\dg +
\chi U^\dg + \chi^\dg U \rangle + e^2 C \langle QUQU^\dg\rangle
- \frac{1}{4} F_{\mu\nu} F^{\mu\nu} + \cL_{\rm g.f.}~.
\label{eq:L2Q}
\eeq
The covariant derivative
\beq
D_\mu U = \partial_\mu U - i(r_\mu - e Q A_\mu)U +
i U(l_\mu - e QA_\mu)
\eeq
now includes the dynamical photon field. $\cL_{\rm g.f.}$ is the gauge--fixing
term for the photon.

The one--loop functional for the Lagrangian (\ref{eq:L2Q}) was recently
analysed by Urech (1995) who calculated in particular its divergent part.
Here, we are interested in the terms of $O(e^2 p^2)$ that must be added
to the lowest--order functional of $O(e^2 p^0)$ [corresponding
to taking the
Lagrangian (\ref{eq:e2p0}) at tree level]. The most general effective
chiral Lagrangian of $O(e^2 p^2)$ can be constructed in a straightforward
way using the spurion formalism described in Sect.~\ref{subsec:chisb}.
It can be cast into the following form (Urech, 1995; Neufeld and
Rupertsberger, 1994):
\beqa
\cL_{(e^2 p^2)} &=&
F^2 e^2 \{  K_1 \; \langle Q^2\rangle \; \langle D_\mu U
D^\mu U^\dg \rangle + K_2 \; \langle U Q U^\dg Q \rangle \; \langle
D_\mu U D^\mu U^\dg \rangle \no \\
&& \mbox{} + K_3 \; [\langle Q U^\dg D_\mu U\rangle \; \langle Q
U^\dg D^\mu U \rangle + \langle Q U D_\mu U^\dg \rangle \;
\langle Q U D^\mu U^\dg \rangle ] \no \\
&& \mbox{} + K_4 \; \langle D_\mu U Q U^\dg \rangle \;
\langle D^\mu U^\dg Q U \rangle
+ K_5 \; \langle Q^2 (D_\mu U^\dg D^\mu U + D_\mu U
D^\mu U^\dg) \rangle \no \\
&& \mbox{} + K_6 \; \langle U Q U^\dg Q D_\mu U D^\mu U^\dg +
Q U Q U^\dg D_\mu U D^\mu U^\dg \rangle \no \\
&& \mbox{} +  K_7 \; \langle Q^2 \rangle \; \langle \chi U^\dg
+ \chi^\dg U \rangle
+ K_8\; \langle U Q U^\dg Q \rangle \; \langle \chi
U^\dg + \chi^\dg U \rangle
\no \\
&& \mbox{}+ K_9 \; \langle Q^2 ( U^\dg \chi + \chi^\dg U +
\chi U^\dg + U \chi^\dg ) \rangle  \label{eq:LE2P2} \\
&& \mbox{} + K_{10}\; \langle Q U^\dg Q \chi + U Q U^\dg Q U
\chi^\dg + U^\dg Q U Q U^\dg \chi + Q U Q \chi^\dg
\rangle \no \\
&& \mbox{} - K_{11} \; \langle Q U^\dg Q \chi - U Q U^\dg Q U
\chi^\dg - U^\dg Q U Q U^\dg \chi + Q U Q \chi^\dg
\rangle \no \\
&& \mbox{}+ K_{12}\; \langle D_\mu U^\dg [D^\mu Q_R,Q] U +
D_\mu U [D^\mu Q_L,Q] U^\dg \rangle \no \\
&& \mbox{}+ K_{13} \; \langle U D_\mu Q_L U^\dg D^\mu Q_R \rangle
+ K_{14} \; \langle D_\mu Q_L D^\mu Q_L + D_\mu Q_R D^\mu Q_R
\rangle  \} ~,\no
\eeqa
with 14 dimensionless LECs $K_i$ and with
\beq
D_\mu Q_L = - i [l_\mu,Q]~, \qquad D_\mu Q_R = - i[r_\mu,Q]~.
\eeq
In the usual way, renormalization at $O(e^2 p^2)$ is performed through
the decomposition
\beq
K_i = K_i^r(\mu) + \Sigma_i \Lambda(\mu)~,
\eeq
with $\Lambda(\mu)$ given in Eq.~(\ref{eq:L4div}). The coefficients
$\Sigma_i$ are chosen to cancel the divergent part of the one--loop
functional of $O(e^2 p^2)$ (Urech, 1995):
\beq
\ba{rclrclrcl}
\Sigma_1 &=& \dfrac{3}{4} & \Sigma_2 &=& Z &
\Sigma_3 &=& - \dfrac{3}{4} \\[7pt]
\Sigma_4 &=& 2Z & \Sigma_5 &=& - \dfrac{9}{4} \qquad &
\Sigma_6 &=& \dfrac{3}{2} Z \\[7pt]
\Sigma_7 &=& 0 & \Sigma_8 &=& Z & \Sigma_9 &=& - \dfrac{1}{4} \\[7pt]
\Sigma_{10} &=& \dfrac{1}{4} + \dfrac{3}{2} Z \quad &
\Sigma_{11} &=& \dfrac{1}{8} & \Sigma_{12} &=& \dfrac{1}{4} \\[7pt]
\Sigma_{13} &=& 0 & \Sigma_{14} &=& 0 & Z &:=& \dfrac{C}{F^4}~.
\ea
\label{eq:Sigma}
\eeq
The renormalized coupling constants $K_i^r(\mu)$ are measurable LECs.

The divergence and renormalization structure has partly been calculated
in a different and independent manner by Neufeld and Rupertsberger
(1994). They concentrate on corrections of $O(e^2 p^2)$ to meson masses,
meson decay constants and $P_{\ell 3}$ form factors that depend on
eight linear combinations of the $K_i$:
\beq
\ba{llll}
S_1 = K_1 + K_2~, & \quad S_2 = K_5 + K_6~, & \quad S_3 = - 2K_3 + K_4~, &
\quad S_4 = K_7 + K_8~, \\[7pt]
S_5 = K_9 + 2K_{10} + K_{11}~, & \quad S_6 = K_8~, & \quad S_7
= K_{10} + K_{11}~, & \quad S_8 = - K_{12}~.
\ea
\eeq
By requiring finiteness of the considered observables, they determine
the divergent parts of the $S_i$. Their results agree with those of
Urech (1995) in Eqs.~(\ref{eq:Sigma}).

At the moment, essentially nothing is known about the values of the
renormalized LECs of $O(e^2 p^2)$. Extracting the $K_i^r(\mu)$ or
$S_i^r(\mu)$ from experimental data is in general difficult, both because
of lack of precise data and because of competition from the isospin
violating quark mass difference $m_u - m_d$. To get some idea about the
values of these LECs, one will have to resort to specific models, e.g.,
including meson resonance exchange. For the time being, one has to content
oneself in most cases with naive chiral dimensional analysis that
suggests $(4\pi)^{-2} = 6 \cdot 10^{-3}$ as a reasonable upper limit
for the LECs. Since there is no a priori argument whether this upper
limit applies to the $|K_i^r(\mu)|$ or $|S_i^r(\mu)|$ or to some other
linear combinations, there is of course some arbitrariness involved.
Nevertheless, the next subsection will show that one can obtain some
understanding of the relevance of $O(e^2 p^2)$ effects even with
such a crude approach. It is an important
achievement to determine the structure of effects of $O(e^2 p^2)$ as
predicted by the Standard Model. Theoretical advances in understanding the
corresponding LECs will in time lead to more quantitative predictions.

\subsection{Applications}
\label{subsec:appvirt}

\subsubsection{Meson masses}
In general, electromagnetic corrections compete with isospin violation
induced by $m_u \neq m_d$. As discussed in Sect.~\ref{subsec:mq},
the electromagnetic corrections to meson masses are needed to obtain
more reliable estimates for the quark mass ratios, in particular for
$m_u/m_d$. Thus, we concentrate here on the corrections of $O(e^2 p^2)$
and keep $m_u = m_d$.

Up to very small corrections of $O[(m_u - m_d)^2]$, the pion mass
difference is purely electromagnetic. Neglecting also terms of $O(e^2
M^2_\pi)$, one finds (Neufeld and Rupertsberger, 1994; Urech, 1995)
\beq
M^2_{\pi^+} - M^2_{\pi^0} = \frac{2 e^2 C}{F^2} + 2 e^2 M^2_K
\left[ 4 S^2_6(\mu) - \frac{16C}{F^4} L_4^r(\mu) -
\frac{C}{(4\pi)^2 F^4} \ln \frac{M^2_K}{\mu^2} \right]~.
\eeq
As it stands, the formula is of little use. Estimating $C$ with
resonance exchange (Das et al., 1967; Ecker et al., 1989a) reproduces
the experimental mass difference already with the lowest--order result
in the chiral limit. However, one can check the consistency of this
prediction by setting the correction of $O(e^2 M^2_K)$ to zero. The
corresponding value (Neufeld and Rupertsberger, 1994)
$S_6^r(M_\rho) = (- 2.1 \pm 1.6) \cdot 10^{-3}$ is compatible
with naive dimensional analysis.

The poorly known, Zweig suppressed LEC $L_4$ drops out in the more
interesting quantity (always for $m_u = m_d$)
\beq
\Delta_{\rm EM} = M^2_{K^+} - M^2_{K^0} - M^2_{\pi^+} + M^2_{\pi^0}~.
\eeq
The result of the one--loop calculation is (Urech, 1995; Neufeld and
Rupertsberger, 1994)
\beqa
\Delta_{\rm EM} &=& - e^2 M^2_K \left[ \frac{1}{(4\pi)^2} \left( 3 \ln
\frac{M^2_K}{\mu^2} - 4 + \frac{2C}{F^4} \ln \frac{M^2_K}{\mu^2}\right)
+ \frac{4}{3} S_2^r(\mu) - 8 S_7^r(\mu) + \frac{16C}{F^4} L_5^r(\mu)
\right] \no \\
&& \mbox{} + O(e^2 M^2_\pi)
\eeqa
or
\beq
\Delta_{\rm EM} = (- 0.26; 0.52; 0.99 \pm 0.13) \cdot 10^{-3}\mbox{ GeV}^2
+ \Delta M^2_{\rm ct}(\mu)~,
\eeq
where the entries in brackets contain both the loop contributions and
$L_5^r(\mu)$ for $\mu = 0.5$, 0.77 and 1~GeV, respectively.
The $\mu$ dependence of the first term is of course
cancelled by the scale dependence of $\Delta M^2_{\rm ct}(\mu)$
containing the LECs $S_2^r(\mu)$, $S_7^r(\mu)$. To appreciate the strong scale
dependence of the two terms, one should recall
\beq
\left. M^2_{\pi^+} - M^2_{\pi^0} \right|_{\rm exp} = 1.26 \cdot 10^{-3}
\mbox{ GeV}^2~.
\eeq
Turning to chiral dimensional analysis for an estimate of
$\Delta M^2_{\rm ct}$, one finds
\beqa
| \Delta M^2_{\rm ct}| \; \lets \; 2.6 \cdot 10^{-3} \mbox{ GeV}^2 &&
\qquad ({\rm Urech}, 1995) \\
- 0.6 \; \lets \; \Delta_{\rm EM} \cdot 10^3 \mbox{ GeV}^{-2} \; \lets \;
2.0 && \qquad ({\rm Neufeld~and~Rupertsberger}, 1994)~.
\eeqa
The conclusion is that corrections to Dashen's theorem {\em could} be sizable.
Moreover, a comparison with Sect.~\ref{subsec:mq} shows that a
positive $\Delta_{\rm EM}$ decreases the ratio $Q^2$ of quark masses
defined in (\ref{eq:Qratio}). For example, $\Delta_{\rm EM} = 2 \cdot
10^{-3}$~GeV$^2$
lowers $Q$ by 15\% compared to the Dashen limit value of $Q \simeq 24$.

More quantitative estimates of $\Delta_{\rm EM}$ require additional
assumptions. Two approaches have been pursued to estimate
$\Delta_{\rm EM}$. Although both models are not restricted to $O(e^2p^2)$
they amount essentially to estimates of the combination
$S_2^r(\mu) - 6 S_7^r(\mu)$. Donoghue et al. (1993b) argue that the
electromagnetic mass differences are dominated by long--distance
contributions. They extend the chiral amplitude for meson Compton
scattering entering the Cottingham formula to intermediate distances
using resonance saturation. For the electromagnetic mass differences,
the model is identical to the one used by Ecker et al. (1989a) to calculate
$C$ in the chiral limit. With the physical masses for the pseudoscalar
and the $V,A$ mesons, Donoghue et al. (1993b) find
\beq
\Delta_{\rm EM} = 1.2 \cdot 10^{-3}\mbox{ GeV}^2~, \label{eq:DDon}
\eeq
where the effect is mainly due to $M^2_K$ vs. $M^2_\pi$ in the meson
propagators. The uncertainty of the prediction (\ref{eq:DDon}) can
be estimated by noting that their model gives a value for
$M^2_{\pi^+} - M^2_{\pi^0}$ that is 20\% too high.

With a different approach, Bijnens (1993c) estimates both
short-- and long--distance contributions to $(M^2_{K^+} - M^2_{K^0})_{\rm em}$
to leading order in $1/N_c$. The same approach gave a satisfactory
result for the pion mass difference (Bardeen et al., 1989) in the
chiral limit. Matching the short-- and long--distance contributions
at the point of minimal sensitivity to the matching scale (600 $\div$
800 MeV), Bijnens finds
\beq
\Delta_{\rm EM} = (1.3 \pm 0.4) \cdot 10^{-3} \mbox{ GeV}^2
\eeq
in agreement with (\ref{eq:DDon}). The overall conclusion is that
$\Delta_{\rm EM}$ is most probably positive. The model estimates suggest
that $Q$ is about 10\% smaller than the value based on Dashen's
theorem. This goes in the right direction to reconcile the chiral
prediction for $\eta \ra 3\pi$ with experiment (see
Sect.~\ref{subsec:phenp4}).

\subsubsection{$P_{l3}$ form factors}
The relatively large corrections of $O(e^2 p^2)$ to Dashen's theorem
are motivation enough to investigate other isospin violating observables.
Neufeld and Rupertsberger (1994) have performed a study of elec\-tromagnetic
corrections to $K_{l3}$ and $\eta_{l3}$ form factors.

For the ratio $r_{K\pi}$ of $K_{l3}$
form factors they find the following expression:
\beq
r_{K\pi} = \frac{f_+^{K^+\pi^0}(0)}{f_+^{K^0 \pi^-}(0)} =
1 + \sqrt{3} \left( \ve - \frac{M^2_{\pi^0 \eta}}
{M_\eta^2 - M_\pi^2} \right) + \frac{3e^2}{4(4\pi)^2} \ln
\frac{M_K^2}{M_\pi^2}~.
\eeq
Here, $\ve$ is the $\pi^0 - \eta$ mixing angle and $M^2_{\pi^0
\eta}$ is the off--diagonal element of the $\pi^0 -
\eta$ mass matrix in the basis of the tree--level mass eigenfields.
It contains both the QCD contribution proportional to $\ve$ and the
electromagnetic contributions of $O(e^2 p^2)$.
In the limit $e = 0$, the formula for $r_{K \pi}$ coincides with
the previous result of Leutwyler and Roos (1984). The
data on the decays $K^+ \ra \pi^0 e^+ \nu_e$ and $K^0 \ra \pi^- e^+ \nu_e$
show clear evidence for the presence of isospin breaking (Gasser
and Leutwyler, 1985b):
\beq
\left|\frac{f_+^{K^+\pi^0}(0)}{f_+^{K^0 \pi^-}(0)}\right|^2 = 1.057 \pm 0.019~,
\label{eq:EXPRAT}
\eeq
which implies
\beq
(r_{K \pi} - 1)_{\rm exp} = (2.8 \pm 0.9) \cdot 10^{-2}~.
\label{eq:RKPIEXP}
\eeq
With the values for the low--energy constants $L_7, L_8^r$
in Table \ref{tab:Li}, the QCD contribution to
$r_{K \pi} - 1$ is given by
\beq
(r_{K \pi} - 1)_{\rm QCD} = 2.1 \cdot 10^{-2}~.
\eeq
On the basis of naive chiral dimensional analysis, the electromagnetic
contributions are expected to be small (Neufeld and Rupertsberger,
1994) compared to the QCD part,
\beq
0 \; \lets \; (r_{K \pi} - 1)_{\rm em} \; \lets \; 0.2 \cdot 10^{-2}~.
\eeq

Neufeld and Rupertsberger (1994) have also calculated  the $\eta_{l3}$ form
factors $f_{\pm}^{\eta\pi}(t)$ at the one--loop level
including the electromagnetic contributions of $O(e^2 p^2)$. In
this case, there is no lowest--order contribution because the
two form factors vanish for $e=0$ in the chiral limit. There is an
interesting low--energy theorem relating  $f_+^{\eta\pi}(0)$  to
the ratio of $K_{l3}$ form factors (Neufeld and Rupertsberger, 1994):
\beq
f_+^{\eta\pi}(0) =
\frac{1}{\sqrt{3}}[r_{K\pi} - 1 - \frac{3\;e^2}{4(4\pi)^2}
\ln \frac{M_K^2}{M_\pi^2}]~.
\eeq
Inserting the experimental value for $r_{K\pi}-1$ given in
(\ref{eq:RKPIEXP}), one obtains the prediction
\beq
f_+^{\eta\pi}(0) = (1.6 \pm 0.5) \cdot 10^{-2}~.
\label{eq:fplusexp}
\eeq
In view of the large error in (\ref{eq:fplusexp}), a more promising
strategy is to employ the same input parameters as before. For
the QCD contribution, one gets
\beq
f_+^{\eta\pi}(0)|_{\rm QCD} = 1.21 \cdot 10^{-2}
\eeq
and the electromagnetic contributions are again expected to be rather small,
\beq
0.02 \cdot 10^{-2} \; \lets \; f_+^{\eta\pi}(0)|_{\rm em} \;
\lets \; 0.15 \cdot
10^{-2}~.
\eeq

With the full expressions for the form factors $f_{\pm}^{\eta\pi}(t)$,
the branching ratios for the decays $\eta \ra \pi \ell \nu$ ($\ell =
e, \mu$) can be computed. Adding all channels, Neufeld and Rupertsberger
(1994) obtain
\beq
1.6 \cdot 10^{-13} \; \lets \;
\sum_{\ell=e,\mu} BR(\eta \ra \pi^{\pm} \ell^{\mp}
\stackrel{(-)}{\nu_\ell}) \; \lets \; 2.0 \cdot 10^{-13}~.
\label{eq:THB}
\eeq
Thus, the decays $\eta \ra \pi l \nu$ are still out of reach
for present experimental facilities.
On the other hand, the observation of a decay rate
considerably larger than the upper bound in (\ref{eq:THB}) would be
a clear signal for a deviation from the Standard Model.

The decay $\tau \ra \eta \pi \nu$ is governed by the same hadronic
form factors as $\eta \ra \pi l \nu$. However, most of the Dalitz plot
is outside the domain of applicability of CHPT.
In order to obtain reasonable theoretical results also in this
intermediate energy range, the dominant contributions of the
lowest--lying resonance states [$\rho(770)$ and
$a_0(980)$ in this case] have to be taken into account.
Taking the vector decay constant $\left| F_{a_0} \right| = 1.28 \mbox{ MeV}$
from a QCD sum rule analysis (Narison, 1987), Neufeld and Rupertsberger
arrive at the prediction
\beq
BR(\tau \ra \eta \pi \nu) \simeq 1.2 \cdot 10^{-5},
\eeq
to be compared with the present experimental bound (Artuso et al., 1992)
\beq
BR(\tau \ra \eta \pi \nu)|_{\rm exp} < 3.4 \cdot 10^{-4}~.
\eeq

\setcounter{equation}{0}
\setcounter{subsection}{0}
\setcounter{table}{0}
\setcounter{figure}{0}

\section{CHPT WITH BARYONS}
\label{sec:baryons}
\subsection{Relativistic formulation}
\label{subsec:relform}
In most of this section, I will restrict the discussion to
chiral $SU(2)$. More precisely, the local chiral symmetry considered
will be $SU(2)_L\times SU(2)_R\times U(1)_V$.
At the QCD level, the starting Lagrangian is [compare with
Eq.~(\ref{eq:QCD})]
\beq
\cL = \cL^0_{\rm QCD} + \bar q \gamma^\mu \left(v_\mu + \frac{1}{3}
v^{(s)}_\mu +  a_\mu \gamma_5\right) q - \bar q (s - i\gamma_5 p)q~, \qquad
q = { u \choose d} ~. \label{eq:QCDf2}
\eeq
The isotriplet vector and axial--vector fields $v_\mu, a_\mu$ are traceless.
The isosinglet vector field $v^{(s)}_\mu$ is included to generate the
electromagnetic current. At the effective level of pions and nucleons,
$v^{(s)}$ couples directly only to nucleons because the pions have zero
baryon number. The normalization of the vector fields in (\ref{eq:QCDf2})
implies
\beqa
v_\mu & = & - {e \over 2}   A_\mu^{\rm ext} ~\tau_3 + \dots \\*
v^{(s)}_\mu & = &  - {e \over 2}  A_\mu^{\rm ext} ~{\bf 1} \no
\eeqa
for external photons.

As discussed in Sect.~\ref{sec:chi}, the nucleon doublet $\Psi$ transforms as
\beq
\Psi = { p \choose n} \toG \Psi' = h(g,\vp)\Psi \label{eq:psi}
\eeq
under chiral transformations. The covariant derivative
\beq
\nabla_\mu \Psi = (\partial_\mu + \Gamma_\mu - i v^{(s)}_\mu)\Psi
\eeq
is defined in terms of the connection (\ref{eq:conn}) and of the isosinglet
field $v^{(s)}$.
The effective Lagrangian for the pion--nucleon system
\beq
\cL_{\rm eff} = \cL_M + \cL_{\pi N} \label{eq:LeffpN}
\eeq
contains the purely mesonic Lagrangian $\cL_M$ of
Sect.~\ref{sec:strong} for chiral $SU(2)$.
Although we shall consider the general meson--baryon Lagrangian
in the following subsection for setting up the chiral power counting,
only the part of $\cL_{\pi N}$ that is bilinear in the nucleon fields
will be needed here. Because of the different Lorentz structure of meson and
baryon fields, the chiral expansion of $\cL_{\pi N}$ contains terms of
$O(p^n)$ for each positive integer $n$ (Gasser et al., 1988):
\beq
\cL_{\pi N} = \cL_{\pi N}^{(1)} + \cL_{\pi N}^{(2)} + \cL_{\pi N}^{(3)}
+ \ldots~,   \label{eq:LpiN}
\eeq
with $\cL_{\pi N}^{(1)}$ given in (\ref{eq:piN1}). The higher--order
pion--nucleon Lagrangians will be considered in the following subsections
in the framework of heavy baryon CHPT. In the construction of those
Lagrangians, one encounters a basic difference between
Goldstone and non--Goldstone fields. Since the nucleon mass remains
finite in the chiral limit, the four--momentum of a nucleon can never be
``soft''. This complicates the chiral counting considerably. For instance,
both $\Psi$ and $\nabla_\mu \Psi$ count as fields of $O(1)$, whereas
$(i \not\!\nabla - m)\Psi$ is $O(p)$ (Gasser et al., 1988; Krause, 1990).

The generating functional of Green functions $Z[j,\eta,\bar
\eta]$ can be defined by the path integral (Gasser et al., 1988)
\beq
e^{iZ[j,\eta,\bar\eta]} =  \int [du d\Psi d \bar \Psi]
\exp [i\{ S_M + S_{\pi N} + \int d^4 x (\bar \eta \Psi + \bar \Psi \eta)\}]~.
\label{eq:ZpiN}
\eeq
The action $S_M + S_{\pi N}$ corresponds to the effective Lagrangian
(\ref{eq:LeffpN}), the external fields $v,a,s,p$ are denoted collectively
as $j$ and $\eta,\bar \eta$ are fermionic sources. Since the
generating functional (\ref{eq:ZpiN}) contains closed nucleon
loops, the question of anomalies arises. On the other hand, the anomalous
Ward identity (\ref{eq:anomaly}) for $Z$ is already saturated by the
Wess--Zumino--Witten functional (\ref{eq:WZW}). Thus, any addition
to the generating functional coming from the meson--baryon sector must be
chiral invariant. This is indeed the case. The nucleon determinant is chiral
invariant and therefore free of chiral anomalies (Leutwyler,
1988, unpublished notes). The reason is very simple: in the non--linear
realization of chiral symmetry, the left-- and right--chiral components
of the nucleon field transform in the same way. The transformation
(\ref{eq:psi}) for $\Psi$ is vector--like. As is well known, the
situation is different for the linear sigma model (Steinberger, 1949;
Bell and Jackiw, 1969).

The presence of the nucleon mass $m$, which stays finite in the chiral limit,
complicates the chiral power counting. In the relativistic formulation
of the pion--nucleon system (more generally, the meson--baryon system),
the correspondence between the loop and the chiral expansion
valid in the meson sector is lost. An
amplitude with given chiral dimension may receive contributions from
diagrams with an arbitrary number of loops (Gasser et al., 1988).
In particular, the coupling
constants in the pion--nucleon Lagrangian (\ref{eq:piN1}) get
renormalized in every order of the loop expansion. In contrast,
the LECs $F$, $B$ of the lowest--order mesonic Lagrangian (\ref{eq:L2})
are not renormalized in a mass--independent regularization scheme.
Likewise, the degree of homogeneity of the baryonic amplitude
is no more related to its chiral dimension as in the
mesonic case [cf. Eq.~(\ref{eq:DF})].
The nucleon mass is comparable to the intrinsic scale
$4\pi F \simeq 1.2$~GeV of CHPT. This suggests to set up baryon CHPT
in such a way as to expand in
$$
\dfrac{p}{4\pi F} \qquad \mbox{and} \qquad \dfrac{p}{m}
$$
simultaneously, where $p$ is a small momentum. In the relativistic formulation
of CHPT, there is a basic difference between $F$ and $m$ in a generic
loop amplitude: $F$ appears only in the vertices, whereas the nucleon
mass is contained in the propagator.

\subsection{Heavy baryon CHPT}
\label{subsec:HBCHPT}
Heavy baryon CHPT (HBCHPT) (Jenkins and Manohar, 1991a, 1992c) can be viewed
as a clever choice of variables for performing the fermionic path integral in
(\ref{eq:ZpiN}). By shifting the dependence on the nucleon mass $m$ from
the nucleon propagator to the vertices of the effective Lagrangian,
the integration over the new fermionic variables produces a systematic
low--energy expansion.

In terms of velocity dependent fields $N_v,H_v$ (Georgi, 1990) defined as
\footnote{Following standard nomenclature,
both the external isotriplet vector matrix field and the four--velocity
are denoted by the same symbol $v_\mu$.}
\beqa
N_v(x) &=& \exp[i m v \cdot x] P_v^+ \Psi(x) \label{eq:vdf} \\
H_v(x) &=& \exp[i m v \cdot x] P_v^- \Psi(x) \no \\
P_v^\pm &=& \frac{1}{2} (1 \pm \not\!v)~, \qquad v^2 = 1 ~, \no
\eeqa
the pion--nucleon action $S_{\pi N}$ takes the form
\beqa
S_{\pi N} &=& \int d^4 x \{ \bar N_v A N_v + \bar H_v B N_v +
\bar N_v \gamma^0 B^\dg \gamma^0 H_v - \bar H_v C H_v\} \label{eq:SpiN} \\
A &=& iv \cdot \nabla + g_A S \cdot u + A_{(2)} + A_{(3)} + \ldots \no \\
B &=& i \not\!\nabla^\perp - \frac{g_A}{2} v \cdot u \gamma_5 +
B_{(2)} + B_{(3)} + \ldots \no \\
C &=& 2m + i v \cdot \nabla + g_A S \cdot u + C_{(2)} + C_{(3)} + \ldots
\no \\
&& \nabla_\mu^\perp = \nabla_\mu - v_\mu v \cdot \nabla~, \qquad
[\not\!v,A] = [\not\!v,C] = 0~, \qquad \{\not\!v,B\} = 0~. \no
\eeqa
The operators $A_{(n)}$, $B_{(n)}$, $C_{(n)}$ are the corresponding projections
of the pion--nucleon Lagrangians $\cL_{\pi N}^{(n)}$.
In $A$ and $C$, the only dependence on Dirac matrices is through
the spin matrix
\beq
S^\mu = \frac{i}{2} \gamma_5 \sigma^{\mu\nu} v_\nu ~, \qquad
S \cdot v = 0~, \qquad S^2 = - \frac{3}{4} {\bf 1}~.\label{eq:spin}
\eeq

Rewriting also the source term in (\ref{eq:ZpiN}) in terms of
$N_v, H_v$,
one can now integrate out the ``heavy'' components $H_v$
 to obtain a non--local action in the ``light''
fields $N_v$ (Bernard et al., 1992f; Ecker, 1994c;
 see also Mannel et al., 1992).
At this point, the crucial approximation of HBCHPT is made:
the action is written as a series of local actions with
increasing chiral dimensions by expanding $C^{-1}$ in a power series
in $1/m$:
\beq
C^{-1} = \frac{1}{2m} - \frac{i v \cdot \nabla + g_A S \cdot u}{(2m)^2}
+ O(p^2)~.
\eeq
The relevant pion--nucleon Lagrangian is then
(Jenkins and Manohar, 1991a; Bernard et al., 1992f)
\beq
{\wh \cL_{\pi N}} = \bar N_v(A + \gamma^0 B^\dg \gamma^0 C^{-1} B)N_v =
\bar N_v \left( A_{(1)} + A_{(2)} + \frac{1}{2m} \gamma^0
B_{(1)}^\dg \gamma^0 B_{(1)}\right) N_v  + O(p^3) \label{eq:LpiN3}
\eeq
$$
A_{(1)} = iv \cdot \nabla + g_A S \cdot u ~, \qquad
B_{(1)} = i \not\!\nabla^\perp - \frac{g_A}{2} v \cdot u \gamma_5 ~.
$$
By construction, the propagator of the field $N_v$
\beq
S_v(k) = \frac{i P_v^+}{v \cdot k + i\ve} \label{eq:HBprop}
\eeq
is independent of the nucleon mass.

The formula (\ref{eq:DL}) for chiral power counting can now be generalized
to the meson--baryon system (Weinberg, 1990, 1991). In HBCHPT, the chiral
dimension $D$ of a connected $L$--loop amplitude with $I_M(I_B)$ internal
meson (baryon) lines and $N_d$ vertices of $O(p^d)$ is given by
\beq
D = 4L - 2I_M - I_B + \sum_d dN_d~.
\eeq
We now distinguish between vertices with different numbers $n_B$ of
baryon lines:
\beq
N_d = \sum_{n_B = 0,2,\ldots} N_{d,n_B}~.
\eeq
Using the general relations ($E_B$ is the number of external baryon lines)
\beqa
L &=& I_M + I_B - \sum_d N_d + 1 \\
2 I_B + E_B &=& \sum_{d,n_B} n_B N_{d,n_B}~,
\eeqa
one arrives at the final result
\beq
D = 2L + 2 - \frac{E_B}{2} + \sum_{d,n_B} \left\{ d - 2 + \frac{n_B}{2}
\right\} N_{d,n_B}~. \label{eq:DMB}
\eeq
Since the quantity in curly brackets is non--negative, this formula is
the starting point for a systematic chiral expansion.

However, the formula (\ref{eq:DMB}) is misleading for $E_B \geq 4$
(Weinberg, 1990, 1991). For instance, there is an obvious conflict with
the existence of nuclear binding. Although the formula (\ref{eq:DMB})
suggests that higher--order diagrams are suppressed, nuclear binding is
a clear manifestation of the breakdown of the perturbation expansion.
The solution of this conflict is provided by seeming infrared
divergences related to the massless baryon propagator (\ref{eq:HBprop}).
Of course, those infrared divergences do not really exist because the
baryons are certainly not massless. However, they invalidate the simple
formula (\ref{eq:DMB}) for $E_B \geq 4$. Weinberg (1990, 1991) has
suggested to use non--relativistic perturbation theory instead where
the problem with chiral power counting is related to small energy
denominators. Those small denominators occur for intermediate states
with baryons only. If one considers instead of the S--matrix an
effective potential defined as the sum of connected,
$N$--baryon--irreducible diagrams, the simple power counting formula
(\ref{eq:DMB}) is applicable for the effective potential (Weinberg,
1990, 1991). This formalism has been applied to various problems of
nuclear physics like the nucleon--nucleon force, the chiral suppression
of three--nucleon forces, meson--exchange currents, pion--nucleus scattering,
etc. [for recent reviews, see Park et al. (1993), Bernard et al. (1995)].

For $E_B \leq 2$, the problem of infrared divergences in HBCHPT does not
arise and the formula (\ref{eq:DMB}) for chiral power counting applies
directly to S--matrix elements. For $n_B = E_B = 0$, one recovers of
course the formula (\ref{eq:DL}) for the mesonic sector. In the
remainder of this section, I consider only processes with a single
incoming and outgoing baryon line ($E_B = 2$, $n_B \leq 2$). Using the
notation
$$
N_{d,0} = N_d^M, \qquad \qquad N_{d,2} = N_d^{MB}~,
$$
we obtain the relevant formula
\beq
D = 2L + 1 + \sum_d (d-2) N_d^M + \sum_d (d-1) N_d^{MB} \geq 2L+1~.
\label{eq:DMB2}
\eeq
Only tree diagrams contribute to $O(p)$ and $O(p^2)$.
Loops are suppressed by powers of $p^{2L}$.

Except at lowest order, the effective chiral Lagrangian of a given
$O(p^d)$ in HBCHPT consists of two parts. The first part $A_{(d)}$
arises from the projection of the relativistic Lagrangian
$\cL_{\pi N}^{(d)}$ on the subspace of the light components $N_v$
defined by the projector $P_v^+$. The second part is due to the
expansion of $C^{-1}$ in a power series in $1/m$. For chiral $SU(2)$
one finds \footnote{Except for a scale factor $1/m$ and for a
different normalization of external gauge fields, I have followed
the conventions of Mei\ss ner (1994c) for the $c_i$.}
 at $O(p^2)$ (Bernard et al., 1992f, 1994f; Mei\ss ner, 1994c)
\beqa
\label{eq:A2}
A_{(2)} &=& \frac{c_1}{m} \langle \chi_+\rangle + \frac{c_2}{m}
(v \cdot u)^2 + \frac{c_3}{m} u \cdot u + \frac{c_5}{m}
\left( \chi_+ - \frac{1}{2} \langle \chi_+\rangle \right) \no \\
&& \mbox{} + \frac{1}{m} \ve^{\mu\nu\rho\sigma} v_\rho S_\sigma
[i c_4 u_\mu u_\nu + c_6 f_{+\mu\nu} + c_7 (\partial_\mu v^{(s)}_\nu
- \partial_\nu v^{(s)}_\mu)]
\eeqa
and
\beqa
\label{eq:B2}
\frac{1}{2m} \gamma^0 B^\dg_{(1)} \gamma^0 B_{(1)} &=& \frac{1}{2m}
\{ (v \cdot \nabla)^2 - \nabla \cdot \nabla - ig_A \{S \cdot \nabla,
v \cdot u\} \no \\
&& \mbox{} - \frac{g_A^2}{4} (v \cdot u)^2 + \frac{1}{2}
\ve^{\mu\nu\rho\sigma} v_\rho S_\sigma [i u_\mu u_\nu + f_{+\mu\nu}
+ 2 (\partial_\mu v^{(s)}_\nu - \partial_\nu v^{(s)}_\mu)]\}~.
\eeqa
The $c_i$ are dimensionless LECs of $O(p^2)$.
At first sight, it may seem surprising that there are terms in
(\ref{eq:B2}) that have no counterparts in the general Lagrangian
$A_{(2)}$. After all, the procedure leading to (\ref{eq:B2})
respects chiral and Lorentz invariance. But if $A_{(2)}$ is the most
general structure respecting these symmetries, why are there terms
in (\ref{eq:B2}) that do not also appear in (\ref{eq:A2})?
On the other hand, there are good reasons why the first three
terms in (\ref{eq:B2}) have fixed coefficients in the complete Lagrangian
of $O(p^2)$. For instance, the first two terms in
(\ref{eq:B2}) determine the Compton scattering amplitude in the Thomson
limit (Bernard et al., 1992f; Ecker and Mei\ss ner, 1994d). Obviously,
those coefficients are fixed by gauge invariance. The solution of this
seeming puzzle has to do with the difference between covariance and
invariance. The four--vector $v^\mu$ singles out a special reference
frame. The projection of the general Lorentz invariant
Lagrangian $\cL_{\pi N}^{(d)}$ onto this reference frame will not
produce all possible covariant local terms with
$v^\mu$ transforming like a Lorentz four--vector. On the other
hand, such terms can and do occur in the expansion of the non--local
part $\gamma^0 B^\dg \gamma^0 C^{-1} B$ in powers of $1/m$. The two
different structures (\ref{eq:A2}) and (\ref{eq:B2}) are a manifestation
of this difference that is not always respected in the literature.

Except for the constant $c_5$ that would appear for instance in isospin
violating amplitudes \footnote{In a different notation, the term
with coupling constant $c_5$ has recently been considered by Weinberg
(1994).} for $m_u \neq m_d$, all LECs of $O(p^2)$ in the
pion--nucleon system are now known phenomenologically. In Table
\ref{tab:ci} that is adapted from recent lecture notes by Mei\ss ner (1994c),
the phenomenological values of these LECs are collected.
The LECs $c_1$, $c_3$ and
$c_4$ have also been determined in a fit to deuteron properties and $NN$
phase shifts at low energies in an investigation of the $NN$ interaction
to one--loop accuracy (Ord\'o\~nez et al., 1994; van Kolck, 1994). The
fitted values disagree with those in Table \ref{tab:ci}.
I refer to the lecture notes of Mei\ss ner (1994c) for a
discussion of this discrepancy.
\begin{table}
\caption{Numerical values of the LECs of $O(p^2)$ in the pion--nucleon
system (except $c_5$). The values are adapted from a recent compilation
of Mei\ss ner (1994c). Note that the conventions for the external
vector fields are different from those of Mei\ss ner. Here, $c_6=
\stackrel{\circ}{\kappa}_v$/4, $c_7=\stackrel{\circ}{\kappa}_s$/2,
where $\stackrel{\circ}{\kappa_v}$ ($\stackrel{\circ}{\kappa_s}$)
is the isovector (isoscalar) anomalous magnetic moment of the nucleon
in the chiral limit. The errors do not account for higher--order
chiral corrections.}
\label{tab:ci}
$$
\begin{tabular}{|c|c|c|} \hline
LEC & value & source \\ \hline
$c_1$ & $\quad  - 0.8 \pm 0.1 \quad$ &  $\sigma_{\pi N}$ \\
$c_2$ & $ 3.1 \pm 0.2$   & $\quad \pi N \ra \pi N \quad $ \\
$c_3$ & $ - 4.9 \pm 0.2$ & $\pi N \ra \pi N$ \\
$c_4$ & $ 3.9 \pm 0.1$   & $\pi N \ra \pi N$ \\
$c_6$ & $ 1.4 \pm 0.1$ & $\kappa_{p,n}$ \\
$c_7$ & $ - 0.06 \pm 0.0$ & $\kappa_{p,n}$ \\ \hline
\end{tabular}
$$
\end{table}

Can one understand the values of the LECs in Table \ref{tab:ci}?
Naive chiral dimensional analysis suggests $c_i = O(1)$ in rough
agreement with the phenomenological values. A more ambitious approach
is to extend the successful notion of resonance saturation in the
mesonic sector (cf. Sect.~\ref{subsec:Li}) to the meson--baryon
system. An instructive example is the case of $c_3$ with the biggest
absolute value in Table \ref{tab:ci}. This LEC receives contributions
from $\Delta(1232)$ and $N^*(1440)$ exchange, but also from scalar
meson exchange. Within the uncertainties of the corresponding resonance
couplings, $c_3^R$ due to resonance exchange is compatible with its
phenomenological value (Mei\ss ner,
1994a,c). A more detailed discussion of resonance contributions to the
$c_i$ can be found in the review of Bernard et al. (1995).

The dominant contribution to $c_3$ is due to $\Delta$ exchange. This is
partly due to the proximity of the $\Delta$, but also to its strong
coupling to the pion--nucleon system $(g_{\pi \Delta N} \simeq
2g_{\pi NN}$). It is a much debated issue in HBCHPT how to properly
account for the $\Delta$ resonance [or the whole decuplet in the
$SU(3)$ case]. While there is no controversy about the importance of
$\Delta$ exchange for the LECs $c_i$ of $O(p^2)$, the main question is
whether one should view the $\Delta$ as having been integrated out so
that its remnants appear in the appropriate LECs only, or whether one
should include the decuplet fields as dynamical fields in the
effective Lagrangian. To some extent, the two approaches correspond
in the meson sector to considering on one side chiral $SU(2)$ where
kaons and the $\eta$ meson appear only in LECs, while they are dynamical
degrees of freedom in chiral $SU(3)$. The difference between the meson
and the meson--baryon sectors is that $K,\eta$ become degenerate with
the pions in the chiral limit, whereas the nucleons and the $\Delta$
(octet and decuplet baryons) become degenerate in the large--$N_c$
limit (Witten, 1979).

A small drawback of the scenario with dynamical $\Delta$ fields is its
incompatibility with the chiral counting formula (\ref{eq:DMB2}).
This is due to the decuplet propagator in HBCHPT,
\beq
\frac{i P_{\mu\nu}}{v \cdot k - \Delta + i \ve}~,
\eeq
where $P_{\mu\nu}$ is an appropriate projector and $\Delta \simeq
\frac{5}{2} F_\pi$ is the octet--decuplet mass splitting in the chiral
limit. The situation is similar to the relativistic formulation of
meson--baryon interactions of the previous subsection.
In addition to modifications in the renormalization program, a term of
$O(p^d)$ in the standard counting will look schematically like (no external
gauge fields in this example)
\beq
M^d F(p/M; M/\Delta)
\eeq
with $p$ a generic (small) momentum and $M$ a pseudoscalar meson mass.
The chiral counting is spoiled because $M/\Delta$ has chiral dimension one.
Although the chiral and the large--$N_c$ limits do not commute, there is no
problem of principle (Jenkins and Manohar, 1991a, 1992c). Consider as an
instructive example the octet baryon mass splittings. Taking the chiral limit
first, only the octet intermediate states contribute to the leading
non--analytic corrections (Gasser, 1981). On the other hand, taking the
large--$N_c$ limit first, an entire tower of states (degenerate for $N_c
\ra \infty$) contributes (Jenkins and Manohar, 1991a, 1992c; Dashen et al.,
1994a,b).

Of course, neither of the two limits corresponds to the real world.
Since CHPT is applicable as long as $M$ and $\Delta$ are both much
smaller than $4\pi F_\pi$, irrespective of the size of $M/\Delta$,
the treatment of the decuplet baryons becomes mainly a question of
effectiveness. In most of the work with two light flavours (where
$M/\Delta < 1$), the baryon resonances have been taken into account
through their contributions to LECs (Bernard et al., any paper). For
chiral $SU(3)$ (where $M/\Delta > 1$ for $M = M_K,M_\eta$), most
authors have followed Jenkins and Manohar (1991a, 1992c) incorporating
the decuplet baryons as dynamical fields. We shall come back to the
three--flavour case in Sect.~\ref{subsec:SU3}.

\subsection{Renormalization to $O(p^3)$}
\label{subsec:piNren}
With the effective pion--nucleon Lagrangian (\ref{eq:LpiN3}) of the last
subsection, all Green functions and amplitudes with a single incoming
and outgoing nucleon can be calculated in a systematic chiral expansion:
nucleon form factors, $\pi N \ra \pi \ldots \pi N$, $\gamma^* N \ra \pi
\ldots \pi N$, $W^* N \ra \pi \ldots \pi N$.

Up to and including $O(p^2)$, only tree--level amplitudes contribute.
At $O(p^3)$, loop diagrams of the type shown in Fig.~\ref{fig:piNloop}
must be taken into account. Those diagrams are in general divergent requiring
regularization and renormalization. The divergences will
be cancelled by counterterms of $O(p^3)$ which are
part of the general chiral invariant pion--nucleon Lagrangian
(\ref{eq:LpiN3}). The sum of the reducible diagrams c, d is automatically
finite because the mesonic one--loop functional \footnote{For
diagrams c and d, the mesonic one--loop functional for chiral $SU(2)$ (Gasser
and Leutwyler, 1984) is needed.} has already been
rendered finite in Sect.~\ref{subsec:p4}.

\begin{figure}
\centerline{\epsfig{file=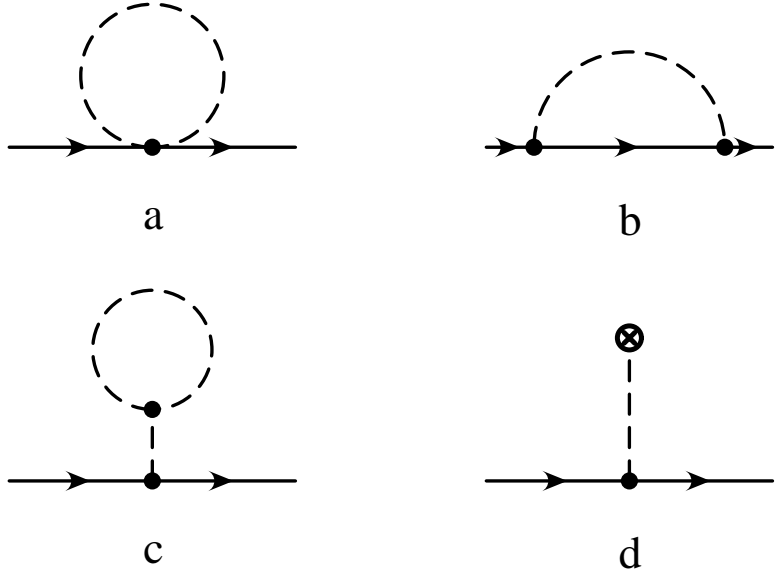,height=8cm}}
\caption{One--loop diagrams for pion--nucleon Green functions (a,b:
irreducible diagrams; c,d: reducible diagrams).
The full (dashed) lines denote the nucleon (meson) propagators.
The propagators and the vertices have the full tree--level
structure attached to them as functionals of the external fields.
The crossed circle denotes a vertex from the meson Lagrangian $\cL_4$
of $O(p^4)$ [for chiral $SU(2)$].}
\label{fig:piNloop}
\end{figure}

The divergent part of the one--loop functional of $O(p^3)$ can be
calculated in closed form (Ecker, 1994c) by using the heat kernel method
[for a general review, see Ball (1989)] for the meson and nucleon propagators
in the presence of external fields.
In this way, one can not only renormalize all single--nucleon Green functions
in a compact way, but one also obtains the chiral
logs for all of them. In a more precise formulation, one obtains
the scale dependence of the renormalized LECs of $O(p^3)$.
The details of the calculation can be found in Ecker (1994c).
A non--trivial part of that calculation consists in finding a heat kernel
representation for the inverse of the differential operator
\beq
iv \cdot \nabla + g_A S \cdot u~,
\eeq
the nucleon propagator in the presence of external fields appearing
in diagram b of Fig.~\ref{fig:piNloop}.

The renormalization program at $O(p^3)$ can be summarized in the
following way, in complete analogy to the mesonic case
at $O(p^4)$ (Sect.~\ref{subsec:p4}). The decomposition of
the one--loop functional into a divergent and a finite part introduces
an arbitrary scale $\mu$: although the total one--loop functional is
independent of $\mu$, the two separate parts are not. The divergent part is
then cancelled by a corresponding piece in the general effective Lagrangian of
$O(p^3)$,
\beqa
{\wh \cL^{(3)}_{\pi N}} & = & \bar N_v \left(A_{(3)} + \gamma^0 B^\dg \gamma^0
C^{-1} B|_{O(p^3)}\right) N_v  \nl
A_{(3)} & = & \frac{1}{(4\pi F)^2} \sum_i b_i O_i ~, \label{eq:Lct}
\eeqa
through the decomposition
\beq
b_i = b_i^r(\mu) + (4\pi)^2 \beta_i \Lambda(\mu) \label{eq:beta}
\eeq
of the dimensionless LECs $b_i$. The quantity $\Lambda(\mu)$ is
defined in Eq.~(\ref{eq:L4div}). The coefficients $\beta_i$ [analogous
to the $\Gamma_i$ in (\ref{eq:L4div})] are chosen such that
the divergent part of (\ref{eq:Lct}) cancels the divergent
piece of the one--loop functional. The generating functional
of $O(p^3)$ then contains both the finite part of the one--loop functional and
a tree--level functional due to (\ref{eq:Lct}),
with the couplings $b_i$ replaced by the renormalized coupling constants
$b_i^r(\mu)$. The complete functional of $O(p^3)$ is finite and independent
of $\mu$ by construction.

The list of operators $O_i$ in (\ref{eq:Lct}) with non--zero $\beta_i$
can be found in Ecker (1994c). Some of those terms ($O_{15}$,$\dots$,
$O_{22}$) contain the differential operator $v \cdot \nabla$
acting on the nucleon field. By making a field transformation on
the nucleon field, those terms in (\ref{eq:Lct}) can be
transformed away. This transformation reduces the number of monomials
of $O(p^3)$ with non--vanishing $\beta$--functions to 13. Of
course, the $\beta_i$ of the original basis are in general modified.
Moreover, the transformation will also affect the higher--order
Lagrangians. At $O(p^4)$, the new terms with (in general) divergent
coefficients are generated by applying the field transformation
to the $O(p^2)$ Lagrangian
\beq
{\wh \cL^{(2)}_{\pi N}} = \bar N_v\left(A_{(2)} + \dfrac{1}{2 m}
\gamma^0 B_{(1)}^\dg \gamma^0 B_{(1)}\right)N_v  \label{eq:LpiN2}
\eeq
with $A_{(2)}$ and $\gamma^0 B_{(1)}^\dg \gamma^0 B_{(1)}/2m$ given
in (\ref{eq:A2}) and (\ref{eq:B2}). The explicit form of the field
transformation and the induced terms of $O(p^4)$ will be given
elsewhere. Here, I will confine myself to the complete renormalization
at $O(p^3)$ for chiral $SU(2)$. In Table \ref{tab:beta}, the complete
list of operators $O_i$ with non--vanishing coefficients
$\beta_i$ in the new basis of nucleon fields is given.

\begin{table}
\caption{All counterterms of $O(p^3)$ with non--vanishing $\beta$--functions
in a suitably defined basis.
Compared to the list of counterterms in Ecker (1994c), a field
transformation has been applied to the nucleon field to
eliminate some of the original terms.}\label{tab:beta}
$$
\begin{tabular}{|r|c|c|} \hline
i  & $O_i$ & $\beta_i$ \\ \hline
1  & $i[u_\mu,v \cdot \nabla u^\mu]$ & $- g^4_A/6$ \\
2  & $i[u_\mu,\nabla^\mu v \cdot u]$ & $- (1 + 5 g^2_A)/12$ \\
3  & $i[v \cdot u, v \cdot \nabla v \cdot u]$ & $(3 + g^4_A)/6$ \\
4  & $S \cdot u \langle u \cdot u \rangle$ & $g_A (1 + 5 g^2_A +
4 g^4_A)/2$ \\
5  & $u_\mu \langle u^\mu S \cdot u\rangle$ & $g_A (3 - 9 g^2_A
+ 4 g^4_A)/6$ \\
6 & $S \cdot u \langle(v \cdot u)^2\rangle$ & $- g_A (2 + g^2_A
+ 2 g^4_A)$ \\
7  & $v \cdot u \langle S \cdot u v \cdot u\rangle$ & $g^3_A + 2 g^5_A/3$ \\
8  & $[\chi_-,v \cdot u]$ & $(1 + 5 g^2_A)/24$ \\
9  & $S \cdot u \langle \chi_+\rangle$ & $g_A/2 + g^3_A$ \\
10 & $\nabla^\mu f_{+\mu\nu} v^\nu$ & $- (1 + 5 g^2_A)/6$ \\
11 & $i S^\mu v^\nu [f_{+\mu\nu},v \cdot u]$ & $g_A + g^3_A$ \\
12 & $i S^\mu [f_{+\mu\nu}, u^\nu]$ & $- g^3_A$ \\
13 & $v_\lambda \ve^{\lambda\mu\nu\rho} S_\rho \langle (v \cdot \nabla u_\mu)
u_\nu\rangle$ & $ g^4_A/3$ \\ \hline
\end{tabular}
$$
\end{table}

The renormalized low--energy constants $b_i^r(\mu)$ are measurable
quantities satisfying the renormalization group equations
\beq
\mu \frac{d}{d\mu} b_i^r(\mu) = - \beta_i
\eeq
implying
\beq
b_i^r(\mu) = b_i^r(\mu_0) - \beta_i \ln \frac{\mu}{\mu_0}~.
\eeq
It remains to extract the LECs $b_i^r(\mu)$ in a systematic way from
pion--nucleon data, together with the other scale independent
LECs of $O(p^3)$. Another important task will be to understand
the actual values of these parameters, in particular to investigate
systematically the effect of meson and baryon resonances.

\subsection{Pion--nucleon scattering}
\label{subsec:piN}
The pion--nucleon system was investigated in the
relativistic formulation of baryon CHPT by Gasser et al. (1988).
Recent years have witnessed the solution of the long--standing puzzle of the
$\pi N$ sigma term. Following recent summaries (Sainio, 1994; H\"ohler,
1994), I briefly review the present status of the $\sigma$--term.

The $\sigma$--term is related to the scalar form factor of the nucleon:
\beq
\sigma(t) \bar u(p') u(p) = \langle p'|\hat m(\bar u u + \bar d d)|p\rangle
\eeq
$$
\sigma := \sigma(0), \qquad t = (p - p')^2~.
$$
It can be rewritten in the form [with the normalization $\bar u(p) u(p)
= 2m$]
\beqa
\sigma &=& \frac{\hat \sigma}{1 - y} \\
\hat \sigma &=& \frac{\hat m}{2m} \langle p| \bar u u + \bar d d - 2 \bar s s
|p\rangle \\
y &=& \frac{2 \langle p|\bar s s|p\rangle}{\langle p| \bar u u + \bar d d
|p \rangle} ~,
\eeqa
with $y$ measuring the strange quark contribution to the nucleon mass.
Since $\hat \sigma$ involves the octet breaking piece in the
QCD Hamiltonian, one finds to first order in $SU(3)$ breaking
\beq
\hat \sigma \simeq \frac{\hat m}{m_s - \hat m} (m_\Xi + m_\Sigma - 2 m_N)
\simeq 26 \mbox{ MeV}
\eeq
with the standard values of quark mass ratios (cf. Sect.~\ref{subsec:mq}).
There are however significant corrections to $\hat \sigma$
of $O(p^3) = O(m_q^{3/2})$ (Gasser, 1981; Gasser and Leutwyler, 1982),
\beq
\hat \sigma = 35 \pm 5 \mbox{ MeV}~, \label{eq:sighat}
\eeq
based on the original calculation of Gasser (1981) employing an ultraviolet
cutoff. Recent one--loop calculations in CHPT (Dentin, 1990; B\"urgi,
1993) suggest somewhat bigger values: $\hat \sigma = 40 \div 50$~MeV. The
leading $m_s^{3/2}$ corrections from the meson octet and from both
octet and decuplet baryons (Jenkins 1992a; Jenkins and Manohar, 1992b)
yield $\hat \sigma \simeq 50$~MeV, if one calculates to leading order in
the octet--decuplet mass splitting $\Delta$. This procedure was criticized
by Bernard et al. (1993b) who claim that it does not lead to a
consistent picture of the scalar sector if one considers also the
$KN$ $\sigma$--terms (cf. Sect.~\ref{subsec:SU3}). Almost everybody
agrees that a full $O(p^4)$ calculation is necessary, including estimates of
the relevant LECs, to obtain a trustworthy value for $\hat \sigma$.

Before one can extract a value for the ratio $y$ or, equivalently, for the
strange quark contribution to the nucleon mass,
\beq
\Delta m(m_s) = \frac{m_s}{2 \hat m} (\sigma - \hat \sigma) \simeq
13 (\sigma - \hat \sigma)~, \label{eq:Dms}
\eeq
one has to relate $\sigma$ to the $\pi N$ scattering amplitude at the
unphysical Cheng--Dashen (1971) point $s = u$, $t = 2M_\pi^2$. The
corresponding quantity $\Sigma$ has been extracted from $\pi N$ phase
shifts by the Karlsruhe group (Koch, 1982; H\"ohler, 1994) and
by Gasser et al. (1991c) to give (Sainio, 1994)
\beq
\Sigma = 60 \pm 7 \mbox{ MeV}~.
\eeq

The final step consists in relating $\sigma$ to $\Sigma$:
\beq
\Sigma = \sigma + \Delta_\sigma + \Delta_R, \qquad
\Delta_\sigma = \sigma(2M_\pi^2) - \sigma(0)~,
\eeq
where the residual correction $\Delta_R \simeq 0.35$~MeV (Gasser et al.,
1988) is small. A main ingredient for the solution of the $\sigma$--term
puzzle is the dispersion theory result (Gasser et al., 1991c)
\beq
\Delta_\sigma = 15.2 \pm 0.4 \mbox{ MeV}~, \label{eq:Dsigma}
\eeq
much bigger than the one--loop result of about 5~MeV in the relativistic
formulation (Gasser et al., 1988) and about 7.5~MeV in HBCHPT
(Bernard et al., 1992f). This large increase is due to the strong
$S$--wave $\pi N$ amplitude in the $t$--channel. In HBCHPT, the difference
between (\ref{eq:Dsigma}) and the leading--order value can be
attributed to $\Delta$ exchange, with $K$-- and $\eta$--loop
contributions being small (Bernard et al., 1993b).
An interesting alternative method to determine $\Delta_\sigma$ has been
suggested by Bernard et al. (1994a). At $O(p^3)$, the scalar nucleon
form factor $\sigma(t)$ appears in one of the $S$--wave multipole
amplitudes for neutrino--induced pion production. Although an unknown
counterterm enters at the same order, chiral dimensional analysis suggests
that $\sigma(t)$ dominates the multipole in question in the threshold
region. It remains to be seen whether this multipole can be separated
experimentally and whether higher--order corrections are under control.

Putting everything together, one obtains
\beq
\sigma = 45 \pm 7 \mbox{ MeV}~. \label{eq:sigbest}
\eeq
Looking at Eq.~(\ref{eq:Dms}), one observes that the strange quark
contribution to $m$ is extremely sensitive to the error of
$\sigma - \hat \sigma$. Taking the mean values of $\hat \sigma$,
$\sigma$ from (\ref{eq:sighat}) and (\ref{eq:sigbest}) yields
\beq
y \simeq 0.2~, \qquad \Delta m(m_s) \simeq 130 \mbox{ MeV}~.
\eeq
In view of the considerable uncertainty of $\hat \sigma$ [probably
underestimated by the error in (\ref{eq:sighat})], $y$ and $\Delta m(m_s)$ are
consistent with zero. Whatever their
precise values may finally turn out to be, the $\sigma$--term problem
has been put to rest once and for all. From this observable, there
is no evidence for a large strange--quark component in the nucleon.

The $\pi N$ scattering amplitude in the forward direction is governed
by two invariant amplitudes $T^\pm(\omega)$ as a function of
$\omega = v \cdot q$, with $q$ the pion momentum. At threshold
($\omega = M_\pi$), the amplitudes can be expressed in terms of the
$S$--wave scattering lengths $a^\pm$:
\beq
a^\pm = \frac{1}{4\pi (1 + M_\pi/m)} T^\pm(M_\pi)~.
\eeq
The scattering lengths $a_{1/2}$, $a_{3/2}$ of definite isospin $I = 1/2,3/2$
are related to $a^\pm$:
\beq
a_{1/2} = a^+ + 2a^-, \qquad a_{3/2} = a^+ - a^-~.
\eeq
The old current algebra predictions (Weinberg, 1966; Tomozawa, 1966)
\beq
a^- = \frac{M_\pi}{8 \pi F_\pi^2} = 8.8 \cdot 10^{-2}/M_\pi~, \qquad
a^+ = 0
\eeq
are remarkably close to the experimental values (Koch, 1986)
\beq
a^- = 9.2 \pm 0.2~, \qquad a^+ = - 0.4 \pm 0.4 \label{eq:aexp}
\eeq
in units $10^{-2}/M_\pi$. It is then interesting to check whether this
agreement survives higher--order chiral corrections. Bernard et al. (1993c)
have calculated the $S$--wave scattering lengths to $O(p^3)$ in HBCHPT.
It turns out that $T^-(M_\pi)$ does not receive any corrections
of order $M_\pi^2$ and $M_\pi^4$. The loop contribution is divergent
entailing a scale dependent counterterm contribution. Fortunately, the
factor of $O(M_\pi^3)$ multiplying the LECs of $O(p^3)$ is comparatively
small. Estimating this contribution with $\Delta$ exchange and setting
the renormalization scale $\mu = m_\Delta$, Bernard et al. (1993c) find
to $O(p^3)$
\beq
a^- = (\underbrace{8.76}_{O(M_\pi)} + 0.40)
\cdot 10^{-2}/M_\pi = 9.16 \cdot 10^{-2}/M_\pi
\eeq
in impressive agreement with experiment. The situation is much less
favourable for $a^+$. There are contributions of both $O(M_\pi^2)$
and $O(M_\pi^3)$ to $T^+(M_\pi)$. In particular, the contribution
$c_2 + c_3 - 2c_1$ of the $O(p^2)$ LECs defined in (\ref{eq:A2}) enters
the leading contribution of $O(M_\pi^2)$. There is no real prediction, but
resonance exchange for the LECs yields a value of $a^+$ compatible
with the experimental result (\ref{eq:aexp}). The convergence of the
chiral expansion appears to be much slower for $a^+$ than for $a^-$.

One can turn the argument around and use $a^+$ to fix the combination
$c_2 + c_3 - 2c_1$ (Bernard et al., 1993c). In fact, $a^+$ has been
used as one of three observables to obtain the values of $c_1,c_2,c_3$
in Table \ref{tab:ci} (Mei\ss ner, 1994c; Bernard et al., 1995).
The LEC $c_3$ determines the $O(p^2)$ contribution to the so--called
axial polarizability $\alpha_A$ entering the crossing--even amplitude
$T^+$ away from the forward region. The chiral prediction to $O(p^3)$
(Bernard et al., 1995)
\beq
\alpha_A = - \frac{2c_3}{m F_\pi^2} - \frac{g_A^2 M_\pi}{8 \pi F_\pi^4}
\left( \frac{77}{48} + g_A^2\right)
\eeq
can be confronted with the experimental result \footnote{Following a
suggestion of Sainio, the error of $\alpha_A$ has been scaled up
 (Mei\ss ner, 1994a).} (H\"ohler, 1983)
$\alpha_A = (2.28 \pm 0.10) \cdot M_\pi^{-3}$ to extract the value for
$c_3$ given in Table \ref{tab:ci}. Finally, the $\pi N$ $\sigma$--term
is given to $O(p^3)$ in HBCHPT by (Bernard et al., 1992f)
\beq
\sigma = - \frac{4 M_\pi^2}{m} c_1 - \frac{9 g_A^2 M_\pi^3}
{64\pi F_\pi^2}~.
\eeq
The present best value (\ref{eq:sigbest}) for $\sigma$ implies the value
for $c_1$ shown in Table \ref{tab:ci}. Thus, the quantities $\sigma$,
$a^+$ and $\alpha_A$ determine the phenomenological values of the LECs
$c_1,c_2,c_3$.

Traditionally, the reactions $\pi N \ra \pi \pi N$ have been studied to get
information on $\pi \pi$ scattering by extrapolating to the pion
pole in the $t$--channel.
However, there is now a large sample of data in the threshold region
that should be analysed on their own merits [see Bernard et al. (1994b)
for the relevant references]. At threshold,
the processes $\pi^a N \ra \pi^b \pi^c N$ are described by the S--matrix
element
\beq
T_{a,bc} = i \vec \sigma \cdot \vec k ~[D_1(\tau^b \delta^{ac} + \tau^c
\delta^{ab}) + D_2 \tau^a \delta^{bc}]
\eeq
where $k$ is the momentum of the incoming pion. The invariant amplitudes
$D_1, D_2$ are related to the more commonly used amplitudes
$A_{2I,I_{\pi\pi}}$, where $I$ is the isospin of the initial state and
$I_{\pi \pi}$ the isospin of the produced two--pion system:
\beqa
A_{32} &=& \sqrt{10}\; D_1 \\
A_{10} &=& - 2D_1 - 3 D_2~.
\eeqa
The experimental values for these amplitudes are (Burkhardt
and Lowe, 1991)
\beq
A_{32} = 2.07 \pm 0.10 ~M_\pi^{-3}, \qquad
A_{10} = 6.55 \pm 0.16 ~M_\pi^{-3}~. \label{eq:Aexp}
\eeq
A more recent fit of near--threshold data only by Bernard et al.
(N. Kaiser, private communication) gives instead
\beq
A_{32} = 2.53 \pm 0.14 ~M_\pi^{-3}, \qquad
A_{10} = 8.01 \pm 0.64 ~M_\pi^{-3}~. \label{eq:ABKM}
\eeq
In the relativistic formulation, the amplitudes for two--pion production
were calculated by Beringer (1993) at tree level. Within HBCHPT, the
amplitudes $A_{32}, A_{10}$ were recently calculated to $O(p^2)$ by
Bernard et al. (1994b):
\beq
A_{32}  =  \frac{g_A \sqrt{10}}{8 F_\pi^3} \left(1 + \frac{7 M_\pi}{2m}
\right) + O(M_\pi^2) \label{eq:A32}
\eeq\beq
A_{10}  =  \frac{7 g_A}{8 F_\pi^3} \left(1 + \frac{37 M_\pi}{14 m}
\right) + O(M_\pi^2)~.\label{eq:A10}
\eeq
Only the kinematical part (\ref{eq:B2}) of the $O(p^2)$ Lagrangian
contributes. In other words, none of the LECs $c_i$ in (\ref{eq:A2})
appears in the amplitudes (\ref{eq:A32}), (\ref{eq:A10}). The numerical values
\beq
A_{32} = 2.7 ~M_\pi^{-3}, \qquad A_{10} = 5.5 ~M_\pi^{-3}
\eeq
compare reasonably well with the experimental results
(\ref{eq:Aexp}) or (\ref{eq:ABKM}). Note
that the corrections of $O(p^2)$ amount to about 50 and 40\% for the
amplitudes $A_{32}$ and $A_{10}$, respectively. One therefore has to
worry about the corrections of $O(p^3)$ corresponding to $O(M_\pi^2)$
in (\ref{eq:A32}), (\ref{eq:A10}). At that order, both loop and counterterm
amplitudes enter. The full $O(p^3)$ calculation is not yet available, but
Bernard et al. (1994b) have estimated the size of $O(p^3)$ corrections by
calculating the unambiguous absorptive parts of the one--loop diagrams.
Provided the absorptive parts are representative for the size of $O(p^3)$
corrections, Bernard et al. (1994b) expect $A_{32}$ to change only little
and $A_{10}$ by some 30\%. The physics behind these estimates is
related to the properties of the $\pi \pi$ final states: small
corrections for $I = 2$, but sizable ones for $I = 0$.

\subsection{Photo--nucleon reactions}
\label{subsec:phnuc}
As in the meson sector, reactions with real or virtual photons are
a fertile field for CHPT, with both chiral symmetry and electromagnetic
gauge invariance playing an important role. There are many experimental
and theoretical developments in this field, bringing together nuclear
and particle physicists. For an up--to--date account of these
activities, I refer to the Proceedings of a recent Workshop at MIT
(Bernstein and Holstein, 1995). Here, I want to review recent work
on nucleon Compton scattering and on single and double pion production
by real or virtual photons off nucleons.

The spin--averaged Compton amplitude \footnote{The spin--flip amplitude
was calculated to $O(p^3)$ in HBCHPT by Bernard et al. (1992f).} in the
forward direction (in the Coulomb gauge $\ve \cdot v = 0$) is given by
\beq
e^2 \ve^\mu \ve^\nu \frac{1}{4} \mbox{ tr } [(1 + \not\!v) T_{\mu\nu}
(v,k)] = e^2[\ve^2 U(\omega) + (\ve \cdot k)^2 V(\omega)]
\eeq
with $\omega = v \cdot k$ ($k$ is the photon momentum) and
\beq
T_{\mu\nu}(v,k) = \int d^4k \; e^{ikx} \langle N(v)|T j_\mu^{\rm em}(x)
j_\nu^{\rm em}(0)|N(v)\rangle~.
\eeq
The invariant amplitude $V(\omega)$ does not contribute for on--shell
photons, but it determines the magnetic polarizability
$\bar \beta$ via
\beq
\bar \beta = - \frac{e^2}{4 \pi} V(0)~.
\eeq
The electric polarizability $\bar \alpha$ is obtained from $U(\omega)$:
\beq
\bar \alpha + \bar \beta = - \left.\frac{e^2}{8 \pi} \frac{d^2U}{d \omega^2}
\right|_{\omega=0}~.
\eeq

Let me concentrate on the actual Compton amplitude $U(\omega)$. With
$D_F = 2$ external photons, the degree of homogeneity $D_L$ defined in
(\ref{eq:DF}) is obtained from (\ref{eq:DMB2}) as
\beq
D_L = 2L - 1 + \sum_d (d-2)N_d^M + \sum_d (d-1)N_d^{MB} \geq - 1~.
\eeq
Inserting the explicit dependence on $M_\pi$, the chiral expansion of
$U(\omega,M_\pi)$ takes the general form (Ecker and Mei\ss ner, 1994d)
\beq
U(\omega,M_\pi) = \frac{1}{\omega} f_{-1}(\omega/M_\pi) + f_0(\omega/M_\pi)
+ \sum_{D_L \geq 1} \omega^{D_L} f_{D_L}(\omega/M_\pi)~.
\label{eq:Uomega} \eeq
Only tree diagrams can contribute to the first two terms. Since the
relevant tree diagrams do not contain pion lines, $f_{-1}$ and $f_0$ must
actually be constants. Explicit calculation (Bernard et al., 1992f)
reproduces the Thomson limit ($Z$ is the nucleon charge in units of $e$)
\beq
f_{-1} = 0~, \qquad f_0 = U(0,M_\pi) = \frac{Z^2}{m}
\eeq
since (Ecker and Mei\ss ner, 1994d)
\beq
\lim_{\omega \ra 0} \omega^{n-1} f_n(\omega/M_\pi) = 0 \qquad
(n \geq 1)~.
\eeq
The amplitude $U(\omega)$ has been calculated to $O(p^4)$ corresponding to
$D_L = 2$ (Bernard et al., 1992f, 1994d) and was found to agree well with
experiment for $\omega \leq 110$~MeV.

Here, I concentrate on the polarizabilities $\bar \alpha,\bar \beta$.
{}From the representation (\ref{eq:Uomega}) and the corresponding one for
$V(\omega)$ one concludes that the
contribution of $O(p^n)$ $(D_L = n-2)$ to the polarizabilities is of
the form $c_n M_\pi^{n-4}$ $(n \geq 3)$ with constant $c_n$. At leading
order, $O(p^3)$, the polarizabilities are given by (Bernard et al., 1991a,
1992f)
\beqa
\bar \alpha_p = \bar \alpha_n &=& \frac{5 e^2 g_A^2}{384 \pi^2 F_\pi^2
M_\pi} = 12 \cdot 10^{-4} \mbox{ fm}^3 \\
\bar \beta_p = \bar \beta_n &=& \frac{\bar \alpha_p}{10} =
1.2 \cdot 10^{-4} \mbox{ fm}^3~.
\eeqa
Comparison with the experimental values in Table \ref{tab:polar} shows
that the long--ranged pion cloud clearly accounts for the main features
of the data. The corresponding calculation in the three--flavour case
(Bernard et al., 1992g; Butler and Savage, 1992) generates in particular
a small positive
contribution $\bar \alpha_p - \bar \alpha_n$ due to the kaon cloud
around the nucleons. The decuplet contribution that
comes in at $O(p^4)$ interferes destructively with the kaon contribution,
but the overall result (with a sizable error) for $\bar \alpha_p -
\bar \alpha_n$ still comes out positive (Butler and Savage, 1992).

Following their general philosophy of including the $\Delta$ via
its contribution to LECs, Bernard et al. (1993d, 1994d) have
performed a full calculation to $O(p^4)$ in chiral $SU(2)$.
In this case, both the kaons and the decuplet baryons generate effects
of $O(p^4)$, but there are additional ones. The one--loop diagrams with
a single vertex from (\ref{eq:A2}) or (\ref{eq:B2}) are well under
control as far as the couplings $c_1,c_2,c_3$ are concerned that appear
in the amplitudes (cf. Sect.~\ref{subsec:HBCHPT}). The difficulty
is to estimate the LECs of $O(p^4)$ that also renormalize the divergent
loop contributions with a vertex of $O(p^2)$. At this stage, both the
$\Delta$ resonance and the kaons enter the game and their effect can
be estimated as in the $SU(3)$ calculation (Butler and Savage, 1992;
Bernard et al., 1992g). Estimating the uncertainties arising to
$O(p^4)$ (but not accounting for higher--order corrections), Bernard
et al. (1993d, 1994d) arrive at the polarizabilities displayed in Table
\ref{tab:polar}. With the possible exception of $\bar\beta_n$, the
agreement between theory and experiment is very good. The influence of
other than $\Delta$ and $K$ contributions can for instance be seen in
the difference $\bar \alpha_p - \bar \alpha_n$ which, albeit with a big
error, turns out to be negative in the calculation of Bernard et al.
Calculations have also been performed for the electromagnetic
polarizabilities of hyperons (Bernard et al., 1992g).
\begin{table}
\caption{Comparison between theory (Bernard et al., 1993d, 1994d) and
experiment (Nathan, 1994) for the electromagnetic polarizabilities of the
nucleons in units of $10^{-4}$~fm$^3$. The experimental errors
are anticorrelated because in the analysis a model independent dispersion sum
rule (Baldin, 1960; Damashek and Gilman, 1970) is used that determines
$\bar\alpha + \bar\beta$ with a relatively small error.}
\label{tab:polar}
$$
\begin{tabular}{|c|c|c|} \hline
polarizability & CHPT to $O(p^4)$ & data \\ \hline
$\bar \alpha_p$ & $10.5 \pm 2.0$ & $\qquad 12.0 \pm 0.8 \pm 0.4\qquad $ \\
$\bar \beta_p$  & $3.5 \pm 3.6$  & $2.2 \mp 0.8 \mp 0.4$ \\
$\bar \alpha_n$ & $13.4 \pm 1.5$ & $12.5 \pm 1.5 \pm 2.0$ \\
$\bar \beta_n$   & $7.8 \pm 3.6$  & $3.5 \mp 1.8 \mp 2.0$ \\ \hline
\end{tabular}
$$
\end{table}

Single pion photo-- and electroproduction off nucleons has been
the focus of intensive experimental and theoretical work. For recent
summaries of the experimental situation, see Walcher (1994) and
Mei\ss ner and Schoch (1994b). I will first summarize recent
theoretical developments for $\pi^0$ production, following
the talk of Bernard (1994g) at the MIT Workshop on Chiral Dynamics.

Near threshold, the photoproduction amplitude in the center--of--mass system
can be described by the
leading multipoles, the $S$--wave $E_{0+}$ and three $P$--waves
$E_{1+}$, $M_{1+}$, $M_{1-}$:
\beq
\frac{m}{4 \pi \sqrt{s}}~T \cdot \ve = i \vec \sigma \cdot \vec \ve~
(E_{0+} + \hat k \cdot \hat q P_1) + i \vec \sigma \cdot \hat k ~\vec \ve
\cdot \hat q P_2 + (\hat q \times \hat k) \cdot \vec \ve P_3
\eeq
$$
\gamma(k) + N(p_1) \ra \pi^0(q) + N(p_2)~.
$$
The $P_i$ $(i = 1,2,3)$ are linear combinations of the $P$--wave
multipoles and $\hat k$, $\hat q$ are 3--dimensional unit vectors.
Much of the excitement in $\pi^0$ photoproduction has
concentrated on the $S$--wave $E_{0+}$. After a reanalysis of the new
data (Mazzucato et al., 1986; Beck et al., 1990), the experimentally
extracted value for $E_{0+}$ seemed to agree with an old
prediction of Vainshtein and Zakharov (1970, 1972) and de Baenst (1970).
It therefore came as a  surprise when Bernard et al. (1991b) showed
that the old prediction was in conflict with the Standard Model to $O(p^3)$ in
the chiral expansion [corresponding to $O(M_\pi^2)$ for $E_{0+}$ at
threshold]. This calculation in the relativistic formulation also
indicated that the chiral expansion of $E_{0+}$ seems to
converge very slowly. The recent calculation of $E_{0+}$ to $O(p^4)$
(Bernard et al., 1994h) confirms this indication making $E_{0+}$ rather
unsuitable for testing the Standard Model.

\begin{figure}
    \begin{center}
       \setlength{\unitlength}{1truecm}
       \begin{picture}(8.0,8.0)
       \put(-2.2,-3.8){\includegraphics{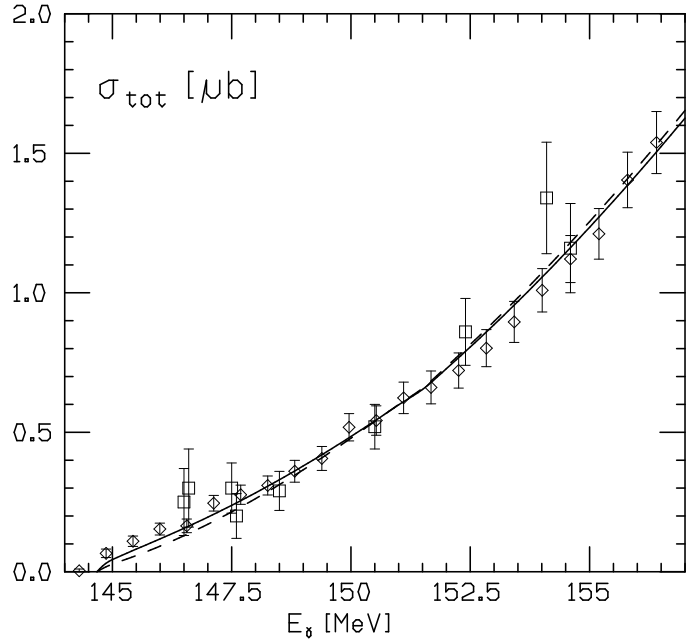}}
       \end{picture}
    \end{center}
\caption{Total cross section for $\gamma p \to \pi^0 p$ as
function of the photon energy $E_\gamma$ in the lab frame. The two
curves correspond to different treatments of LECs (Bernard et
al., 1994h). The data are from Mazzucato et al. (1986, squares) and
Beck et al. (1990, diamonds).} \label{fig:photo}
\end{figure}

On the other hand, as shown in Fig.~\ref{fig:photo}, CHPT accounts very
well for the total cross section for $\gamma p \ra \pi^0 p$ for small
photon energies. The reason is that, except very close to threshold
where $E_{0+}$ must dominate on kinematical grounds, the cross section
is mainly sensitive to the $P$--waves. It turns out that the chiral
expansion, to $O(p^3)$ at least, is much better behaved for the
$P$--waves than for $E_{0+}$.
Up to and including $O(p^3)$, $E_{0+}$, $P_1$ and $P_2$ depend only on
the anomalous magnetic moments $\kappa_p, \kappa_n$ (related to
$c_6,c_7$), but neither on the other LECs of $O(p^2)$ in
(\ref{eq:A2}) nor on any LECs of $O(p^3)$. On the other hand, $P_3$
depends on a LEC of $O(p^3)$ that can be estimated with resonance exchange
(Bernard et al., 1994h). It is actually completely dominated by this
LEC because the Born term contribution is very small and loops do not
contribute at this order. On the other hand, loops do contribute to
$E_{0+}$, $P_1$, $P_2$. Since no LECs of $O(p^3)$ appear in the loop
amplitudes, those amplitudes are necessarily finite. Referring to
Bernard et al. (1994h) for the complete multipoles as functions of
$\omega$ (the center--of--mass energy of the $\pi^0$), I reproduce here
the expansions of $P_i/\sqrt{\omega^2 - M_{\pi^0}^2}$ at threshold in
powers of $M_\pi/m$ (for the proton case only):
\beqa
\left. \frac{P_1}{|\vec q|} \right|_{\rm thresh} &=&
\frac{e g_A}{8 \pi F_\pi m} \left\{1 + \kappa_p + \frac{M_\pi}{m}
\left[ -1 - \frac{\kappa_p}{2} + \frac{g_A^2 m^2(10 - 3\pi)}{48\pi F_\pi^2}
\right] + O(M_\pi^2)\right\} \\
\left. \frac{P_2}{|\vec q|} \right|_{\rm thresh} &=&
\frac{e g_A}{8 \pi F_\pi m} \left\{-1 - \kappa_p + \frac{M_\pi}{2m}
\left[3 + \kappa_p - \frac{g_A^2 m^2}{12\pi F_\pi^2}
\right] + O(M_\pi^2)\right\}~.
\eeqa
Unlike for $E_{0+}$, the corrections of $O(M_\pi)$ are quite small for the
$P_i$ $(i = 1,2)$ at threshold: 6\% for $P_1$ and less than 0.1\% for
$P_2$. Up to unknown higher--order corrections, the $P$--wave multipoles
$P_1$ and $P_2$ seem therefore much better suited than $E_{0+}$ for
testing chiral dynamics. In addition to their intrinsic interest, the
CHPT predictions for $P_1,P_2$ can be used to constrain the data when
trying to extract the elusive $E_{0+}$. For instance, the non--vanishing
combination $P_1 + P_2$ at next--to--leading order implies that the
multipole $E_{1+}$ is non--zero at this order. However, this multipole
has commonly been set to zero for analysing the threshold data.

New data are also forthcoming for electroproduction of pions. In
Fig.~\ref{fig:electro}, preliminary data for $\pi^0$ electroproduction
on protons  for $k^2 = - 0.1$~GeV$^2$ (Walcher, 1994; Blomqvist et al., 1995;
Distler, 1995) are compared with
a CHPT calculation of $O(p^3)$ (Bernard et al., 1993a, 1994e). The angular
distribution in Fig.~\ref{fig:electro} clearly favours CHPT over a
pseudovector Born term model.

\begin{figure}
\centerline{\epsfig{file=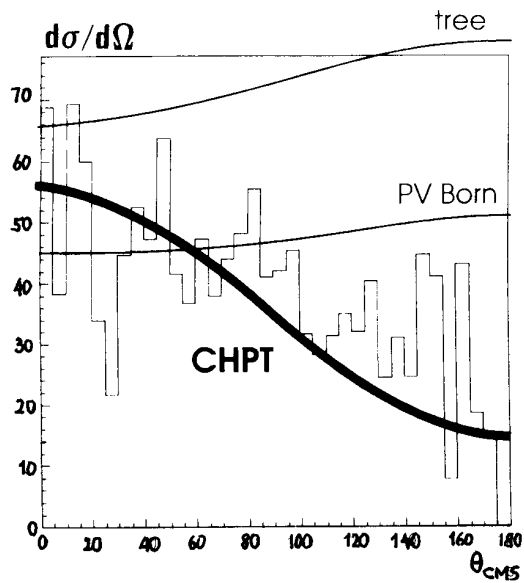,height=8cm}}
\caption{Differential cross section for $\pi^0$
electroproduction on protons for $k^2 = - 0.1$~GeV$^2$ (Walcher,
1994; Blomqvist et al., 1995; Distler, 1995) in comparison with
the CHPT prediction (Bernard et al., 1993a, 1994e).}
\label{fig:electro}
\end{figure}

Another example of a successful explanation of experimental data by CHPT
is charged pion electroproduction. A measurement of both $\gamma^* p \ra
\pi^+ n$ and $\gamma^* n \ra \pi^- p$ allows to extract the
isospin--odd dipole amplitude $E_{0+}^{(-)}$. According to a low--energy
theorem (Nambu and Luri\'e, 1962a; Nambu and Shrauner, 1962b), this amplitude
depends on the axial radius of the nucleon, $r_A$:
\beq
E_{0+}^{(-)}(M_\pi = 0,k^2) = \frac{e g_A}{8\pi F_\pi}
\left\{ 1 + \frac{k^2}{6} r_A^2 + \frac{k^2}{4m^2} \left(\kappa_v +
\frac{1}{2}\right) + O(k^4)\right\}~.\label{eq:E0p}
\eeq
It turns out that the axial radius extracted from electroproduction
$(r_A = 0.59 \pm 0.04$~fm) is a little smaller than the value found
in neutrino nucleon reactions $(r_A = 0.65 \pm 0.03$~fm). Although
the values differ only by about one standard deviation, it is comforting
that chiral corrections for $E_{0+}^{(-)}$ go in the right direction
to bring them into agreement. The leading chiral correction due to
pion loops (Bernard et al., 1992e) amounts to a shift of $r_A^2$ in
(\ref{eq:E0p}),
\beq
r_A^2 \ra r_A^2 + \frac{3}{64 F_\pi^2} \left(1 - \frac{12}{\pi^2}\right)
= r_A^2 - 0.046\mbox{ fm}^2~,
\eeq
closing the gap between electro-- and neutrino--production.

Finally, Bernard et al. (1994f) have started an investigation of two--pion
photoproduction at threshold. The scattering amplitude at threshold can
be expressed in terms of three multipoles $M_1,M_2,M_3$. As usual,
CHPT gives rise to an expansion in $M_\pi$ for these multipoles. At
leading order, $O(p^2)$, one finds (Bernard et al., 1994f)
\beq
M_2 = - 2M_3 = \frac{e}{4m F_\pi^2} (2g_A^2 - 1 - \kappa_v)~, \qquad
M_1 = 0~.
\eeq
The contribution proportional to the isovector magnetic moment $\kappa_v$
was overlooked in earlier treatments. The leading--order results imply
that the amplitude for $\pi^0\pi^0$ production vanishes at this order for both
proton and neutron targets ($A_p \sim M_1 + M_2 + 2M_3$,
$A_n \sim M_1 - M_2 - 2M_3)$. At $O(p^3)$, Bernard et al. (1994f) obtain
a simple result for $M_1$,
\beq
M_1 = \frac{e g_A^2 M_\pi}{4 m^2 F_\pi^2}~,
\eeq
while $M_2,M_3$ receive both loop and counterterm contributions, in
particular due to $\Delta$ exchange. However, $2\pi^0$ production is
not affected by $\Delta$ exchange at this order. Using the correct
phase--space and approximating the amplitudes near threshold
essentially by their threshold values, one finds a surprisingly big
cross section for the $2\pi^0$ case. It is not only the biggest cross
section of all $2\pi$ channels (both for $p$ and $n$ targets), but
there is in addition a kinematic window of $10 \div 12$~MeV above
threshold where practically only $2\pi^0$ should be produced.
There is preliminary evidence from an experiment at
Mainz (Walcher, 1994) for $2\pi^0$ events close to threshold.

\subsection{Chiral $SU(3)$}
\label{subsec:SU3}
The mesonic chiral Lagrangian of lowest order (\ref{eq:L2}) has the same
form for $N_f = 2$ and 3. In the meson--baryon sector, the lowest--order
Lagrangians (\ref{eq:piN1}) and (\ref{eq:MB1}) are different. There is a
group theoretical reason for this difference: whereas the mesons are
always in the adjoint representation of $SU(N_f)$, the nucleons are in
the fundamental representation of $SU(2)$, but the baryons are in the
adjoint representation of $SU(3)$.

Most of the work in the three--flavour case has concentrated on the
so--called non--analytic chiral corrections. In other words, there is
in general no systematic discussion of the LECs of a given chiral order.
Since it would be totally misleading to disregard the LECs completely,
most authors have followed Jenkins and Manohar (1991a) by including the
decuplet baryons as dynamical fields in the effective Lagrangian. The
motivation is at least two--fold: the octet--decuplet mass difference
$\Delta$ is smaller than $M_K,M_\eta$ and it vanishes in the large--$N_c$
limit (Witten, 1979). As shown by Dashen et al. (1994a,b, and references
therein), many approximate relations in the baryon system can be
understood by a systematic expansion in $1/N_c$. For that purpose, one
must include the whole tower of states that become degenerate for
$N_c \ra \infty$, in particular the octet and decuplet of baryons.

A good example for this mechanism is provided by the baryon axial--vector
currents. In the original CHPT calculation of Jenkins and Manohar
(1991b), large cancellations between the octet and decuplet contributions
to the leading non--analytic corrections of $O(M^2 \ln M^2)$ were
found. A priori, the octet contributions alone grow with $N_c$ because
the pion--baryon couplings diverge like $\sqrt{N_c}$. However, the
total $M^2 \ln M^2$ correction is of $O(1/N_c)$ (Dashen et al., 1994a)
as in the meson sector. The level of accuracy of such large--$N_c$
relations can be estimated from a comparison of the $f/d$ ratio of
the baryon axial currents (Dashen et al., 1994a) [cf. Eq.~(\ref{eq:MB1})],
\beq
\frac{f}{d} = \frac{2}{3} + O(1/N_c^2)~, \label{eq:fdratio}
\eeq
with the experimental value $f/d = 0.58 \pm 0.04$ (Jaffe and Manohar,
1990). The same large--$N_c$ relation applies for the $f/d$ ratio of
the baryon magnetic moments (Dashen et al., 1994a; Jenkins and
Manohar, 1994), where the experimental value is 0.72.

The large--$N_c$ expansion is a very useful tool that should be combined
with the chiral expansion but it is certainly not a substitute for it.
In the original calculation of Jenkins and Manohar (1991b) for the
baryon axial--vector form factors, only the leading terms in
$\Delta/M_K$ and $\Delta/M_\eta$ were kept. Keeping all terms, Luty
and White (1993b) confirm the cancellation between octet and decuplet
contributions, but they find that the $SU(6)$ values of the axial couplings
such as (\ref{eq:fdratio}) are strongly disfavoured when
$M^2 \ln M^2$ corrections are included. The conclusion is that one
cannot reliably calculate $f$ and $d$ from the data with the non--analytic
corrections only. The corrections of $O(M^2)$, in particular due to LECs,
cannot be neglected.

The situation is similar for the baryon masses. Here, the leading
non--analytic corrections are of $O(M^3) = O(m_q^{3/2})$. In the
large--$N_c$ counting (Dashen et al., 1994a), the terms of $O(N_c)$
are pure $SU(3)$ singlet, $O(1)$ is octet, so that the first
``non--trivial'' corrections are $O(1/N_c)$. Thus, corrections to
Gell-Mann--Okubo relations and other mass relations derived from broken
$SU(3)$ are actually of order $m_s^{3/2}/N_c$. Consequently, the baryon
masses can be strongly non--linear functions of $m_s$ and still satisfy
the Gell-Mann--Okubo formula. One can give a list of baryon mass relations
that hold up to corrections of $O(1/N_c^2)$ (Dashen et al., 1994a;
Manohar, 1994). Those relations that can also be obtained from broken
$SU(3)$ (octet symmetry breaking) work extremely well because effects
violating those relations must break both symmetries.

As for the axial currents, the non--analytic corrections to the
baryon masses are not the whole story, however. Bernard et al. (1993b)
have reexamined the issue by including
pion-- and kaon--nucleon sigma terms in their analysis. In her
calculation, Jenkins (1992a) had included the decuplet contributions
to the baryon masses to leading order in $\Delta$ and found a consistent
picture, albeit for rather small values of $f$ and $d$. Although taking
into account all orders in $\Delta$ does not change the picture
drastically, Bernard et al. (1993b) argue that the small values of $f$
and $d$ in the fit of Jenkins (1992a) indicate that other
contributions to the masses of $O(p^4)$ and higher are not negligible.
Chiral corrections for baryon masses have also been calculated by
Mallik (1994) and by Banerjee and Milana (1994).

Recently, Lebed and Luty (1994) have considered corrections of
$O(m_s^2)$ and $O(e^2)$ to the baryon masses. As discussed in
Sect.~\ref{subsec:mq}, a calculation of baryon mass differences
to $O(m_q^{3/2})$ leads to the value (\ref{eq:Rq}) for the quark mass
ratio $R$. In principle, terms of $O(e^2p^0)$ in the meson--baryon
Lagrangian of the form (Lebed and Luty, 1994)
\beq
\cL_{B (e^2p^0)} = d_1 \frac{e^2 \Lambda_\chi}{(4\pi)^2}
\langle \bar B Q_+^2 B\rangle + d_2 \frac{e^2 \Lambda_\chi}{(4\pi)^2}
\langle \bar B B Q_+^2\rangle + \ldots
\eeq
$$
Q_+ = \frac{1}{2} (u Q u^\dg + u^\dg Q u)~, \qquad \Lambda_\chi \simeq
1\mbox{ GeV}
$$
induce a shift in the value of $R$. An estimate of $d_1,d_2$ is
necessary for a more quantitative statement. Lebed and Luty (1994) also
show that there are no local contributions of $O(e^2)$ and $O(m_q^2)$
to the Coleman--Glashow relation (Coleman and Glashow, 1961):
\beq
\Delta_{CG} := \Sigma^+ - \Sigma^- + n - p + \Xi^- - \Xi^0 =
O(\hat m m_s^2) + O(e^2 m_s)~.
\eeq
The particle names denote the corresponding masses and the corrections
refer to local contributions from LECs only. Loop diagrams give
calculable contributions to $\Delta_{CG}$ of $O[(m_u - m_d)m_s]$.
Including decuplet intermediate states in the loops (there is again a
substantial cancellation between octet and decuplet contributions),
Lebed and Luty (1994) obtain from the loops
$\Delta_{CG}^{\rm (theory)} = 0.2 \pm 0.7$~MeV, in agreement with the
experimental value $\Delta_{CG}^{\rm (exp)} = - 0.3 \pm 0.6$~MeV.
On the other hand, the loop corrections to the $\Sigma$ equal--spacing
rule (Coleman and Glashow, 1961)
\beq
\Delta_\Sigma := \Sigma^+ - \Sigma^0 - (\Sigma^0 - \Sigma^-) = 0
\eeq
are of $O(\hat m^2)$ and numerically negligible compared to the experimental
value $\Delta_\Sigma = 1.7 \pm 0.2$~MeV. Here, local terms of $O(e^2)$
contribute and give at least the right order of magnitude.

A CHPT analysis of baryon magnetic moments was performed by Jenkins et al.
(1993a). Seven of the octet moments and the transition moment $\Sigma^0 \ra
\Lambda \gamma$ have been measured. To lowest order, $O(p^2)$, those
moments depend on two couplings $\mu_d,\mu_f$. Without the
large--$N_c$ relation (\ref{eq:fdratio}), there are six linear
relations among the measured moments (Coleman and Glashow, 1961).
Altogether, they are not too well satisfied.
The leading corrections of $O(p^3)$ contain non--analytic corrections
of $O(m_q^{1/2})$ for the moments. Including again the decuplet
contributions, one obtains three linear relations independent of the
coupling constants (Jenkins et al., 1993a), first written down
by Caldi and Pagels (1974) (the experimental values are in units of
nuclear magnetons):
\beqa
2.42 \pm 0.05 = & \mu_{\Sigma^+} = - 2 \mu_\Lambda - \mu_{\Sigma^-} & =
2.39 \pm 0.03 \nl
- 3.81 \pm 0.01 = & \mu_{\Xi^0} + \mu_{\Xi^-} + \mu_n = 2\mu_\Lambda
- \mu_p & =  - 4.02 \pm 0.01 \\
- 3.40 \pm 0.14  = & \mu_\Lambda - \sqrt{3} \; \mu_{\Lambda\Sigma^0} =
\mu_{\Xi^0} + \mu_n & =  - 3.16 \pm 0.01~.\no
\eeqa
These relations are in good agreement with experiment and they
work much better than the tree--level relations.
The remaining three relations depend on the axial meson--baryon couplings.
There is no cancellation between octet and decuplet contributions in this
case and the three additional relations are not fulfilled experimentally.
The baryon magnetic moments are a rather clean case where the leading
non--analytic chiral corrections are not sufficient to describe
all the data. The static electromagnetic moments of the decuplet
have also been computed in CHPT (Butler et al., 1994).

The kaon--nucleon scattering lengths have recently been
calculated by various groups (Lee et al., 1994a, 1994b;
Savage, 1994). The off--shell $KN$ scattering amplitude is of interest
for kaonic atoms and kaon condensation in dense nuclear matter
(e.g., Brown et al., 1994; Lee et al., 1994c).

The nonleptonic weak decays of hyperons are another interesting area of
application for CHPT with three light flavours. Both the non--radiative
decays $B_i \ra B_f \pi$ and the radiative decays $B_i \ra B_f \gamma$
have been studied in the CHPT framework.

The seven measurable nonleptonic hyperon decays $B_i \ra B_f \pi$ are
described by matrix elements
\beq
M(B_i \ra B_f\pi) = G_F M_{\pi^+}^2 \bar u_f (S_{if} + P_{if}
\gamma_5) u_i
\eeq
with $S$ $(P)$--wave amplitudes $S_{if}$ $(P_{if})$. Neglecting the
27--plet part of the nonleptonic weak Hamiltonian (\ref{eq:Hnl}), the
lowest--order $|\Delta S|= 1$ Lagrangian bilinear in the baryon fields
contains two terms (Georgi, 1984; Manohar and Georgi, 1984):
\beq
\cL_{B,0}^{|\Delta S|=1} = w_+ \langle \bar B \{ u \lambda u^\dg,B\}\rangle
+ w_- \langle \bar B [u \lambda u^\dg,B]\rangle + {\rm h.c.}~, \qquad
\lambda = \frac{1}{2} (\lambda_6 - i \lambda_7)~. \label{eq:B0}
\eeq
There are three isospin relations among the amplitudes $S$, $P$ and
thus four independent amplitudes (separately for $S$ and $P$).
The $S$--wave amplitudes of lowest order are tree--level amplitudes from
(\ref{eq:B0}). Eliminating the couplings $w_+,w_-$ in the four independent
$S$--waves leads to $S(\Sigma^+ \ra n \pi^+) = 0$ and to the
Lee--Sugawara relation (Lee, 1964; Sugawara, 1964)
\beq
S(\Lambda \ra p \pi^-) + 2 S(\Xi^- \ra \Lambda \pi^-) +
\sqrt{\frac{3}{2}}\; S(\Sigma^- \ra n \pi^-) = 0~.\label{eq:LS}
\eeq
Both relations agree reasonably well with experiment. The $P$--wave
amplitudes are given by pole diagrams to lowest order and involve
in addition to $w_+,w_-$ the strong meson--baryon couplings $f,d$ in
(\ref{eq:MB1}) and octet baryon mass differences. It is a long--standing
problem that the $P$--waves disagree strongly with experiment if one uses the
values of $w_+,w_-$ extracted from the $S$--waves (Georgi, 1984;
Manohar and Georgi, 1984).

The leading non--analytic corrections to these amplitudes were calculated
by Jenkins (1992d) within HBCHPT including the decuplet contributions.
For the $S$--wave amplitudes, the decuplet contributions are bigger than
the octet ones in all cases and the results are in good agreement with
experiment. Looking more closely, one finds that the experimentally
observed deviation from the Lee--Sugawara relation (\ref{eq:LS}) seems
to require a big decuplet contribution because the octet part alone
gives the wrong sign. The situation is less obvious for
$S(\Sigma^+ \ra n \pi^+)$, but the amplitude is small anyway.

The non--analytic corrections do not improve the situation for the
$P$--waves. Since the tree--level amplitudes consist in all seven
cases of two terms that interfere destructively, the one--loop
corrections are of the same magnitude as the leading terms
(Jenkins, 1992d). In addition to the non--analytic corrections of
$O(m_s \ln m_s)$, the terms of $O(m_s)$ coming from both
loops and counterterms must be taken into account. There are eight
possible terms of $O(p)$ with unknown LECs (Neufeld, 1993). To get an
idea of the order of magnitude of those LECs, Neufeld employed the
``weak deformation model'' (Ecker et al., 1990a) mentioned in
Sect.~\ref{subsec:Ni}. He finds that the resulting contributions
to the $S$--wave amplitudes are about 30\% of the measured values,
but of the same size as the experimental values for the $P$--waves.
Without progress towards understanding the LECs of $O(p)$, the
amplitudes $P_{if}$ are not predictable in CHPT.

The weak radiative baryon decays $B_i(p_i) \ra B_f(p_f) + \gamma(q)$
are described by matrix elements
\beq
M(B_i \ra B_f \gamma) = i \bar u (p_f) \sigma^{\mu\nu} (A_{if} +
B_{if} \gamma_5) u(p_i) \ve_\mu^*(q) q_\nu
\eeq
with parity conserving (violating) amplitudes $A_{if}$ $(B_{if})$.
They were calculated by Neufeld (1993) in the relativistic formulation
and by Jenkins et al. (1993c) in HBCHPT to one--loop accuracy.
All loop amplitudes turn out to be finite. However,
at the same order in the chiral expansion, there are a number of local
couplings of $O(p^2)$. From the most general such Lagrangian,
one finds two relations for $A_{if}$ and five relations for $B_{if}$
that had been derived before under stronger assumptions (Hara, 1964;
Lo, 1965).

There are six measurable radiative decays, five of which have been
observed: $\Sigma^+ \ra p \gamma$, $\Lambda \ra n \gamma$, $\Xi^- \ra
\Sigma^- \gamma$, $\Xi^0 \ra \Sigma^0 \gamma$ and $\Xi^0 \ra \Lambda
\gamma$. In addition to the decay rates, the asymmetry parameter
\beq
\alpha = \frac{2\mbox{ Re }(AB^*)}{|A|^2 + |B|^2}
\eeq
has been measured for $\Sigma^+ \ra p \gamma$, $\Xi^0 \ra \Sigma^0 \gamma$
and $\Xi^0 \ra \Lambda \gamma$. The one--loop diagrams for the radiative
decay amplitudes contain vertices that describe the previously discussed
non--radiative decays at tree level. The parity violating amplitudes
$B_{if}$ are related to the $S$--wave amplitudes $S_{if}$, while the
parity conserving $A_{if}$ are sensitive to $P_{if}$. Since the
tree--level amplitudes for $P_{if}$ are nowhere near the data, it
clearly does not make much sense to use the corresponding vertices to
calculate the $A_{if}$ at the one--loop level. Therefore,
the following strategy was employed (Neufeld, 1993; Jenkins et
al., 1993c): calculate $B_{if}^{\rm (loop)}$
with the couplings $w_+,w_-$ in (\ref{eq:B0}), but use the measured
$P$--waves $P_{if}$ for the calculation of $\Im m A_{if}^{\rm (loop)}$.
Leaving $\Re e A_{if}$ and $B_{if}^{\rm (counter)}$ as free parameters,
one can establish allowed domains for the rates and asymmetry
parameters.

For $\Xi^0 \ra \Sigma^0 \gamma$ and $\Xi^0 \ra \Lambda \gamma$, the
theoretical constraints are compatible with experiment. On the other
hand, even such a general analysis cannot explain the data for
$\Sigma^+ \ra p \gamma$: the asymmetry parameter $\alpha$ is constrained
in theory to be small while experiments find $\alpha = -0.76 \pm 0.08$
(Review Part. Prop., 1994). It turns out that the unambiguously
calculable absorptive parts are much smaller than the experimental
amplitudes so that the most important contributions must come from other
than the one--loop diagrams. It has been suggested [cf. Neufeld (1993)
and Jenkins et al. (1993c) for the relevant literature] that
$\frac{1}{2}^-$ intermediate hyperon
states are mostly responsible for $B(\Sigma^+ \ra p \gamma)$. Finally,
predictions can be made for the asymmetry parameters in the decays
$\Xi^- \ra \Sigma^- \gamma$, $\Lambda \ra n \gamma$. For an update of
those predictions, taking into account new experimental information,
I refer to a forthcoming article by Neufeld (1995).

\vspace{2.5cm}

\noindent {\bf Acknowledgements}

\noindent
For general instruction and helpful comments and/or for
providing me with partly unpublished material including some
of the figures, I am especially grateful to V. Bernard, J. Bijnens,
G. Colangelo, G. D'Ambrosio, J. Gasser, G. H\"ohler,
G. Isidori, J. Kambor, H. Leutwyler, U.-G. Mei\ss ner, A.M. Nathan,
H. Neufeld, A. Pich, E. de Rafael, M. Sainio, J. Stern
and T. Walcher. I am also indebted to M. Moi\v zi\v s for checking
some of the calculations in Sect.~\ref{subsec:piNren}. Finally, I
want to thank N. Kaiser, U.-G. Mei\ss ner, M. Moi\v zi\v s and H. Neufeld for
suggesting improvements of the manuscript.

\end{document}